\documentclass{aa}  

\usepackage{graphicx}
\usepackage{txfonts}
\usepackage{hyperref}
\hypersetup{
	breaklinks=true,
	colorlinks=true,
	linkcolor=red, 
	urlcolor=blue,
	citecolor=blue,
	bookmarks=true,
}

\begin{document}

\title{V4142\,Sgr: a Double Periodic Variable with an accretor surrounded by the accretion-disk's atmosphere}

%\subtitle{}

\author{Rosales, J. A.\inst{1,2,3} \and Mennickent, R.\,E. \inst{2} \and Djura\v{s}evi\'{c}, G. \inst{4,5} \and	Araya, I. \inst{6} \and	Cur\'{e}, M. \inst{7} \and Schleicher, D.\,R.\,G. \inst{2} \and Petrovi\'{c}, J.\inst{4}}

\institute{
	Main Astronomical Observatory, National Academy of Sciences of Ukraine, 27 Akademika Zabolotnoho St, 03680 Kyiv, Ukraine.
	\and
	Departamento de Astronom\'{i}a, Universidad de Concepci\'{o}n, Casilla 160-C, Concepci\'{o}n, Chile.
	\and
	Department of Physics, North-West University, Private Bag X2046, Mmabatho 2735, South Africa.
	\and
	Astronomical Observatory, Volgina 7, 11060 Belgrade 38, Serbia.
	\and
	Isaac Newton Institute of Chile, Yugoslavia Branch, 11060 Belgrade, Serbia.
	\and
    Centro de Investigación DAiTA Lab, Facultad de Estudios Interdisciplinarios, Universidad Mayor, Alonso de Córdova 5495, Santiago, Chile.
	\and
	Instituto de F\'{i}sica y Astronom\'{i}a, Facultad de Ciencias, Universidad de Valpara\'{i}so, Chile.
	}

\date{Received December XX, 2021; accepted December XX, 2022}

% \abstract{}{}{}{}{} 
% 5 {} token are mandatory
 
  \abstract
   {A detailed study of the close interacting binary V4142\,Sgr based on photometric and spectroscopic analysis is presented.This system belongs to the enigmatic class of Algol-like variables showing a long photometric cycle of unknown nature.}
   {Performing photometric data-mining and spectroscopic observations covering the orbital cycle, we obtain the orbital parameters and the stellar properties of the binary system, along with the physical properties of the
   accretion disk located around the hot star. Insights on the evolutive path of the system are obtained.
   %Modeling the V-band light curve and performing a fitting method based on an iterative simple algorithm, which includes the donor and gainer stars and an accretion disk, we obtain the fundamental parameters of the system and the accretion disk located around the hot star, together with the normalized flux contribution of each stellar component. In addition, from a collection of 16 spectra obtained with the CHIRON spectrograph, covering an orbital cycle, and through the combined use of these spectra, the radial velocities were measured and separated from the composed spectra of V4142\,Sgr. The radial velocity curves, the mass ratio, and orbital parameters were obtained. Through the use of synthetic spectra and the comparison with these, we estimate some physical parameters of the donor star and the temperature of the hot star.
   }
   {The light curve was modeled through an inverse modeling method using a theoretical light curve of the binary system, considering the light curve contribution of both stars and the accretion disk of the hot star to obtain the fundamental parameters. To constrain the main stellar parameters the mass ratio was fixed, as well as the donor  temperature using the obtained values from our spectroscopic analysis including deblending methods to isolate the spectral lines of the stellar components. The system parameters were compared with a grid of binary star evolutive models in order to get insights on the evolutionary history of the system.}
   {The orbital period and the long cycle were re-calculated and found to be of $30\fd633 \pm 0\fd002$ and $1201 \pm 14 ~\mathrm{days}$. The spectral analysis reveals H$\alpha$ double emission  with a persistent $V \leq R$ asymmetry which is considered evidence of a possible wind emergin from the hotspot region. In addition, a cold and evolved donor star of $M_\mathrm{d}=1.11 \pm 0.2 ~\mathrm{M_{\odot}}$, $T_\mathrm{d}= 4500 \pm 125 ~\mathrm{K}$ and a $R_\mathrm{d}=19.4 \pm 0.2 ~\mathrm{R_{\odot}}$ and a rejuvenated B-dwarf companion of $M_\mathrm{g}=3.86 \pm 0.3 ~\mathrm{M_{\odot}}$, $T_\mathrm{g}=14380 \pm 700 ~\mathrm{K}$ and $R_\mathrm{g}=6.35 \pm 0.2 ~\mathrm{R_{\odot}}$ were found.  The gainer is surrounded by a concave and geometrically thick disk creating its own atmosphere around the main component of radial extension ${\cal R}_{\rm d}= 22.8 \pm 0.3 ~\mathrm{R_{\odot}}$, contributing $\sim 1.4$ percent of the total luminosity of the system at the $V$-band at orbital phase 0.25. The disk is characterized by a hot-spot roughly placed where the stream hits the disk and an additional bright-spot separated $102.\!\!^{\circ}5 \pm 0.\!\!^{\circ}04$ degrees along the disk edge rim in the direction of the orbital motion. The system is seen under inclination $81.\!\!^{\circ}5 \pm 0.\!\!^{\circ}3$ and at a distance of $d_\mathrm{Gaia}=1140 \pm 35 ~\mathrm{pc}$. Doppler maps of the emission lines reveal sites of enhanced line emission in the 2nd and 3rd velocity quadrants. The first one would correspond to a hotspot at the next one to a bright spot detected by the light curve analysis. We find that the system comes from an initially shorter orbital period binary,
	that inverted its mass ratio due to mass exchange. A plaussible model indicates that at present,  the K-type giant should has depleted of hydrogen its core while the companion would has gained about 2 M$_{\odot}$ in a process lasting about 2 Myr. 
   %In addition, we find that the Balmer line emissivity forms a horseshoe structure in the circumstellar environment.
   }
   {}


   \maketitle

\section{Introduction}
\label{Sec: Sec. 1}

Since their discovery, the algol-like close binaries dubbed Double Periodic Variables (DPVs) have acquired a predominant role in the study of different mechanisms that have been proposed to explain the second photometric variability reported  by \citet{2003A&A...399L..47M}. Being the long cycle length on average about 33 times longer than the orbital period \citep{2017SerAJ.194....1M} and observed for the first time globally in a group of blue variable stars in the Magellanic Clouds, the long-cycle phenomenon was, however, already reported in some historical Algols. A famous example is $\beta$\,Lyrae with an orbital period of 12\fd9 increasing at a rate of $19 ~\mathrm{s\,yr^{-1}}$ \citep{1993A&A...279..131H} and with a long cycle of 275 days \citep{1989SSRv...50...35G}. 

Further research highlighted as a property of the DPVs the constancy of their orbital period, a feature observed in most of these binaries, which is not occurring in Algol stars undergoing strong mass transfer driven by Roche Lobe Overflow (RLOF) \citep{2013MNRAS.428.1594G}. $\beta$\,Lyrae is almost unique among the DPVs, and seems to contradict the almost universal constancy of the orbital period in DPVs. This constancy of the orbital period is puzzling considering that under conservative mass transfer due to RLOF, a change of period should be detected. A possible explanation is that these binaries are in a mild non-conservative mass transfer stage \citep{2008MNRAS.389.1605M,2008A&A...487.1129V,2011A&A...528A..16V,2013MNRAS.428.1594G}, as it is also suggested by some models so far \citep{2019MNRAS.483..862R,2019BAAA...61..107S}. 

A second photometric variability in Algol-type stars, is not something new, though already reported in $\beta$\,Per by \citet{1980A&A....89..100S}. Years later \citet{1997MNRAS.286..209S,2004MNRAS.352..416M,2010Natur.463..207P} tried to explain the second observed photometric variability through a mechanism of cyclical magnetic activity produced in the outer convective zone of the cold star, which would cause changes in the orbital period on a 10-year scale. Since then new studies started to consider the possibility of the existence of dynamos in some Algol-type variables. Following this line of reasoning, \citet{2017A&A...602A.109S} proposed an explanation for the long cycles of the DPV based on magnetic cycles of the most evolved star, considering that the Applegate mechanism \citep{1987ApJ...322L..99A} should induce cyclic changes in the donor equatorial radius and therefore in the mass transfer rate. Evidence for such a dynamo was reported in the DPV V393\,Sco by \citet{2018PASP..130i4203M} and explored in the evolutionary track of V495\,Centauri through numerical simulations \citep{2018MNRAS.476.3039R,2019MNRAS.483..862R}. The first demonstration through 3D MHD simulation of the importance of the MHD dynamo for the stellar structure was provided by \citet{2020MNRAS.491.1043N}.

\begin{figure}
	\begin{center}
		\includegraphics[trim=0.0cm 0.0cm 0.0cm 0.0cm,clip,width=0.5\textwidth,angle=0]{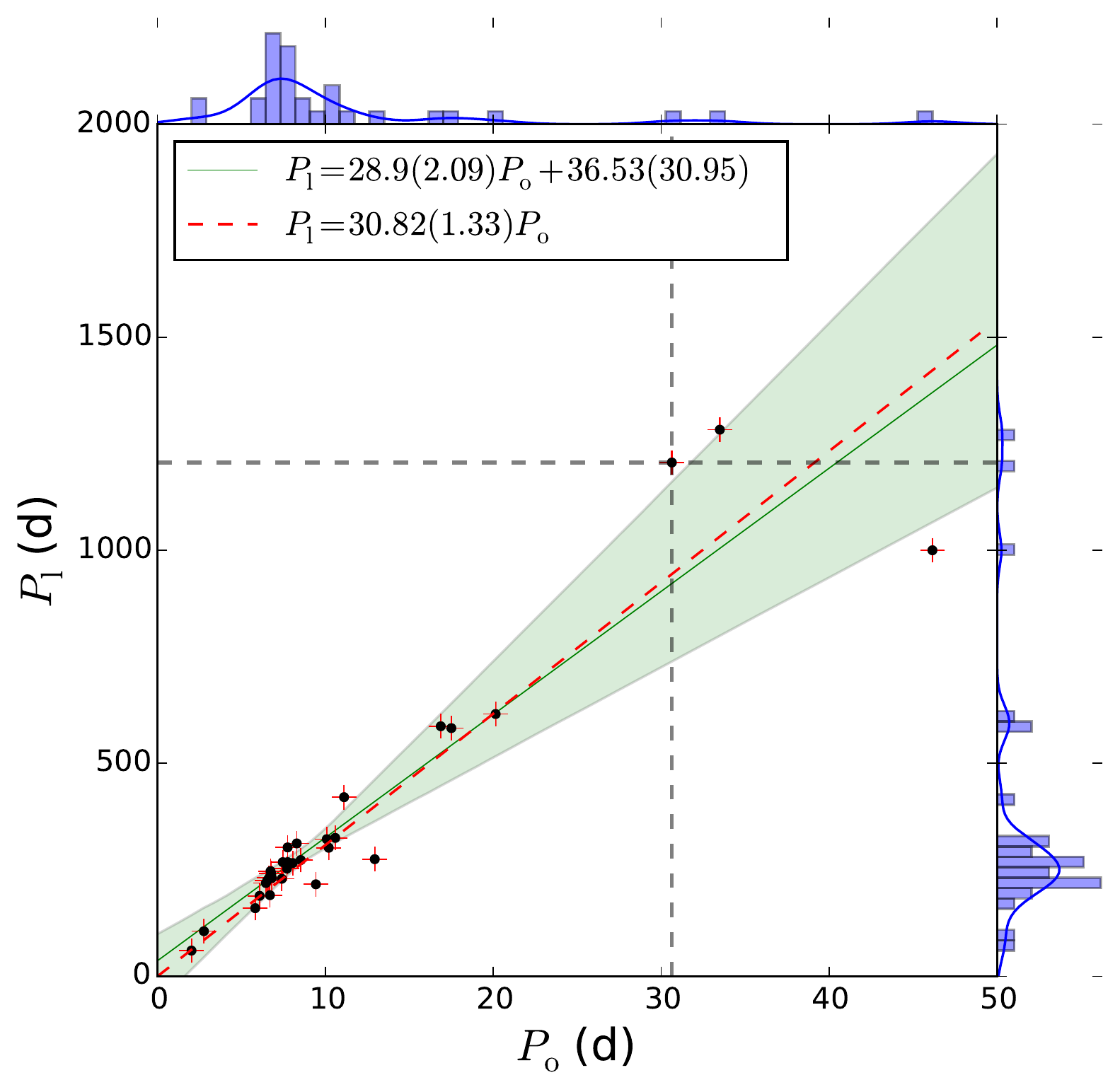}
	\end{center}
	\caption{Orbital and long period of Galactic DPVs, see the text for more details. Shared region dimension the 90\% confidence boundary.}
	\label{fig:Fig. 1}
\end{figure}

The few DPV models that have been studied reveal a semidetached binary system with a hot star of B-type rejuvenated and surrounded by an optically thick accretion disk, dubbed as gainer, that accumulates matter from the less massive and evolved star, dubbed as donor, and that is commonly observed as a giant of A/F/G type \citep{2016MNRAS.455.1728M,2017SerAJ.194....1M}. In addition, a cyclically variable bipolar wind modulated by the long-cycle was reported in V393\,Scorpii by \citet{2012MNRAS.427..607M}, reminiscent of the bipolar wind reported in $\beta$\,Lyrae \citep{1996A&A...312..879H}. These investigations arose the question of the role of mass loss in these systems. Thus, the study of these systems will allow improving the understanding of the complex evolutionary path involving the phase of mass exchange, stellar winds, a mass-loss and stellar dynamo in semidetached stars.

\ \\
In this work, a study of an interacting close binary system that belongs to the group of the DPVs with the longest orbital periods is presented, a sample that so far has not been well studied. The eclipsing interacting binary V4142\,Sgr, according to ASAS\footnote{\url{http://www.astrouw.edu.pl/asas/?page=acvs}} is classified as Eclipsing Algol Detached (EA/DS) binary, and named ID 180745-2824.1, with equatorial coordinates $\alpha_{2000}=18:07:45$ and $\delta_{2000}=-28:24:06$, and with $\mathrm{V}=10.94 \pm 0.04 ~\mathrm{mag}$, $\mathrm{V_{amp}}=2.019 \pm 0.05 ~\mathrm{mag}$ \citep{1997AcA....47..467P}. It was classified as a DPV by \citet{2014IBVS.6116....1M}, who found a long photometric cycle of 1206 days, resulting to be the second one with the longest period among Galactic DPVs until then (Fig.\,\ref{fig:Fig. 1}). In addition, according to SIMBAD \citep{2000A&AS..143....9W}\footnote{\url{http://simbad.u-strasbg.fr/simbad/}} this object is an eclipsing binary of Algol type of $\mathrm{V=10.95 \pm 0.04 ~{mag}}$, and its distance based on GAIA DR3 \citep{2018A&A...616A...1G,2018A&A...616A..11G}\footnote{\url{https://gea.esac.esa.int/archive/}} is $d_\mathrm{Gaia}=1140 \pm 35 ~\mathrm{pc}$. In this study we determine for the first time the fundamental stellar and orbital parameters of this system and contribute to the understanding of the DPVs.

\ \\
In Section \ref{Sec: Sec. 2}, a photometric analysis of V4142\,Sgr is presented. In Section \ref{Sec: Sec. 3}, a spectroscopic analysis is presented, including a Doppler tomography study of emission lines. In Section \ref{Sec: Sec. 4}, the orbital and physical parameters of the system based on the light curve model including the light contribution of both stars and an accretion disk is also revealed. Stellar luminosities, radii, temperatures, surface gravities, masses, and the system's inclination were obtained.
We provide a discussion of our results in Section \ref{Sec: Sec. 5}. Finally, the main results of the present investigation are summarized in Section \ref{Sec: Sec. 6}.

%%%%%%%%%%%%%%%%%%%%%%%%%%%%%%%%%%%%%%%%%%%%%%%%%%%%%%%%%%%%%%%%%%%%%%%%%%%%%%%%%%%%%%%%%%%%%%%%%%%%%%%%%%%%%%%%%%%%%%%%%%%%%%%%%%%%%%%%%%%%%%%%%%%%%%%%%%%%%%
%%%%%%%%%%%%%%%%%%%%%%%%%%%%%%%%%%%%%%%%%%%%%%%%%%%%%%%%%%%%%%%%%%%%%%%%%%%%%%%%%%%%%%%%%%%%%%%%%%%%%%%%%%%%%%%%%%%%%%%%%%%%%%%%%%%%%%%%%%%%%%%%%%%%%%%%%%%%%%
%%%%%%%%%%%%%%%%%%%%%%%%%%%%%%%%%%%%%%%%%%%%%%%%%%%%%%%%%%%%%%%%%%%%%%%%%%%%%%%%%%%%%%%%%%%%%%%%%%%%%%%%%%%%%%%%%%%%%%%%%%%%%%%%%%%%%%%%%%%%%%%%%%%%%%%%%%%%%%
%%%%%%%%%%%%%%%%%%%%%%%%%%%%%%%%%%%%%%%%%%%%%%%%%%%%%%%%%%%%%%%%%%%%%%%%%%%%%%%%%%%%%%%%%%%%%%%%%%%%%%%%%%%%%%%%%%%%%%%%%%%%%%%%%%%%%%%%%%%%%%%%%%%%%%%%%%%%%%
\section{Photometric analysis}
\label{Sec: Sec. 2}

\subsection{All Sky Automated Survey (ASAS)}
\label{Sec: Sec. 2.1}

\begin{figure}
	\begin{center}
		\includegraphics[trim=0.0cm 0.0cm 0.0cm 0.0cm,clip,width=0.5\textwidth,angle=0]{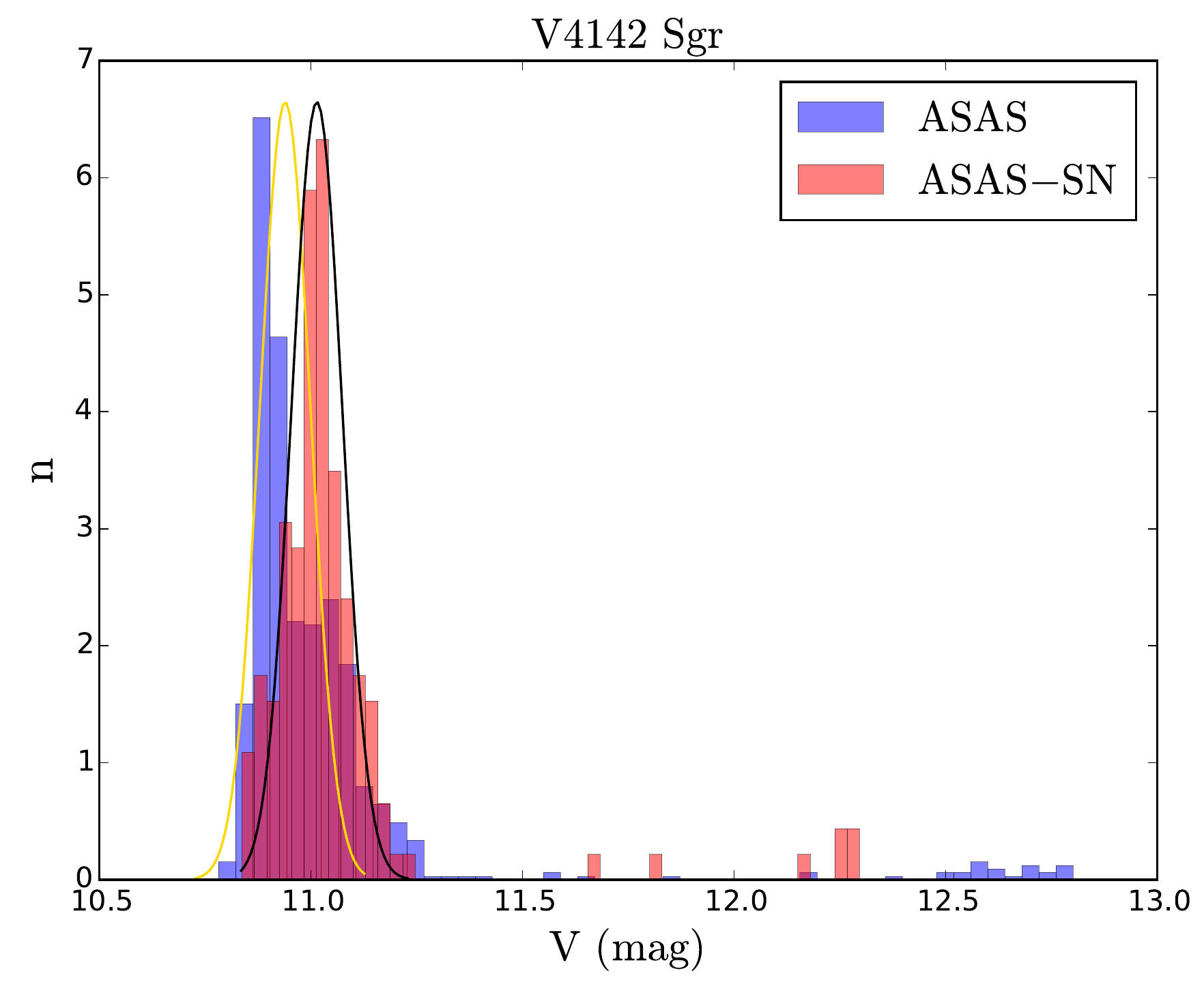}
	\end{center}
	\caption{Histogram of magnitude for V4142\,Sgr using 50 bins for 806 ASAS A-type data, showing a positive skew distribution. The dashed red line corresponds to a normal distribution of 806 random data with a mean $\mu=10.939 ~\mathrm{mag}$ and sigma $\sigma= 0.06 ~\mathrm{mag}$.}
	\label{fig:Fig. 2}
\end{figure}

In this section we re-analyzed the ASAS photometric data for V4142\,Sgr considering the 806 better quality data points labeled as A-type and we rejected all outlier points. For that, the data were grouped in a histogram of 50 bins and it was noted that this photometric time series shows an average apparent magnitude $\mathrm{V_{ASAS}= 10.939 \pm 0.04 ~{mag}}$ with a positive skew distribution and a tail towards faint magnitudes that corresponds to measured magnitudes during the primary and secondary eclipse (see Fig.\,\ref{fig:Fig. 2}), with a range approximately from $11.07$ to $12.80 ~\mathrm{mag}$, i.e. if the primary and secondary eclipses are less deep, then the skews would be much smaller. 

In addition, we have decided to strengthen the analysis for the orbital period using the Phase Minimization Dispersion (\texttt{PDM}) code of \texttt{IRAF} \citep{1978ApJ...224..953S}. Initially, the errors were computed through visual inspection of the light curve phased with trial periods close to the minimum of the periodogram until the light curve decreased its dispersion. Despite that the method gives us reliable results for the orbital period, it lacks the accuracy to estimate the errors. Therefore, we decided to apply the Generalized Lomb-Scargle Periodogram (\texttt{GLS}) \citep{2009A&A...496..577Z} algorithm, in order to strengthen the result and handle adequately the errors; hence we get an orbital period of $30\fd633 \pm 0\fd002$. This value coincides and confirms that obtained by \citet{2014IBVS.6116....1M}. The GLS method has been sucesfully applied to photometric time series in the past, for instance when studying the rotation and differential rotation of active Kepler stars  \citep{2013A&A...560A...4R}.

After obtaining the orbital period more accurately determined, a second analysis was carried out to find the named \textquotedblleft{long cycle}\textquotedblright. For that, an algorithm was used to disentangle multi-periodic light curves, written by Zbigniew Kolaczkowski, which has been used for previous disentangling analyses \citep{2012MNRAS.427..607M,2015MNRAS.448.1137M,2018MNRAS.476.3039R}. The code adjusts the light curve with a Fourier series consisting of fundamental frequencies plus their harmonics. The orbital frequency was computed previously using \texttt{PDM} for our best value. Since the light curve of an eclipsing binary has a more complex nature than a sinusoidal form, it is emphasized that the use of a few harmonics is not applicable, as the fit depends on the form of the light curve. Thus, is needed a great number of harmonics between 12 and 15 to fit adequately the light curve of an eclipsing binary. Further details of the code can be found in \citet{2012MNRAS.421..862M}. 

\begin{figure}
	\begin{center}
		\includegraphics[trim=0.0cm 0.0cm 0.0cm 0.0cm,clip,width=0.5\textwidth,angle=0]{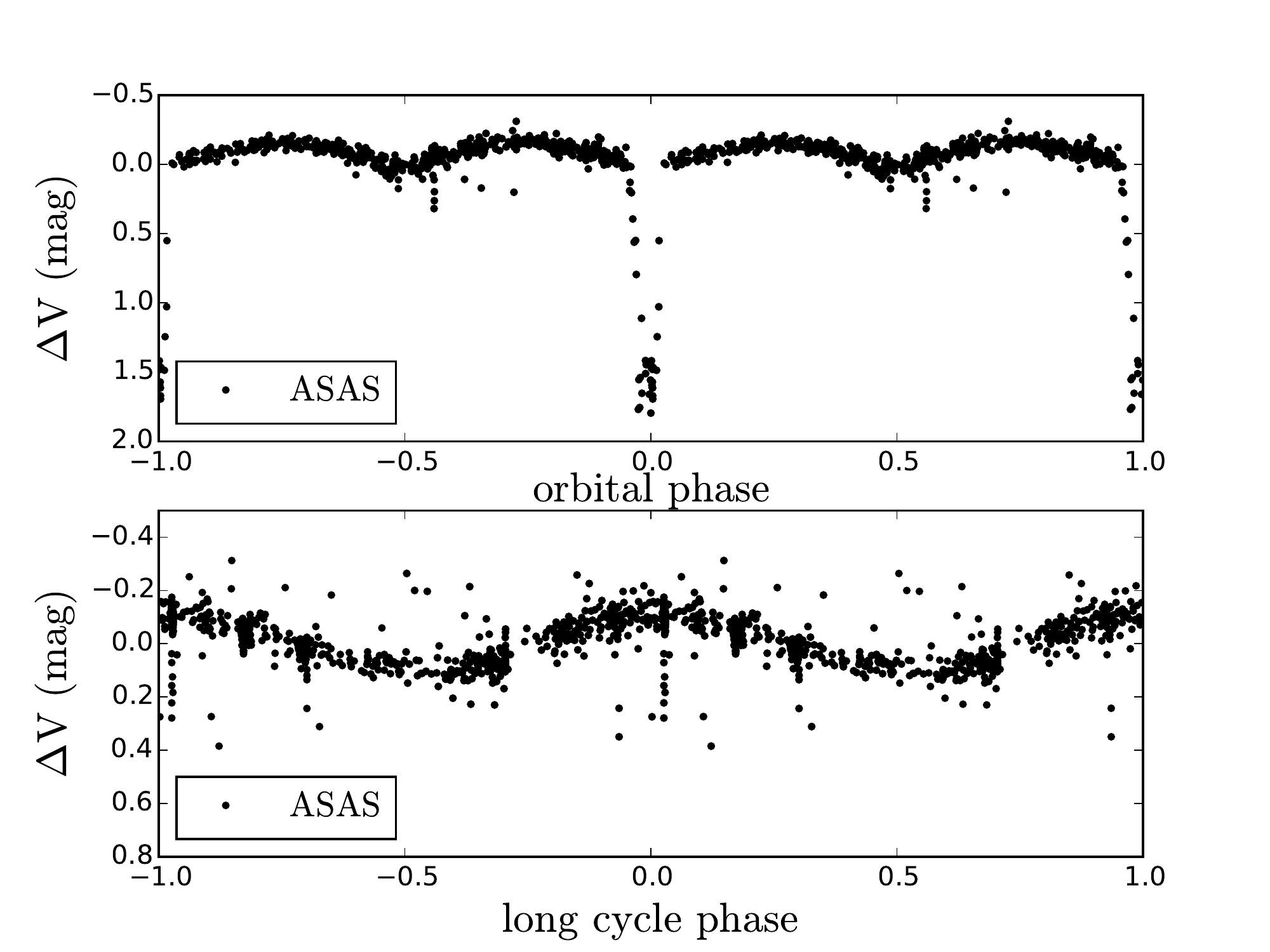}
	\end{center}
	\caption{V4142\,Sgr ASAS V-band light curve after disentangling. (Up) Orbital and long cycle phase (Down) were phased according to times of the light curve minimum and maxima, given by equations \ref{eq: eq. 1} and \ref{eq: eq. 2}}
	\label{fig:Fig. 3}
\end{figure}

The main frequency $f_{1}$ (the orbital one) is used in an algorithm for a least-square fit over the light of a sum of sinus functions of variable amplitudes and phases representing the main frequency and their additional harmonics. After this, we analyzed the residuals to find another periodicity or frequency $f_{2}$. Further, the new harmonics and main frequencies are included in the new fitting procedure to obtain the best light curve based on Fourier component frequencies considering both periods. Thus, in the applied deconvolution method on the light curve we get an orbital frequency of $f_{o}=0.0326445337 ~\mathrm{d^{-1}}$ and other  $f_{l}=0.0008291874 ~\mathrm{d^{-1}}$ for the long cycle. So, in this way we disentangled both light curves as shown in Fig.\,\ref{fig:Fig. 3}. The disentangled light curve with the new orbital period revealed an orbital modulation typical of an eclipsing binary with rounded inter-eclipse regions including proximity effects and a computed amplitude of $2.107 \pm 0.057 ~\mathrm{mag}$, while the long cycle is characterized by a quasi-sinusoidal variability and typical of the rest of DPVs, whose amplitude in the V-band of $0.192 \pm 0.003 ~\mathrm{mag}$ is around of $\sim 9$ percent with respect to the total brightness. The EA classification provided by ASAS is probably due to the long period and large differences between the main and secondary minimum, but the light curve can easily be classified as EB (i.e. $\beta$-Lyrae type) as well, considering the rounded inter-eclipse sections. The aforementioned procedure revealed the following ephemerides for the light curve:

\begin{equation}
HJD_\mathrm{min,orbital}= 2454726.55524 + 30.633(2) \times E,
\label{eq: eq. 1}
\end{equation}

\begin{equation}
HJD_\mathrm{max,long}= 2453547.76585 + 1206\times E.
\label{eq: eq. 2}
\end{equation}

\ \\
\noindent
These are used for spectroscopic analysis in the rest of the present work.

\subsection{All-Sky Automated Survey for Supernovae (ASAS-SN)}
\label{Sec: Sec. 2.2}

\begin{figure}
	\begin{center}
		\includegraphics[trim=0.0cm 0.0cm 0.0cm 0.0cm,clip,width=0.5\textwidth,angle=0]{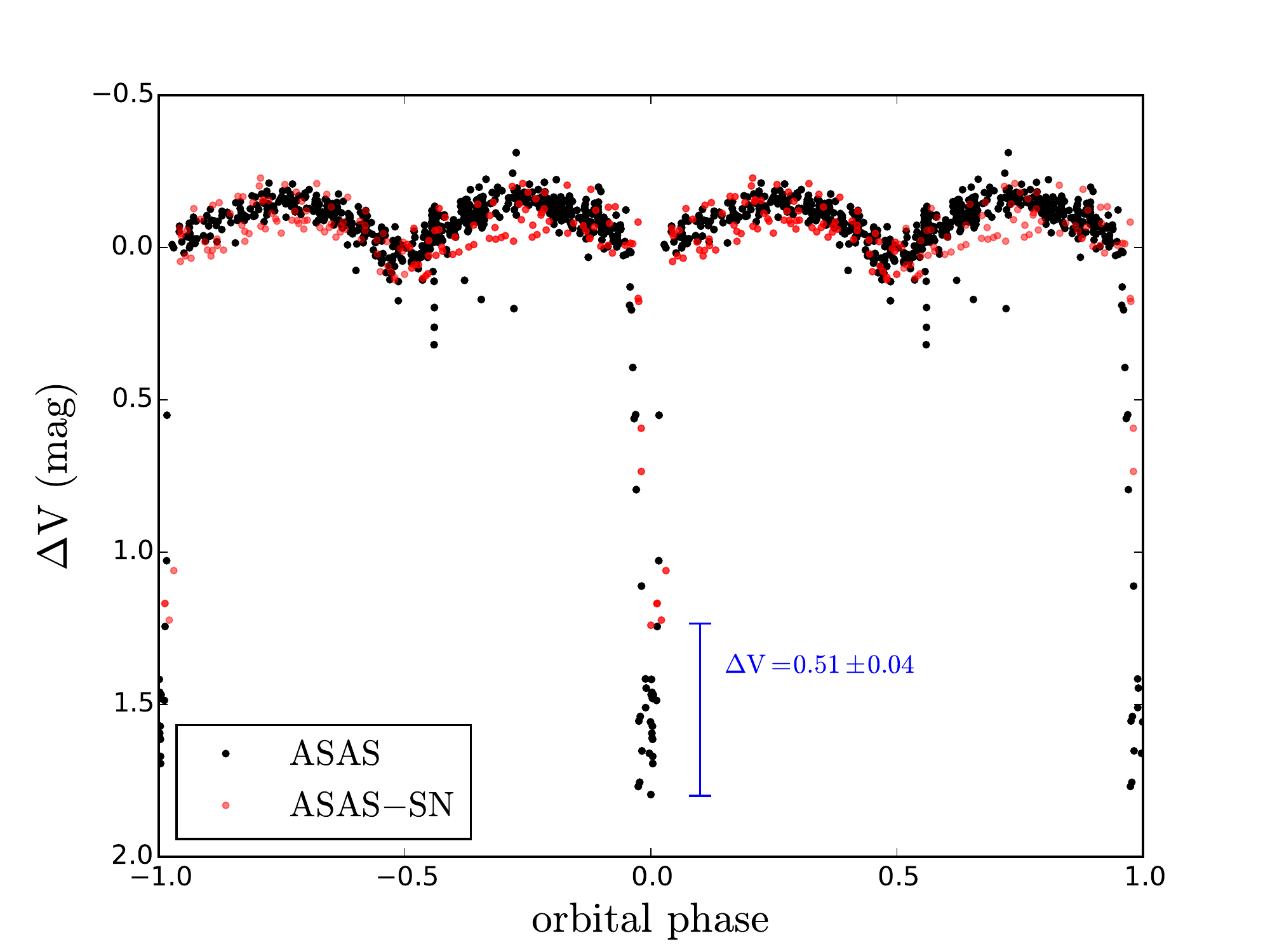}
	\end{center}
	\caption{The ASAS and ASAS-SN V-band light curves comparisons phased using the orbital period $P_\mathrm{o}=30\fd633$, show a morphological difference during the primary eclipse. The primary eclipse of ASAS is $0.51\pm0.04 \,\mathrm{mag}$ deeper than for ASAS-SN.}
	\label{fig:Fig. 4}
\end{figure}

It was decided to reinforce the results obtained in the previous section (Sec.\,\ref{Sec: Sec. 2}), adding a data set from the ASAS-SN Variable Stars Database \footnote{\url{https://asas-sn.osu.edu/}} \citep{2014ApJ...788...48S,2017PASP..129j4502K,2019MNRAS.486.1907J}, whose images in the V-band photometry were obtained through a two-pixel ($16^{\prime\prime}$) radius aperture, different to the 3 pixels ($15^{\prime\prime}$) used by ASAS. For that, a second analysis of the 157 data points was carried out using \texttt{PDM} and their errors were estimated through the same \texttt{GLS} algorithm used in the previous section. In addition, during a comparison between data from ASAS and ASAS-SN we noticed a morphological difference in the shape and amplitude of the light curve during the primary eclipse (see Fig.\,\ref{fig:Fig. 4}). The range covered by the data is 893 days.

For the disentangling method applied on the ASAS-SN light curve, we used the same frequency for the orbital cycle as in the previous section. The obtained results agree and confirm the obtained values with ASAS; we get an orbital period $P_\mathrm{o}=30\fd635 \pm 0\fd002$ and an amplitude $A_\mathrm{o}= 1.47 \pm 0.03 \,\mathrm{mag}$. In addition, the comparison between long cycles showed no morphological changes as occurred during the orbital cycle. The long cycle has an amplitude $A_\mathrm{l}=0.266 \pm 0.006 \,\mathrm{mag}$, corresponding to $\sim 18$ percent of the total brightness. %(Fig.\,\ref{fig:Fig. 4}).

\subsection{Multiband photometry}
\label{Sec: Sec. 2.4}

So far, there are only references to the V-band analysis using ASAS and ASAS-SN. Thus, a search of photometric data in ExoFOP-TESS \footnote{\url{https://exofop.ipac.caltech.edu/tess/}} \citep{2019AAS...23314009A} was performed for other bands and we summarized a series of these in Tab.\,\ref{Tab: Tab. 1}. The observed mean color was computed using the Wide-field Infrared Survey Explorer (WISE) \citet{2010AJ....140.1868W} and obtained that $\mathrm{W2-W3= 0.079 \pm 0.024\,mag}$, and $\mathrm{W3-W4= 0.589 \pm 0.088\,mag}$. We find the system does not show a color excess in the WISE photometry. 

\begin{figure}
	\begin{center}
		\includegraphics[trim=0.0cm 0.0cm 0.0cm 0.0cm,clip,width=0.5\textwidth,angle=0]{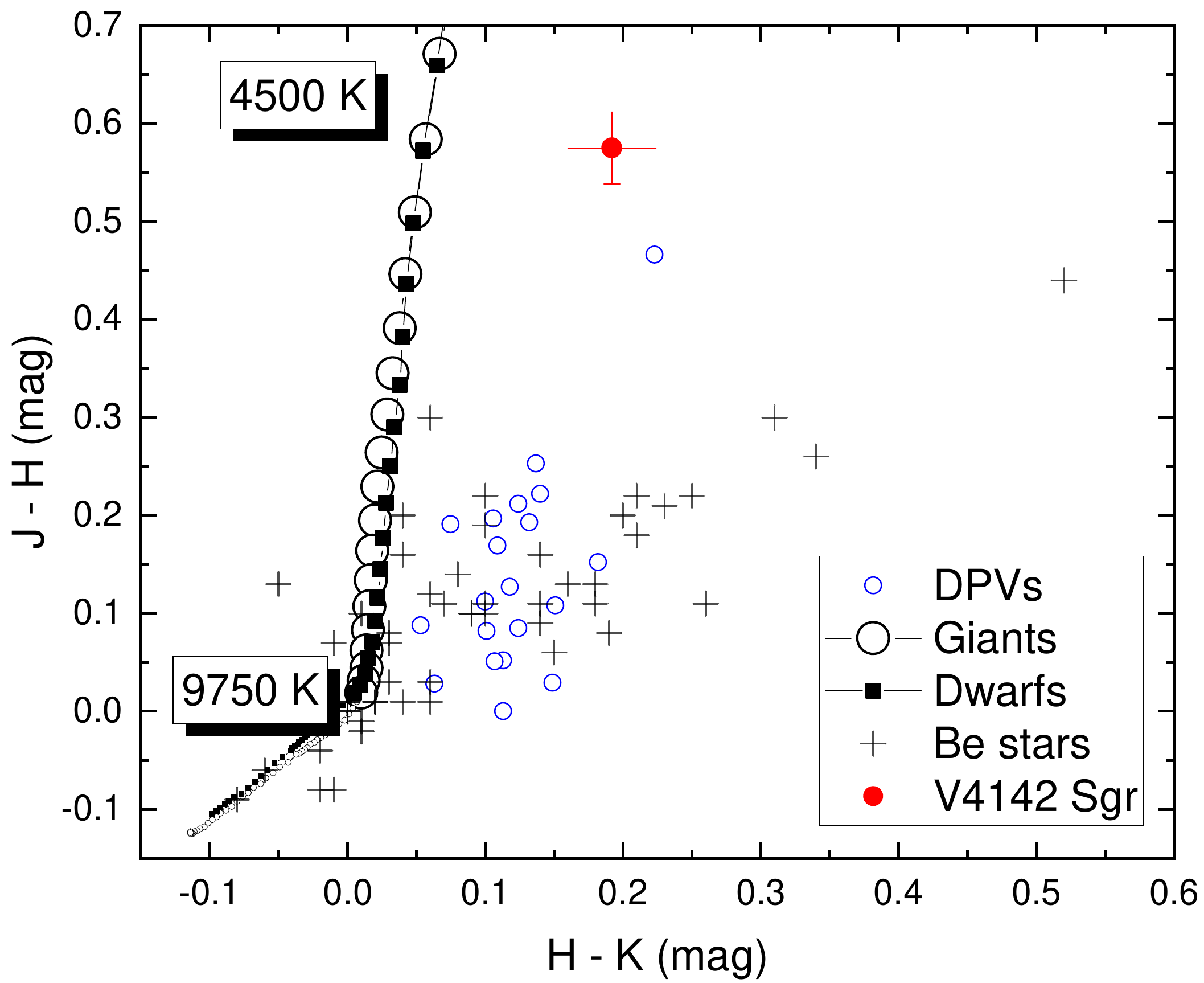}
	\end{center}
	\caption{2MASS $J-H$ versus $H-K$ color-color diagram for V4142\,Sgr and other systems with/out circumstellar matter. For more detail of data collection see \citet{2016MNRAS.455.1728M}. }
	\label{fig:Fig. 5}
\end{figure}

Furthermore, a data comparison was carried out using Two Micron All Sky Survey (2MASS) \citet{2006AJ....131.1163S} and the work by \citet{2016MNRAS.455.1728M} wherein are compared systems with circumstellar envelopes, namely Be-stars, the rapidly rotating B-type with circumstellar disk and included the Double Periodic Variable (DPVs) and others. The computed differences for V4142\,Sgr by means of 2MASS $\mathrm{J-H= 0.575 \pm 0.037\,mag}$ and $\mathrm{H-K= 0.192 \pm 0.032\,mag}$ show a color excess in $JHK$ photometry. The $J-H$ color fits a single star of temperature around $\mathrm{\sim 5000\,K}$ while the $H-K$ excess could be interpreted as evidence for circumstellar material (See Fig.\,\ref{fig:Fig. 5}). In both cases, we have not accounted for extinction and not corrected the color since the correction should be insignificant at these wavelengths.

\ \\
\begin{table}
	\caption{Summary of photometric data for V4142\,Sgr at different band passes.}
	\normalsize
	\begin{center}
		\begin{tabular}{lccrc}
			\hline
			\noalign{\smallskip}
			\textrm{Band}  	&  \textrm{Value (mag)}	& \textrm{error (mag)}& {$\lambda_\textrm{eff}$} (\AA{})\\
			\hline
			\hline
			\textrm{ExoFOP-B}		& \textrm{11.5100} 		& \textrm{0.2670} &\textrm{4400}\\
			\textrm{ExoFOP-V}		& \textrm{11.0960} 		& \textrm{0.0190} &\textrm{5500}\\
			\textrm{Gaia} 	 		& \textrm{11.2099} 		& \textrm{0.0456} &\textrm{5857.56}\\
			\textrm{TESS} 	 		& \textrm{10.8857} 		& \textrm{0.0552} &\textrm{7455.64}\\
			\textrm{2MASS-J} 		& \textrm{ 9.3320} 		& \textrm{0.0290} &\textrm{12350}\\
			\textrm{2MASS-H} 		& \textrm{ 8.7570} 		& \textrm{0.0230} &\textrm{16620}\\
			\textrm{2MASS-K}	 	& \textrm{ 8.5650} 		& \textrm{0.0230} &\textrm{21590}\\
			\textrm{WISE2} 	 		& \textrm{ 7.3990} 		& \textrm{0.0140} &\textrm{46028}\\
			\textrm{WISE3} 	 		& \textrm{ 7.3200} 		& \textrm{0.0190} &\textrm{115608}\\
			\textrm{WISE4} 	 		& \textrm{ 6.7310} 		& \textrm{0.0860} &\textrm{220883.83}\\
			\hline
		\end{tabular}
	\end{center}
	\label{Tab: Tab. 1}
\end{table}

%%%%%%%%%%%%%%%%%%%%%%%%%%%%%%%%%%%%%%%%%%%%%%%%%%%%%%%%%%%%%%%%%%%%%%%%%%%%%%%%%%%%%%%%%%%%%%%%%%%%%%%%%%%%%%%%%%%%%%%%%%%%%%%%%%%%%%%%%%%%%%%%%%%%%%%%%%%%%%
%%%%%%%%%%%%%%%%%%%%%%%%%%%%%%%%%%%%%%%%%%%%%%%%%%%%%%%%%%%%%%%%%%%%%%%%%%%%%%%%%%%%%%%%%%%%%%%%%%%%%%%%%%%%%%%%%%%%%%%%%%%%%%%%%%%%%%%%%%%%%%%%%%%%%%%%%%%%%%
%%%%%%%%%%%%%%%%%%%%%%%%%%%%%%%%%%%%%%%%%%%%%%%%%%%%%%%%%%%%%%%%%%%%%%%%%%%%%%%%%%%%%%%%%%%%%%%%%%%%%%%%%%%%%%%%%%%%%%%%%%%%%%%%%%%%%%%%%%%%%%%%%%%%%%%%%%%%%%
%%%%%%%%%%%%%%%%%%%%%%%%%%%%%%%%%%%%%%%%%%%%%%%%%%%%%%%%%%%%%%%%%%%%%%%%%%%%%%%%%%%%%%%%%%%%%%%%%%%%%%%%%%%%%%%%%%%%%%%%%%%%%%%%%%%%%%%%%%%%%%%%%%%%%%%%%%%%%%

\section{Spectroscopic analysis}
\label{Sec: Sec. 3}

In this section, information about the spectroscopic data acquisition and a detailed description of the analysis is also provided, as well as, the study of radial velocities, the spectral decomposition and information about the obtained results.

\subsection{Spectroscopic observations}
\label{Sec: Sec. 3.1}

To investigate the fundamental properties of the DPV V4142\,Sgr, a series of 16 spectra with resolving power $R\sim 27000$ (fiber mode) was collected with the CHIRON spectrograh \citep{2013PASP..125.1336T,2021AJ....162..176P} (Proposal ID CN2018A-4)  mounted at the SMARTS 1.5 meter telescope, located in Cerro Tololo Interamerican Observatory. The covered spectral region is from 4500\AA{}  to 8500\AA{} with a mean signal to noise ratio of $\sim 92$. The observations were conducted during the maximum of the long cycle and cover relatively well the orbital cycle  (see Table\,\ref{Tab: Tab. 2}).

The correction with flat and bias, wavelength calibration, and order merging was carried out with \texttt{IRAF}. In addition, all spectra to the continuum were normalized and corrected to the heliocentric rest frame. No flux calibration of the spectra was needed since this has no effect for the line strength measurements and radial velocities, since the analyzed spectra were normalized to the continuum, thus the flux in absolute terms is not considered, and every intensity is in units of the intensity of the continuum. In addition, the equivalent width by definition does not depend on the flux calibration and these can be defined in spectra normalized to the continuum and the measure of radial velocities, too. Hence, a flux calibration is not needed for the present purposes. 

\ \\
\begin{table}
	\caption{Summary of spectroscopic observation using CHIRON spectrograph mounted on the SMARTS-1.5m Telescope. The HJD at mid-exposure for the first spectrum series is given, the orbital ($\phi_\mathrm{o}$) and long cycle ($\phi_\mathrm{l}$) phases were calculated using eq.1 and eq. 2 respectively. These spectra have a spectral resolution of $\mathrm{R \sim 27000}$.}
	\normalsize
	\begin{center}
		\resizebox{0.5\textwidth}{3.0cm}{		
			\begin{tabular}{cccccr}
				\hline
				\noalign
				{\smallskip}
				\textrm{UT-date}& \textrm{exptime\,(s)} &\textrm {HJD} & \textrm {$\phi_{o}$}  &\textrm{$\phi_{l}$} &\textrm {S/N} \\
				\hline
				\hline
				\noalign{\smallskip}
				\textrm{2018-06-02} & \textrm{3600} & \textrm{2458271.69881789} & \textrm{0.7296} & \textrm{0.9179} & \textrm{110} \\
				\textrm{2018-06-03} & \textrm{3600} & \textrm{2458272.73360867} & \textrm{0.7633} & \textrm{0.9188} & \textrm{ 81} \\
				\textrm{2018-06-04} & \textrm{3600} & \textrm{2458273.72971956} & \textrm{0.7959} & \textrm{0.9196} & \textrm{118} \\
				\textrm{2018-06-05} & \textrm{3600} & \textrm{2458274.73941726} & \textrm{0.8288} & \textrm{0.9204} & \textrm{ 41} \\
				\textrm{2018-06-05} & \textrm{3600} & \textrm{2458274.78308750} & \textrm{0.8303} & \textrm{0.9205} & \textrm{106} \\
				\textrm{2018-06-07} & \textrm{3600} & \textrm{2458276.71305890} & \textrm{0.8933} & \textrm{0.9221} & \textrm{ 90} \\
				\textrm{2018-06-15} & \textrm{3600} & \textrm{2458284.71360285} & \textrm{0.1544} & \textrm{0.9287} & \textrm{ 93} \\
				\textrm{2018-06-17} & \textrm{3600} & \textrm{2458286.70777669} & \textrm{0.2195} & \textrm{0.9304} & \textrm{ 85} \\
				\textrm{2018-06-18} & \textrm{3600} & \textrm{2458287.68065316} & \textrm{0.2513} & \textrm{0.9312} & \textrm{ 73} \\
				\textrm{2018-06-19} & \textrm{3600} & \textrm{2458288.65402813} & \textrm{0.2831} & \textrm{0.9320} & \textrm{ 78} \\
				\textrm{2018-06-20} & \textrm{3600} & \textrm{2458289.72153509} & \textrm{0.3179} & \textrm{0.9329} & \textrm{153} \\
				\textrm{2018-06-21} & \textrm{3600} & \textrm{2458290.70659200} & \textrm{0.3501} & \textrm{0.9337} & \textrm{ 73} \\
				\textrm{2018-07-30} & \textrm{3600} & \textrm{2458329.64851880} & \textrm{0.6213} & \textrm{0.9660} & \textrm{ 93} \\
				\textrm{2018-08-02} & \textrm{3600} & \textrm{2458332.61173707} & \textrm{0.7180} & \textrm{0.9684} & \textrm{107} \\
				\textrm{2018-08-10} & \textrm{3600} & \textrm{2458340.60944136} & \textrm{0.9791} & \textrm{0.9750} & \textrm{ 57} \\
				\textrm{2018-08-12} & \textrm{3600} & \textrm{2458342.59987233} & \textrm{0.0441} & \textrm{0.9767} & \textrm{109} \\
				\hline
			\end{tabular}
		}
	\end{center}
	\label{Tab: Tab. 2}
\end{table}

%%%%%%%%%%%%%%%%%%%%%%%%%%%%%%%%%%%%%%%%%%%%%%%%%%%%%%%%%%%%%%%%%%%%%%%%%%%%%%%%%%%%%%%%%%%%%%%%%%%%%%%%%%%%%%%%%%%%%%%%%%%%%%%%%%%%%%%%%%%%%%%%%%%%%%%%%%%%%%
%%%%%%%%%%%%%%%%%%%%%%%%%%%%%%%%%%%%%%%%%%%%%%%%%%%%%%%%%%%%%%%%%%%%%%%%%%%%%%%%%%%%%%%%%%%%%%%%%%%%%%%%%%%%%%%%%%%%%%%%%%%%%%%%%%%%%%%%%%%%%%%%%%%%%%%%%%%%%%
%%%%%%%%%%%%%%%%%%%%%%%%%%%%%%%%%%%%%%%%%%%%%%%%%%%%%%%%%%%%%%%%%%%%%%%%%%%%%%%%%%%%%%%%%%%%%%%%%%%%%%%%%%%%%%%%%%%%%%%%%%%%%%%%%%%%%%%%%%%%%%%%%%%%%%%%%%%%%%
%%%%%%%%%%%%%%%%%%%%%%%%%%%%%%%%%%%%%%%%%%%%%%%%%%%%%%%%%%%%%%%%%%%%%%%%%%%%%%%%%%%%%%%%%%%%%%%%%%%%%%%%%%%%%%%%%%%%%%%%%%%%%%%%%%%%%%%%%%%%%%%%%%%%%%%%%%%%%%

\subsection{Radial velocities}
\label{Sec: Sec. 3.2}

The first step to determine the fundamental properties of the DPV V4142\,Sgr was to measure the radial velocities (RVs) of both stellar components. For that, a set of absorption lines that represents the movements of both stars has been identified. Before starting, from now the primary and hotter star and the secondary and cooler star will be named as gainer and donor, respectively.

The O\,I (8446.335 \AA{}) was selected as a characteristic absorption-line of the gainer star (see Table\,\ref{Tab: Tab. 3}) since it represents its movement and it is easily identifiable visualy between other lines such as N\,III (4514.89 \AA{}), Ti\,II (4571.971 \AA{}), H$\alpha$ (6562.817 \AA{}), and O\,I (7771.96 \AA{}). In order to measure the RVs of the donor star, we used the Fourier Cross-Correlation tool (\texttt{fxcor}) of \texttt{IRAF} between the range from 4500 to 6000 \AA{}, in this way we obtained the relative velocities listed in Table\,\ref{Tab: Tab. 4}.

\begin{figure}
	\begin{center}
		\includegraphics[trim=0.0cm 0.0cm 0.0cm 0.0cm,clip,width=0.5\textwidth,angle=0]{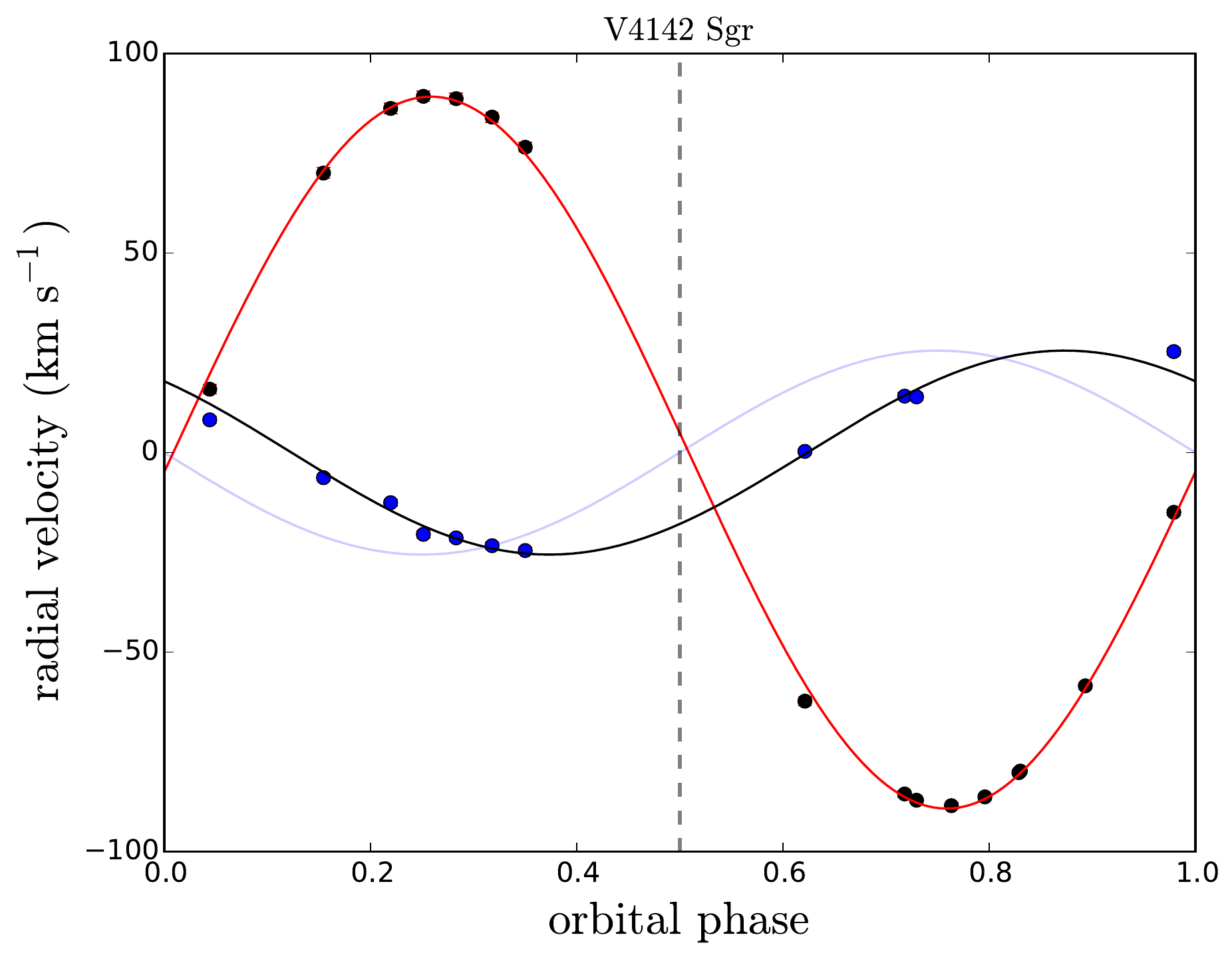}
	\end{center}
	\caption{Radial Velocities for donor (black-dots) through \texttt{fxcor} process and gainer star (blue-dots) using O\,I (8446.35 \AA{}) measured by Gaussian fits. The best sinusoidal fits are shown for both orbits. The smoothed blue line corresponds to the RVs of the gainer star without lag.}
	\label{fig:Fig. 6}
\end{figure}

The RVs of donor and gainer star were fitted using a sine function ($f(x)=a+b\sin(2\pi{x})$) through a Marquart-Levenberg \citep{1963SIAM...11..431} method a non-linear least square fit and their respective errors were obtained through an iterative succession of local linearization. The amplitude of the radial velocity variation of donor $K_\mathrm{d}= 89.2 \pm 0.5 ~\mathrm{km\,s^{-1}}$ and a zero point $0.00 \pm 0.47 ~\mathrm{km\,s^{-1}}$ were computed, whereas that for the gainer we computed has an amplitude $K_\mathrm{g}=25.56 \pm 1.20 ~\mathrm{km\,s^{-1}}$ and a zero point $0.00 \pm 0.86 ~\mathrm{km\,s^{-1}}$ (Fig.\,\ref{fig:Fig. 6}). However, a lag for RVs of the gainer was noted, which could be interpreted as the effect of the brightness contribution from the accretion disk or bright spot. In addition, we get a system mass ratio $q=K_\mathrm{g}/K_\mathrm{c}=0.287 \pm 0.047$.

\ \\
\begin{table}
	\caption{Radial velocities of the gainer and their respective errors, using O\,I (8446.35 \AA{}) line.}
	\normalsize
	\begin{center}
		%\resizebox{0.5\textwidth}{7.5cm}{		
		\begin{tabular}{ccrc}
			\hline
			\noalign
			{\smallskip}
			\textrm {HJD} & \textrm {$\phi_{o}$}  &\textrm{RV (km\,s$^{-1}$)} &\textrm {error (km\,s$^{-1}$)} \\
			\hline
			\hline
			\noalign{\smallskip}
			\textrm{2458342.59987233} & \textrm{0.0441} & \textrm{  8.2} & \textrm{0.9}\\
			\textrm{2458284.71360285} & \textrm{0.1544} & \textrm{ -6.3} & \textrm{0.9}\\
			\textrm{2458286.70777669} & \textrm{0.2195} & \textrm{-12.6} & \textrm{0.9}\\
			\textrm{2458287.68065316} & \textrm{0.2513} & \textrm{-20.5} & \textrm{0.9}\\
			\textrm{2458288.65402813} & \textrm{0.2831} & \textrm{-21.4} & \textrm{0.9}\\
			\textrm{2458289.72153509} & \textrm{0.3179} & \textrm{-23.3} & \textrm{0.9}\\
			\textrm{2458290.70659200} & \textrm{0.3501} & \textrm{-24.5} & \textrm{0.9}\\
			\textrm{2458329.64851880} & \textrm{0.6213} & \textrm{  0.3} & \textrm{0.9}\\
			\textrm{2458332.61173707} & \textrm{0.7180} & \textrm{ 14.2} & \textrm{0.9}\\
			\textrm{2458271.69881789} & \textrm{0.7296} & \textrm{ 14.0} & \textrm{0.9}\\
			\textrm{2458340.60944136} & \textrm{0.9791} & \textrm{ 25.3} & \textrm{0.9}\\
			\hline
		\end{tabular}
		%}
	\end{center}
	\label{Tab: Tab. 3}
\end{table}

\begin{table}
	\caption{Radial velocities of the donor and their respective errors, using Fourier Cross Correlation process.}
	\normalsize
	\begin{center}
		%\resizebox{0.5\textwidth}{7.5cm}{		
		\begin{tabular}{ccrc}
			\hline
			\noalign
			{\smallskip}
			\textrm {HJD} & \textrm {$\phi_{o}$}  &\textrm{RV (km\,s$^{-1}$)} &\textrm {error (km\,s$^{-1}$)} \\
			\hline
			\hline
			\noalign{\smallskip}
			\textrm{2458342.59987233} & \textrm{0.0441} & \textrm{ 15.9} & \textrm{1.2} \\
			\textrm{2458284.71360285} & \textrm{0.1544} & \textrm{ 70.0} & \textrm{1.4} \\
			\textrm{2458286.70777669} & \textrm{0.2195} & \textrm{ 86.3} & \textrm{1.3} \\
			\textrm{2458287.68065316} & \textrm{0.2513} & \textrm{ 89.3} & \textrm{1.3} \\
			\textrm{2458288.65402813} & \textrm{0.2831} & \textrm{ 88.7} & \textrm{1.3} \\
			\textrm{2458289.72153509} & \textrm{0.3179} & \textrm{ 84.1} & \textrm{1.3} \\
			\textrm{2458290.70659200} & \textrm{0.3501} & \textrm{ 76.5} & \textrm{1.2} \\
			\textrm{2458329.64851880} & \textrm{0.6213} & \textrm{-62.3} & \textrm{1.1} \\
			\textrm{2458332.61173707} & \textrm{0.7180} & \textrm{-85.6} & \textrm{0.8} \\
			\textrm{2458271.69881789} & \textrm{0.7296} & \textrm{-87.1} & \textrm{0.5} \\
			\textrm{2458272.73360867} & \textrm{0.7633} & \textrm{-88.5} & \textrm{0.7} \\
			\textrm{2458273.72971956} & \textrm{0.7959} & \textrm{-86.3} & \textrm{0.8} \\
			\textrm{2458274.73941726} & \textrm{0.8288} & \textrm{-80.2} & \textrm{0.6} \\
			\textrm{2458274.78308750} & \textrm{0.8303} & \textrm{-79.8} & \textrm{0.6} \\
			\textrm{2458276.71305890} & \textrm{0.8933} & \textrm{-58.5} & \textrm{0.8} \\
			\textrm{2458340.60944136} & \textrm{0.9791} & \textrm{-15.0} & \textrm{0.8} \\
			\hline
		\end{tabular}
		%}
	\end{center}
	\label{Tab: Tab. 4}
\end{table}

\ \\
\noindent
To obtain the orbital parameters of V4142\,Sgr a public subroutine based on a generic algorithm called \texttt{PIKAIA} \citep{1995ApJS..101..309C} was used. It is a heuristic search technique that incorporates in a computational setting the biological notion of evolution by means of natural selection, i.e. the algorithm determines the single parameter set which minimizes the difference between the model's predictions and the data. Thus, to obtain the solution for radial velocities and Keplerian orbits, the algorithm finds the best parameters through minimization of chi-square:

\small
\begin{equation}
\chi^2_6(P_\mathrm{o},\tau,\omega,e,K_\mathrm{d},\gamma)
= \frac{1}{N-6}\sum_\mathrm{j=1}^{N} \left(\frac{V_\mathrm{j}^\mathrm{obs}-V(t_\mathrm{j};P_\mathrm{o},\tau,\omega,e,K_\mathrm{d},\gamma)}{\sigma_\mathrm{j}}  \right)^2,
\label{eq: eq. 3}
\end{equation}
\\

\noindent
the factor $(N-6)$ of the above equation corresponds to the number of freedom degree of the fit, $V_\mathrm{{j}^{obs}}$ is the observed radial velocities in the dataset, while the $V(t_\mathrm{j};P_\mathrm{o},\tau,\omega,e,K_\mathrm{d},\gamma)$ is the theoretical radial velocities at the time $t_\mathrm{j}$, which depends on the following six parameters; wherein $P_\mathrm{o}$ represents the orbital period, $\tau$ is the time of passing through the periastron, $\omega$ is the periastron longitude, $e$ is the orbital eccentricity, $K_\mathrm{d}$ the half-amplitude of the radial velocity, $\gamma$ the system's radial velocity and the associated error estimates $\sigma_\mathrm{j}$. Thus, the theoretical radial velocities are based and computed through the equation 2.45 given by \citet{2001icbs.book.....H}:

\ \\
\begin{equation}
V(t)=\gamma + K_\mathrm{d}(\cos(\omega+\theta(t)) +e{\cos(\omega)}),
\label{eq: eq. 4}
\end{equation}
\\

\noindent
to solve the equation\,\ref{eq: eq. 4} is necessary to compute the true and eccentricity anomaly $E$. However, $E$ has not an analytic solution and thus it is necessary implement an iterative method to solve it. Instead, it is possible to find a solution for the true anomaly $\theta$ and solve the above equation (eq.\,\ref{eq: eq. 4}). For more details, the steps and equations are described by \citet{2021AJ....162...66R}. 

\begin{figure}
	\begin{center}
		\includegraphics[trim=0.0cm 0.0cm 0.0cm 0.0cm,clip,width=0.47\textwidth,angle=0]{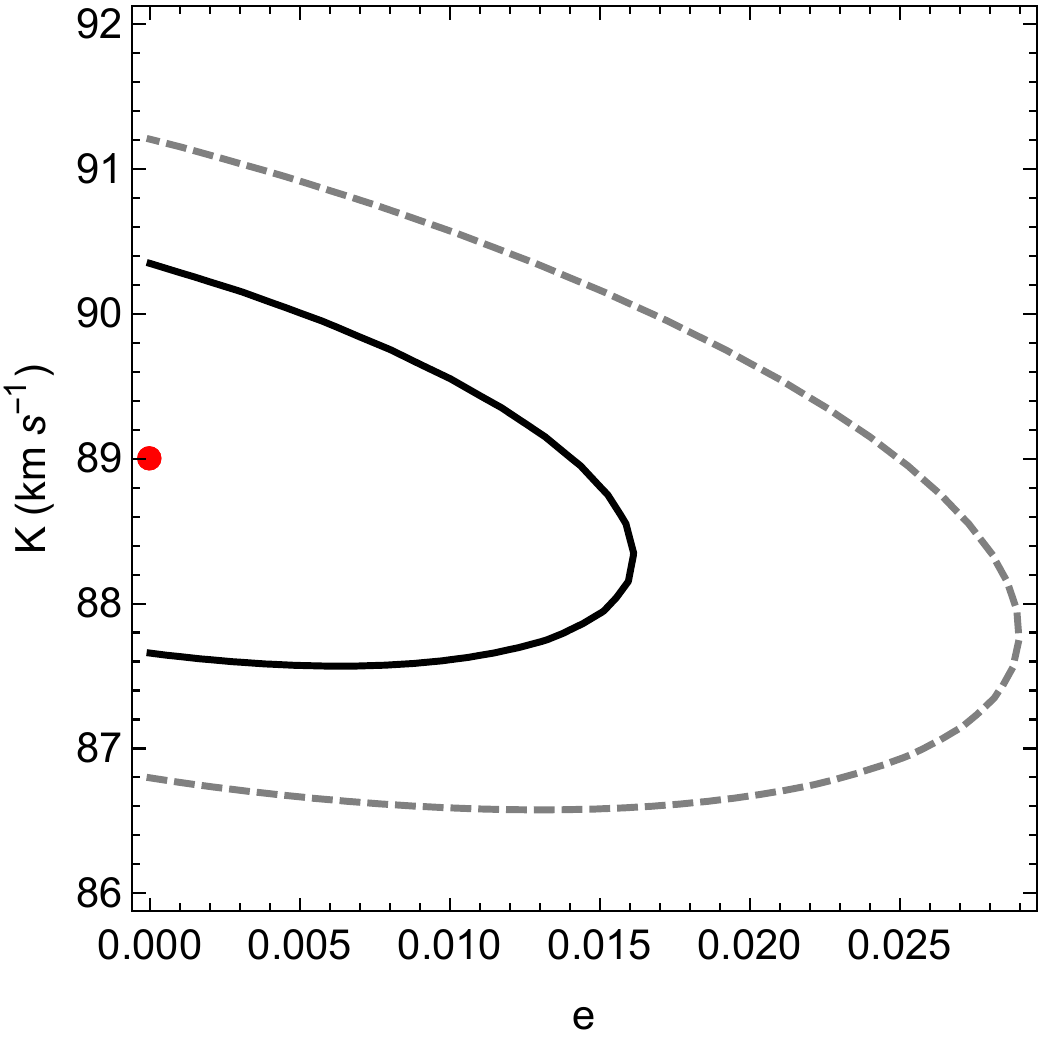}
		\includegraphics[trim=0.0cm 0.0cm 0.0cm 0.0cm,clip,width=0.5\textwidth,angle=0]{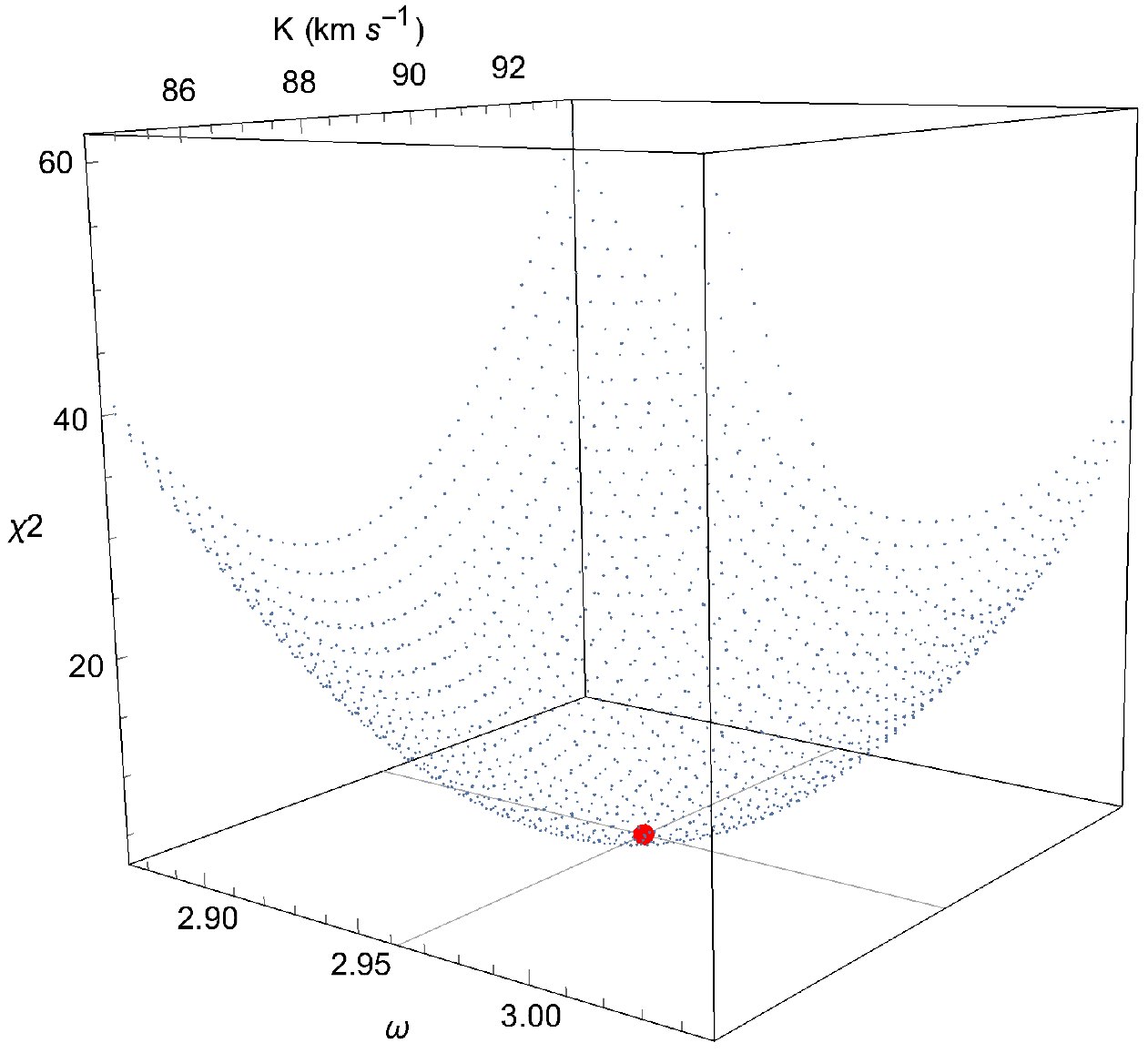}
	\end{center}
	\caption{(Upper) representation of $\chi2$ distribution in the K-e plane computed through Montecarlo simulation. The solid-black and dashed-gray lines correspond to $1\sigma$ and $2\sigma$ iso-contours, and the red dot indicates the minimum $\chi2$ solution.(Down) Contour corresponding to the $\Delta\chi2$ that includes $68.3\%$ of the probability.}
	\label{fig:Fig. 7}
\end{figure}

Since the analysis of the 806 data points of the obtained magnitudes from ASAS photometry provide us a much more accurate orbital period than the analysis of the 16 CHIRON spectra, we decided to fix the orbital period computed from the photometric analysis (Sec.\,\ref{Sec: Sec. 2}) at the moment to use \texttt{PIKAIA} and during the Montecarlo simulation to estimate the errors. It is based and supported because the absolute error for Montecarlo simulation, increases as $\sqrt{1/N}$. Thus, this reveals the importance of a large data set beside the homogeneous sampling of the orbital phase. We present the orbital parameters in Table\,\ref{Tab: Tab. 5}. It was noted that the eccentricity solution for V4142\,Sgr shows a circular orbit but for $1\sigma$ error still allows this solution (Fig.\,\ref{fig:Fig. 7}).

\begin{table}
	\caption{Orbital parameters for the donor star of V4142\,Sgr obtained through minimization of $\chi^2$ given by equation \ref{eq: eq. 4}. The value $\tau^{*}=\tau-2450000$ is given and the maximum and minimum are one isophote of 1$\sigma$.}
	\normalsize
	\begin{center}
		\begin{tabular}{lrrr}
			\hline
			\noalign{\smallskip}
			\textrm{Parameter}  				&  \textrm{Best value}	& \textrm{Low limit}& \textrm{Upper limit} \\
			\hline
			\hline
			{$\tau$}							& \textrm{302.332}	&\textrm{302.189}	&\textrm{302.474}\\
			{$e$}								& \textrm{0.000}	&\textrm{0.000}		&\textrm{0.017}\\
			{$\omega\,\mathrm{(rad)}$}			& \textrm{2.96196}	&\textrm{2.93310}	&\textrm{2.98810}\\
			{$K_{d}\,\mathrm{(km\,s^{-1})}$}	& \textrm{89.0}		&\textrm{87.4}		&\textrm{90.6}\\
			{$\gamma\,\mathrm{(km\,s^{-1})}$}	& \textrm{0.2}		&\textrm{-1.3}		&\textrm{1.7}\\
			{$\chi2$}							& \textrm{3.185}	&\textrm{}			&\textrm{}\\
			\hline
		\end{tabular}
	\end{center}
	\label{Tab: Tab. 5}
\end{table}

%%%%%%%%%%%%%%%%%%%%%%%%%%%%%%%%%%%%%%%%%%%%%%%%%%%%%%%%%%%%%%%%%%%%%%%%%%%%%%%%%%%%%%%%%%%%%%%%%%%%%%%%%%%%%%%%%%%%%%%%%%%%%%%%%%%%%%%%%%%%%%%%%%%%%%%%%%%%%%
%%%%%%%%%%%%%%%%%%%%%%%%%%%%%%%%%%%%%%%%%%%%%%%%%%%%%%%%%%%%%%%%%%%%%%%%%%%%%%%%%%%%%%%%%%%%%%%%%%%%%%%%%%%%%%%%%%%%%%%%%%%%%%%%%%%%%%%%%%%%%%%%%%%%%%%%%%%%%%
%%%%%%%%%%%%%%%%%%%%%%%%%%%%%%%%%%%%%%%%%%%%%%%%%%%%%%%%%%%%%%%%%%%%%%%%%%%%%%%%%%%%%%%%%%%%%%%%%%%%%%%%%%%%%%%%%%%%%%%%%%%%%%%%%%%%%%%%%%%%%%%%%%%%%%%%%%%%%%
%%%%%%%%%%%%%%%%%%%%%%%%%%%%%%%%%%%%%%%%%%%%%%%%%%%%%%%%%%%%%%%%%%%%%%%%%%%%%%%%%%%%%%%%%%%%%%%%%%%%%%%%%%%%%%%%%%%%%%%%%%%%%%%%%%%%%%%%%%%%%%%%%%%%%%%%%%%%%%
\subsection{Spectral disentangling}
\label{Sec: Sec. 3.3}

Due to its binary nature, V4142\,Sgr presents a series of composed absorption lines into spectra. Therefore, it was decided to remove or separate both components in order to analyze the main features of the donor and gainer star. We disentangled every spectrum through an iterative method proposed by \citet{2006A&A...448..283G}. Briefly, using a Doppler correction, each spectrum is corrected for its theoretical radial velocity obtained in the previous section (Sec.\,\ref{Sec: Sec. 3.2}). Then a combination of these spectra is made, generating a template for the flux contribution that comes from a single star but is still slightly contaminated by its partner. This process must be repeated with the other star and thus generates the first templates for each stellar component. These first templates are subtracted from each spectrum to eliminate the individual contribution of each star. In the next iteration, the previous steps will be repeated, generating new templates and subtracting so successively until convergence and a clean average spectrum for both components are successfully obtained. However, it must be emphasized that the disentangling method for the gainer and an accretion disk will not be possible to perform since there are no available models for disk spectral lines.

\begin{figure}
	\begin{center}
		\includegraphics[trim=0.0cm 0.0cm 0.0cm 0.0cm,clip,width=0.5\textwidth,angle=0]{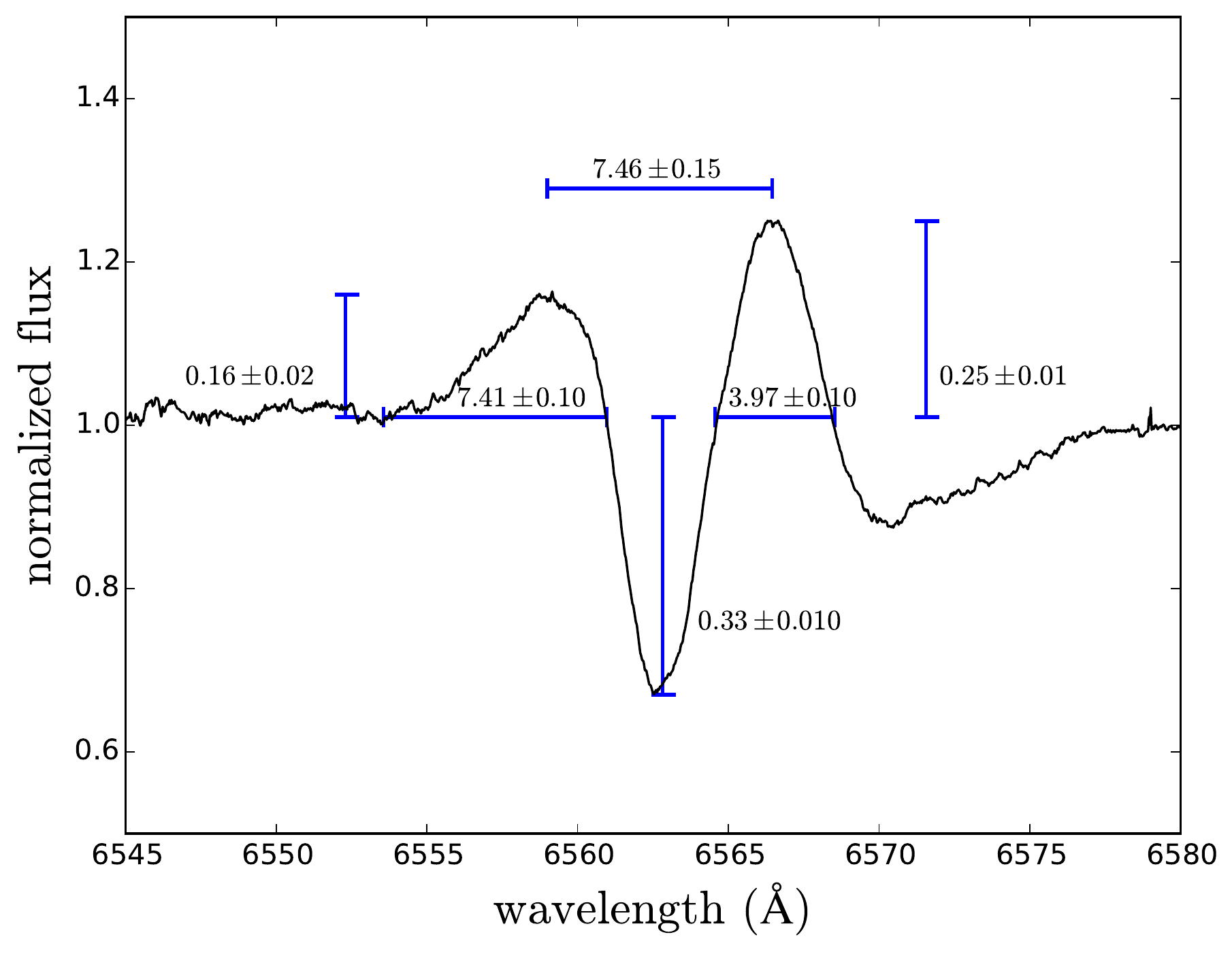}
	\end{center}
	\caption{Average H$\alpha$ profile with their respective sizes. The widths of their emissions peaks and their absorption profile were measured at the height of the continuum.}
	\label{fig:Fig. 8}
\end{figure}

\begin{figure}
	\begin{center}
		\includegraphics[trim=0.0cm 0.0cm 0.0cm 0.0cm,clip,width=0.5\textwidth,angle=0]{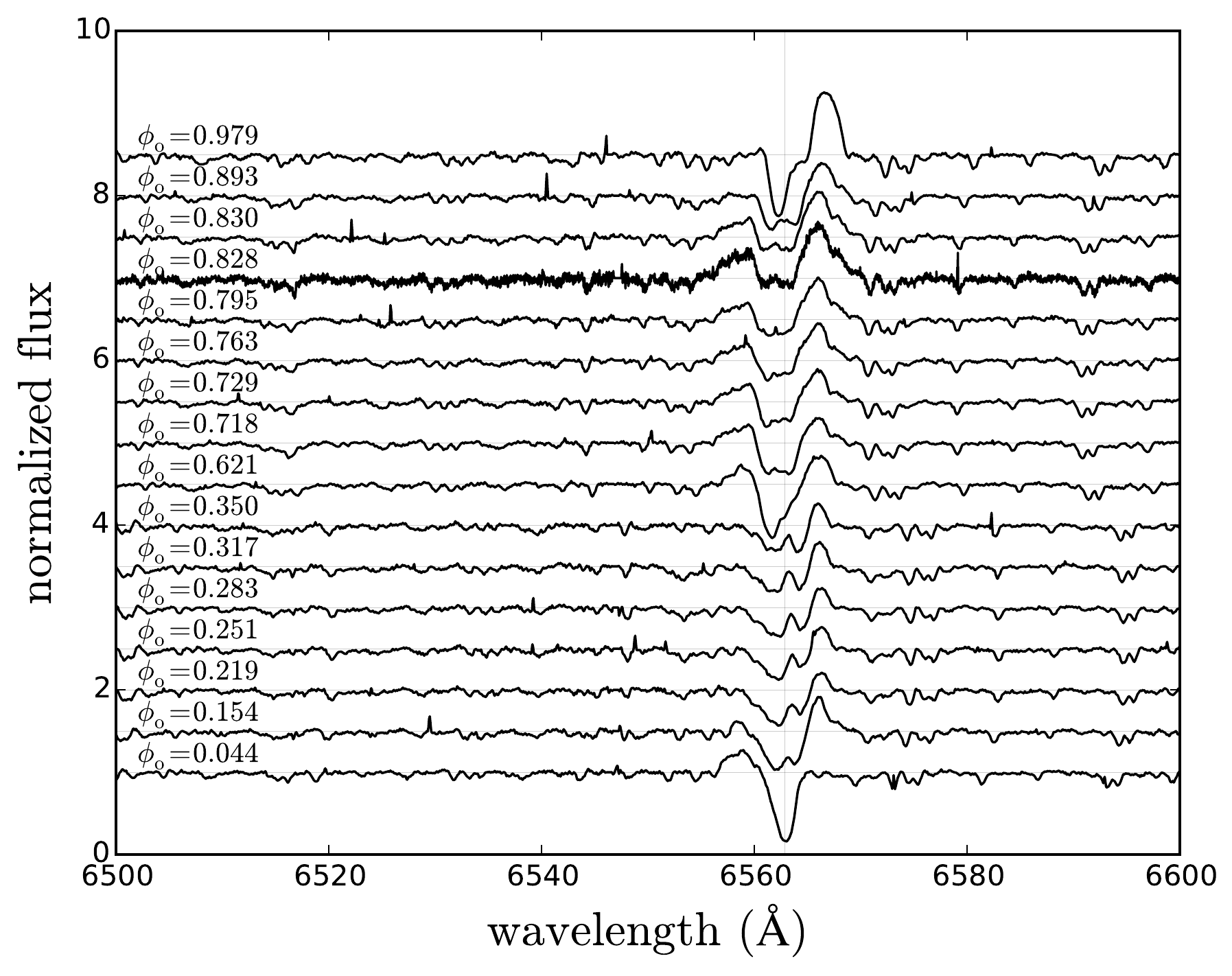}
	\end{center}
	\caption{The Behaviour of H$\alpha$ profile sorted by orbital phases and centered at 6562.817 \AA{} to compare the red-peak emission. The observed spectra show the donor and gainer flux contribution, i.e. these were not disentangled.}
	\label{fig:Fig. 9}
\end{figure}

After removing and obtaining the isolated spectra, the presence of a prominent double peak emission for the H$\alpha$ profile of the gainer star was noted. This evidence suggests us the existence of a circumstellar accretion disk and supports the interacting binary nature of the system (see Fig.\,\ref{fig:Fig. 8}). Using the peak separation we calculated that a particle located on the accretion disk rim would have a velocity $V_\mathrm{p}= 170.4 \pm 45.7 ~\mathrm{km\,s^{-1}}$.
Another interesting but not usual characteristic is the predominant persistence of the red-peak of the emission, during all observed spectra. Normally the interacting binary stars show variations of the relative intensities on the emission peaks. Thus,  this could be interpreted as an effect of the gas stream on the gainer star or a wind emerging probably from the hotspot region (Fig.\,\ref{fig:Fig. 9}). Despite the H$\alpha$ red-peak emission is large compared with the blue side peak, it is not something completely new among DPVs, because this was reported in the DPV V495\,Centauri by \citet{2018MNRAS.476.3039R}. The same $V<R$ characteristic was found in other Balmer lines in V\,4142\,Sgr.

\begin{figure}
	\begin{center}
		\includegraphics[trim=0.0cm 0.0cm 0.0cm 0.0cm,clip,width=0.5\textwidth,angle=0]{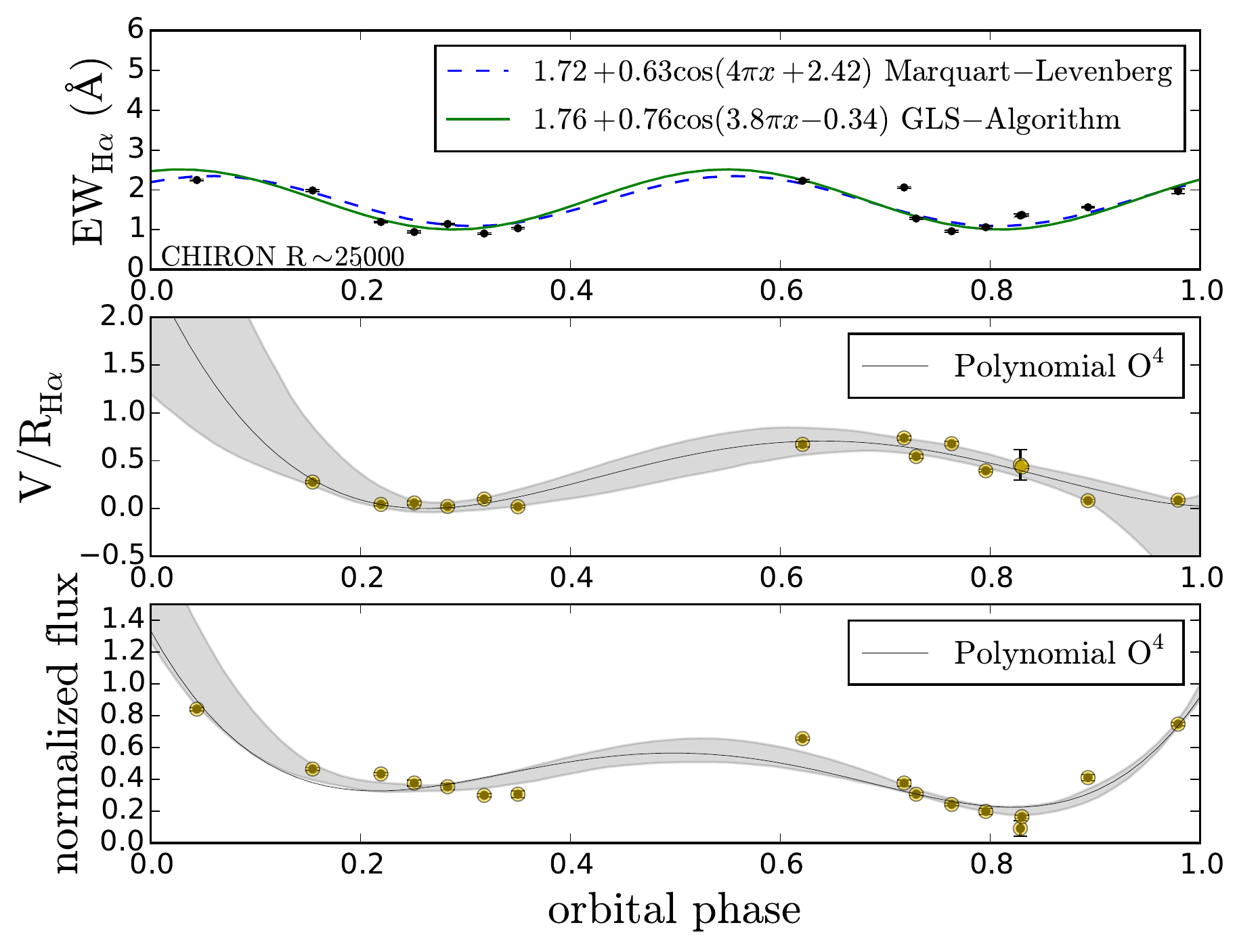}
	\end{center}
	\caption{(Top) Equivalent width of H$\alpha$ profile from gainer star during a complete orbital phase with donor contribution subtracted, using CHIRON spectrograph. The dashed blue line corresponds to a Marquart-Levenberg fit using a fixed period $0.5P_\mathrm{o}$ and the continuum green line represents a GLS fit with a free period, revealing a period of $15\fd95$. (Center) violet and red intensities ($V/R$) ratio of H$\alpha$ profile, measured from the continuum of the gainer star. (Bottom) Variations of the depth of absorption line of H$\alpha$ measured in gainer star.} 
	\label{fig:Fig. 10}
\end{figure}

We quantified the ratio $V/R$ of the H$\alpha$ profile commonly defined as the ratio between the peak relative intensities to the normalized continuum $V/R=(I_\mathrm{v}-1)/(I_\mathrm{r}-1)$, and a non-cyclical variation was noted, where the maximum values are in the second quadrature (Fig.\,\ref{fig:Fig. 10}-center). The equivalent width (EW) shows a cyclical variation, which varies during the orbital period. Thus, through a Marquart-Levenberg non-linear least square, a sinusoidal function was fitted through a fixed period of $0.5P_\mathrm{o}$ and obtained a half amplitude of $0.63 \pm 0.08 ~\AA{}$  (Fig.\,\ref{fig:Fig. 10}- dashed blue line). However, to strengthen the fit and the results we decided to implement the \texttt{GLS} algorithm, and this revealed a period $P_\mathrm{o/EW}= 15\fd95 \pm 0\fd16$ and an amplitude $A_\mathrm{EW}= 0.75 \pm 0.05 ~\AA{}$ (Fig.\,\ref{fig:Fig. 10}-continuum green line).

%%%%%%%%%%%%%%%%%%%%%%%%%%%%%%%%%%%%%%%%%%%%%%%%%%%%%%%%%%%%%%%%%%%%%%%%%%%%%%%%%%%%%%%%%%%%%%%%%%%%%%%%%%%%%%%%%%%%%%%%%%%%%%%%%%%%%%%%%%%%%%%%%%%%%%%%%%%%%%
%%%%%%%%%%%%%%%%%%%%%%%%%%%%%%%%%%%%%%%%%%%%%%%%%%%%%%%%%%%%%%%%%%%%%%%%%%%%%%%%%%%%%%%%%%%%%%%%%%%%%%%%%%%%%%%%%%%%%%%%%%%%%%%%%%%%%%%%%%%%%%%%%%%%%%%%%%%%%%
%%%%%%%%%%%%%%%%%%%%%%%%%%%%%%%%%%%%%%%%%%%%%%%%%%%%%%%%%%%%%%%%%%%%%%%%%%%%%%%%%%%%%%%%%%%%%%%%%%%%%%%%%%%%%%%%%%%%%%%%%%%%%%%%%%%%%%%%%%%%%%%%%%%%%%%%%%%%%%
%%%%%%%%%%%%%%%%%%%%%%%%%%%%%%%%%%%%%%%%%%%%%%%%%%%%%%%%%%%%%%%%%%%%%%%%%%%%%%%%%%%%%%%%%%%%%%%%%%%%%%%%%%%%%%%%%%%%%%%%%%%%%%%%%%%%%%%%%%%%%%%%%%%%%%%%%%%%%%
\subsection{Determination of secondary component physical parameters}
\label{Sec: Sec. 3.4}

After the disentangling between the donor and gainer star described in the previous section (Sec.\,\ref{Sec: Sec. 3.3}), we have determined the physical parameters of both components. We compared the observed with a grid theoretical spectra to find the best fit and identify the physical parameters of this. A grid of synthetic spectra was constructed, modeling the stellar atmosphere through \texttt{SPECTRUM}\footnote{\url{http://www.appstate.edu/~grayro/spectrum/spectrum.html}} code \citep{2001AJ....121.2159G,1994AJ....107..742G} where the model atmospheres are provided by \texttt{ATLAS9} \citep{2003IAUS..210P.A20C} in Local Thermodynamic Equilibrium (LTE).

Therefore, a grid was constructed for the donor star with different free parameters such as effective temperature ($T_\mathrm{eff}$), surface gravity ($\log{g}$), macro-turbulence velocity ($v_\mathrm{mac.}$), micro-turbulence velocity ($v_\mathrm{mic.}$), rotational velocity ($v\sin{i}$), even the veiling factor ($\eta$). This last parameter corresponds to a constant of proportionality between the theoretical spectrum and the observed due to the light contribution from its companion, which has veiled the absorption lines. All spectra were constructed using a metallicity index similar to the sun and a fixed mixing length parameter $l/\mathrm{H}=1.25$ by default from the grid, which corresponds to the disruption and dispersion of the blobs that travel in a convecting fluid during a \textquotedblleft{l}\textquotedblright distance from their equilibrium position in an atmosphere of scale height H. 

A chi-square optimization algorithm was implemented, which consists in the minimization of the deviation between the theoretical normalized and the observed average spectrum of the donor. The theoretical model was produced varying the previous mentioned parameters of the following way; the effective temperature varies from $3500 \leq T_\mathrm{d} \leq 9000 ~\mathrm{K}$ with step of $250 ~\mathrm{K}$, the surface gravity varies from  $0.0 \leq \log{g} \leq 5.0 ~\mathrm{dex}$ with step of $0.5~\mathrm{dex}$, the macro turbulence velocity varies from $0 \leq v_\mathrm{mac.} \leq 10 ~\mathrm{km\,s^{-1}}$ with steps of $1~\mathrm{km\,s^{-1}}$ whereas the micro-turbulence velocity there is 0.0 and 2.0 $\mathrm{km\,s^{-1}}$, the rotational velocity $10 \leq v\sin{i} \leq 150 ~\mathrm{km\,s^{-1}} $ with steps of $10 ~\mathrm{km\,s^{-1}}$ and the veiling factor $0.0 \leq \eta_{5400-5700\,\AA{}} \leq 1.0$ with steps $0.1$ 

\begin{figure}
	\begin{center}
		\includegraphics[trim=4.0cm 1.5cm 2.0cm 3.0cm,clip,width=0.5\textwidth,angle=0]{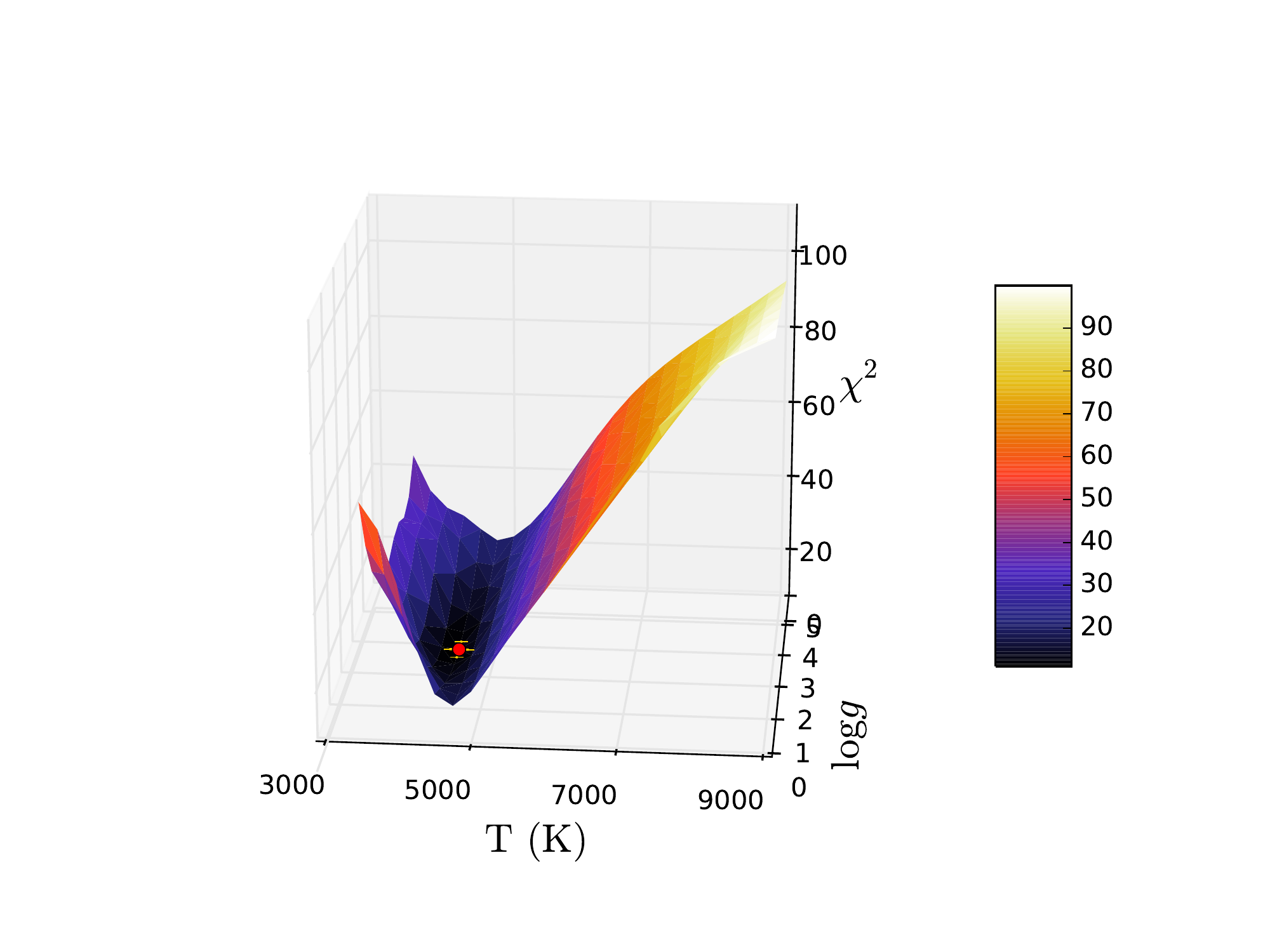}
	\end{center}
	\caption{Surface color map of chi-square analysis for the donor star. The best model converges with a $\chi2_{6}=10.822$ and it is represented by a red dot and their respective errorbars at $T_\mathrm{d}=4500 \pm 125 ~\mathrm{K}$ and $\log{g}=2.5 \pm 0.125 ~\mathrm{dex}$.}
	\label{fig:Fig. 11}
\end{figure}

\begin{figure}
	\begin{center}
		\includegraphics[trim=0.0cm 0.0cm 0.0cm 0.0cm,clip,width=0.5\textwidth,angle=0]{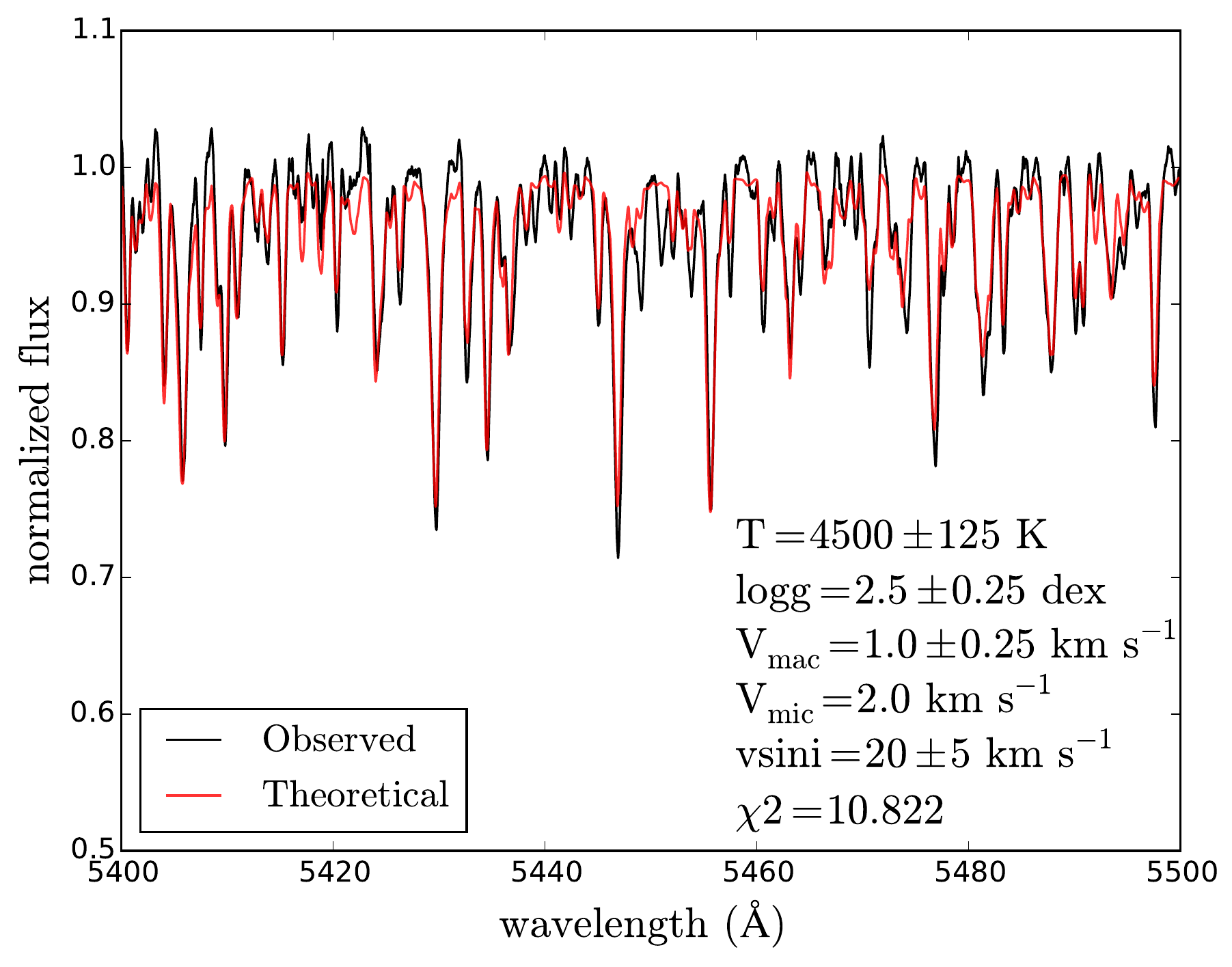}
	\end{center}
	\caption{Comparison between the disentangled observed (black) and theoretical (red) spectrum of the donor star and $\eta_{5400-5500 \AA{}}=0.3$.}
	\label{fig:Fig. 12}
\end{figure}

\begin{figure*}
	\begin{center}
		\includegraphics[trim=0.0cm 0.0cm 0.0cm 0.0cm,clip,width=1.0\textwidth,angle=0]{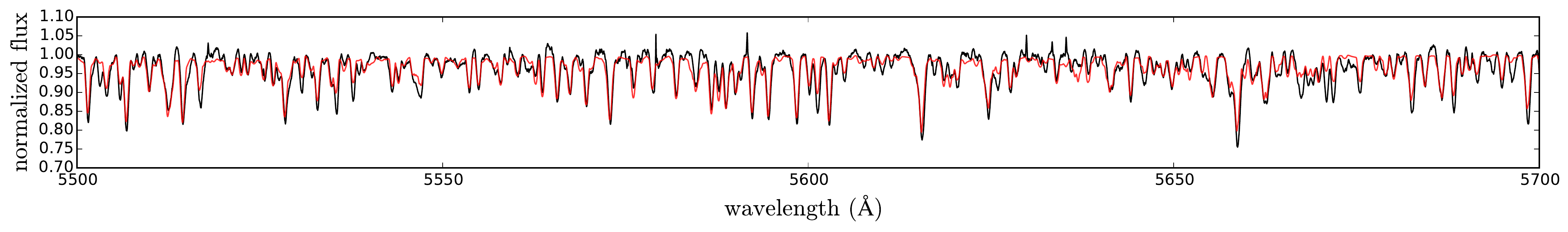}
	\end{center}
	\caption{A detailed comparison between the best theoretical model using a veiling factor {$\eta_{5500-5700\AA{}}=0.3$} (red-line) and the average observed spectrum (black-line) for the donor star.}
	\label{fig:Fig. 13}
\end{figure*}

The implemented method converged successfully to a minimum chi-square at $T_\mathrm{d}=4500 \pm 125 ~\mathrm{K}$, $\log{g}=2.5 \pm 0.125 ~\mathrm{dex}$, $v_\mathrm{mac.}=1.0 \pm 0.25  ~\mathrm{km\,s^{-1}}$, $v_\mathrm{mic.}=2.0 ~\mathrm{km\,s^{-1}}$, $v\sin{i}=20 \pm 5~\mathrm{km\,s^{-1}}$ and $\eta=0.3 \pm 0.05$ for six freedom degrees with a best value of $\chi^2_{6}=10.822$ (see Fig.\,\ref{fig:Fig. 11} and Fig.\,\ref{fig:Fig. 12},\,\ref{fig:Fig. 13}).

\ \\

%%%%%%%%%%%%%%%%%%%%%%%%%%%%%%%%%%%%%%%%%%%%%%%%%%%%%%%%%%%%%%%%%%%%%%%%%%%%%%%%%%%%%%%%%%%%%%%%%%%%%%%%%%%%%%%%%%%%%%%%%%%%%%%%%%%%%%%%%%%%%%%%%%%%%%%%%%%%%%
%%%%%%%%%%%%%%%%%%%%%%%%%%%%%%%%%%%%%%%%%%%%%%%%%%%%%%%%%%%%%%%%%%%%%%%%%%%%%%%%%%%%%%%%%%%%%%%%%%%%%%%%%%%%%%%%%%%%%%%%%%%%%%%%%%%%%%%%%%%%%%%%%%%%%%%%%%%%%%
%%%%%%%%%%%%%%%%%%%%%%%%%%%%%%%%%%%%%%%%%%%%%%%%%%%%%%%%%%%%%%%%%%%%%%%%%%%%%%%%%%%%%%%%%%%%%%%%%%%%%%%%%%%%%%%%%%%%%%%%%%%%%%%%%%%%%%%%%%%%%%%%%%%%%%%%%%%%%%
%%%%%%%%%%%%%%%%%%%%%%%%%%%%%%%%%%%%%%%%%%%%%%%%%%%%%%%%%%%%%%%%%%%%%%%%%%%%%%%%%%%%%%%%%%%%%%%%%%%%%%%%%%%%%%%%%%%%%%%%%%%%%%%%%%%%%%%%%%%%%%%%%%%%%%%%%%%%%%
\subsection{Temperature indicators} 
\label{Sec: Sec. 3.6}

To date, the temperature indicators are completely useful to examine some DPV stars which tend to be biased or veiled by other flux contributions. These indicators are considered indispensable in the spectroscopic investigations, such as: He\,I\,4471\,\AA{}, Mg\,II\,4482\,\AA{}, Fe\,II\, 4128.735\,\AA{}, Si\,II\,4130.884\,\AA{}. However, it is not always possible to find them close to wavelengths around the visual. Despite that, there is other indicators like O\,I\, 7771.96\,\AA{} ($\mathrm{3s^{5}S^{0} - 3p^{5}P}$, multiplet 1) used by \citet{1992PASJ...44..309T,1997PASJ...49..367T,2008JKAS...41...83T,2016PASJ...68...32T} for the study of the behavior of microturbulence velocity, abundances and efficiency rotational in B/A/F-type stars.

\begin{figure}
	\begin{center}
		\includegraphics[trim=0.2cm 0.0cm 0.0cm 0.0cm,clip,width=0.5\textwidth,angle=0]{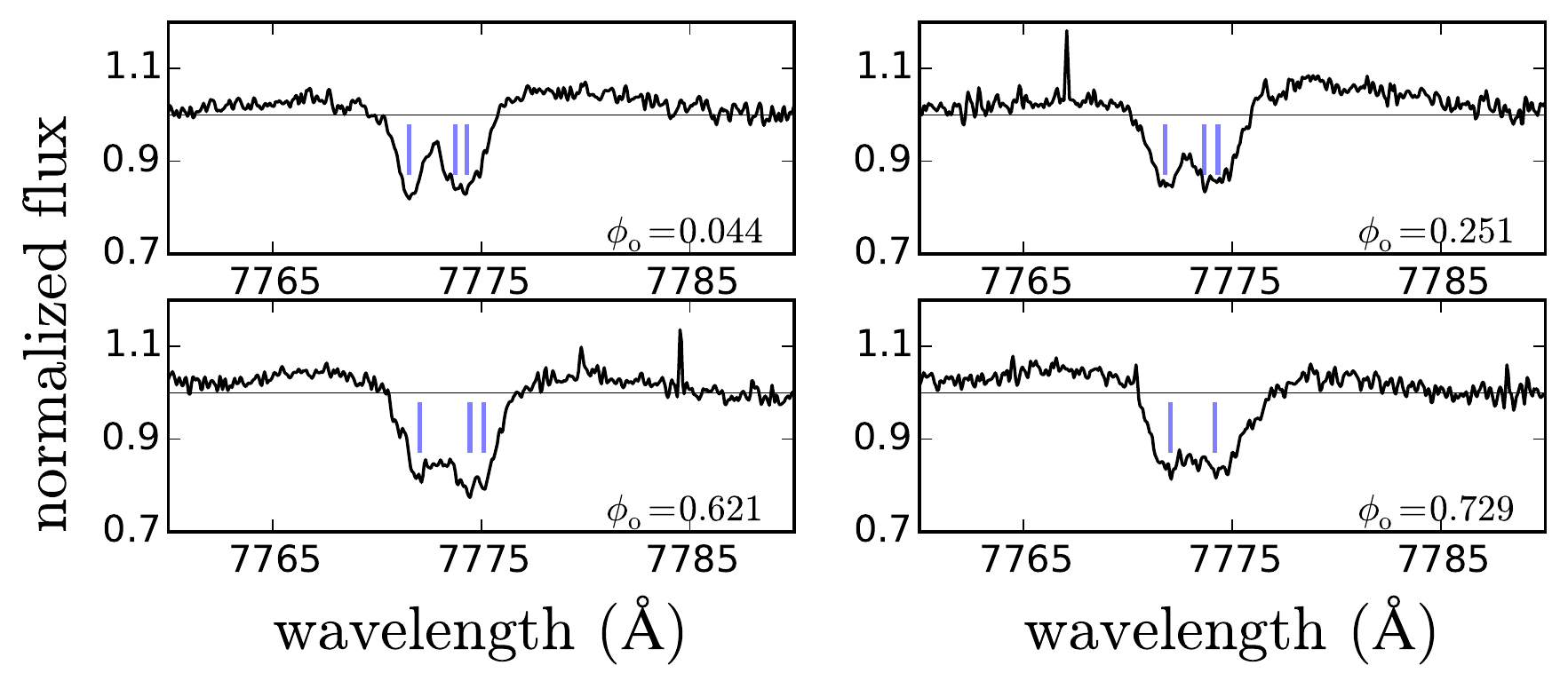}
	\end{center}
	\caption{Oxygen absorption line O\,I\,7771.96\,\AA{} ($\mathrm{3s^{5}S^{0} - 3p^{5}P}$, multiplet 1) of observed spectra (without donor contribution) at different orbital phases, showing the triplet.}
	\label{fig:Fig. 14}
\end{figure}

Based on that, the O\,I\,7771.96,\AA{} multiplet was identified in the present spectra without disentangling (Fig.\,\ref{fig:Fig. 14}). Thus, to estimates the temperature of the gainer star, we performed the procedure of subsection \ref{Sec: Sec. 3.4} to the average spectrum of the gainer star. This consists in the minimization of the deviation between the theoretical LTE normalized and the observed average spectrum of the gainer, performed in the range from 7700\,\AA{} to 7800\,\AA{}. The theoretical models were constructed varying the physical parameters of the following way; the effective temperature is composed for two group $8000 \leq T_\mathrm{g} \leq 11,000 ~\mathrm{K}$ with step of $250\mathrm{K}$ and from $11,000 \leq T_\mathrm{g} \leq 25,000 ~\mathrm{K}$ with step of $1000\mathrm{K}$, while the surface gravity run from $0.0 \leq \log{g} \leq 5.0 ~\mathrm{dex}$ with steps of $0.5 ~\mathrm{dex}$, the macroturbulence velocity varies from $0 \leq v_\mathrm{mac.} \leq 10 ~\mathrm{km\,s^{-1}}$ with steps of $1 ~\mathrm{km\,s^{-1}}$ and the microturbulence velocity was analyzed for 0.0 and 2.0 $\mathrm{km\,s^{-1}}$, while the rotational velocities varies from $10 \leq v\sin{i} \leq 150 ~\mathrm{km\,s^{-1}}$ with steps of $10~\mathrm{km\,s^{-1}}$ and finally the veiling factor varies from $0.0 \leq \eta_{7700-7800\,\AA{}} \leq 1.0 $ with steps $0.1$ (dimensionless).

\begin{figure}
	\begin{center}
		\includegraphics[trim=0.2cm 0.0cm 0.0cm 0.0cm,clip,width=0.5\textwidth,angle=0]{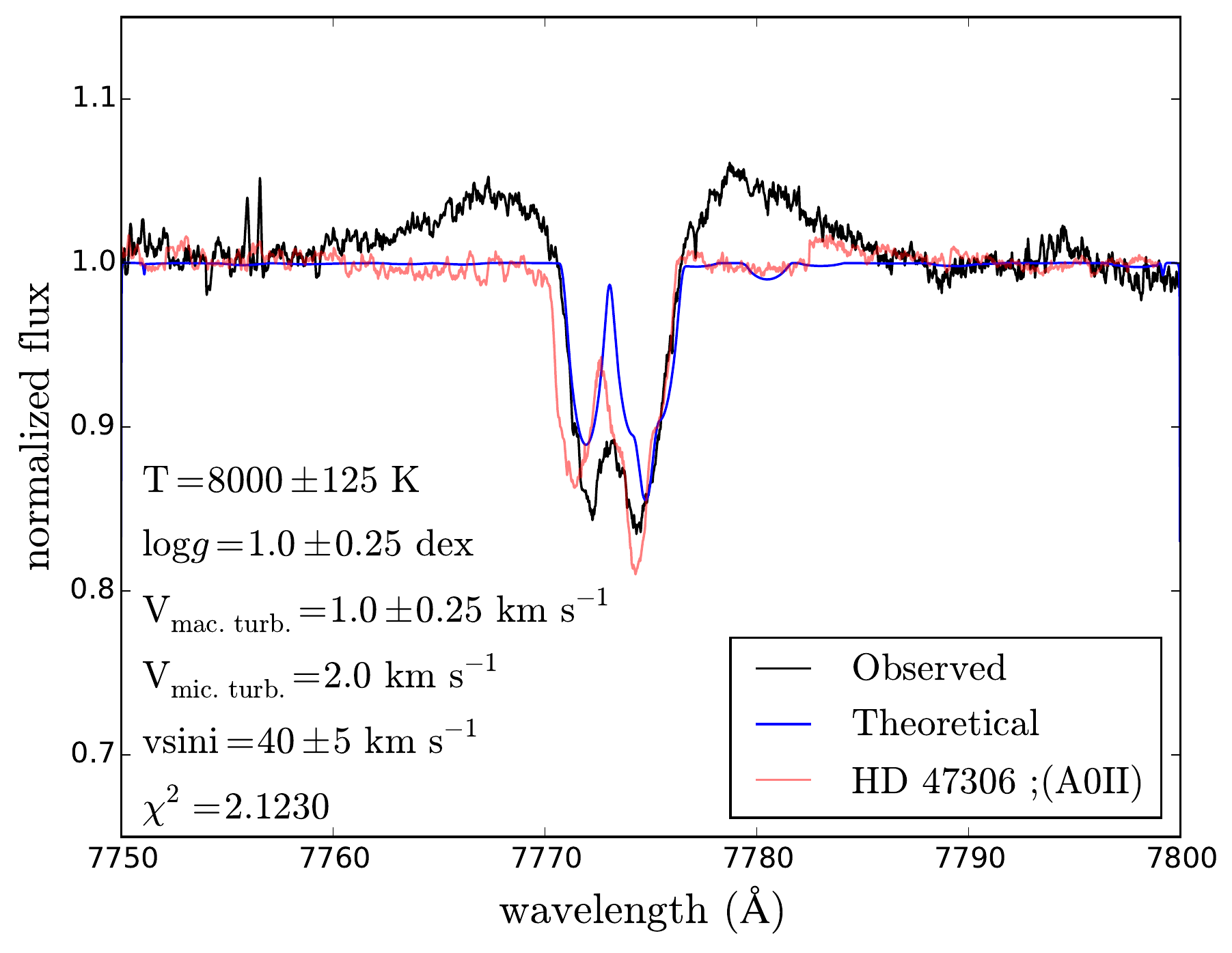}
	\end{center}
	\caption{Comparison between the average disentangled gainer (black line) and theoretical LTE spectrum (blue line) for  O\,I\,7771.96\,\AA{} ($\mathrm{3s^{5}S^{0} - 3p^{5}P}$, multiplet 1) without donor contribution using a veling factor $\eta_{7700-7800\,\AA{}}=0.9$. Red line corresponds to A0II-type star (HD\,47306).}
	\label{fig:Fig. 15}
\end{figure}

\begin{figure}
	\begin{center}	
		\includegraphics[trim=0.0cm 0.0cm 0.0cm 0.0cm,clip,width=0.5\textwidth,angle=0]{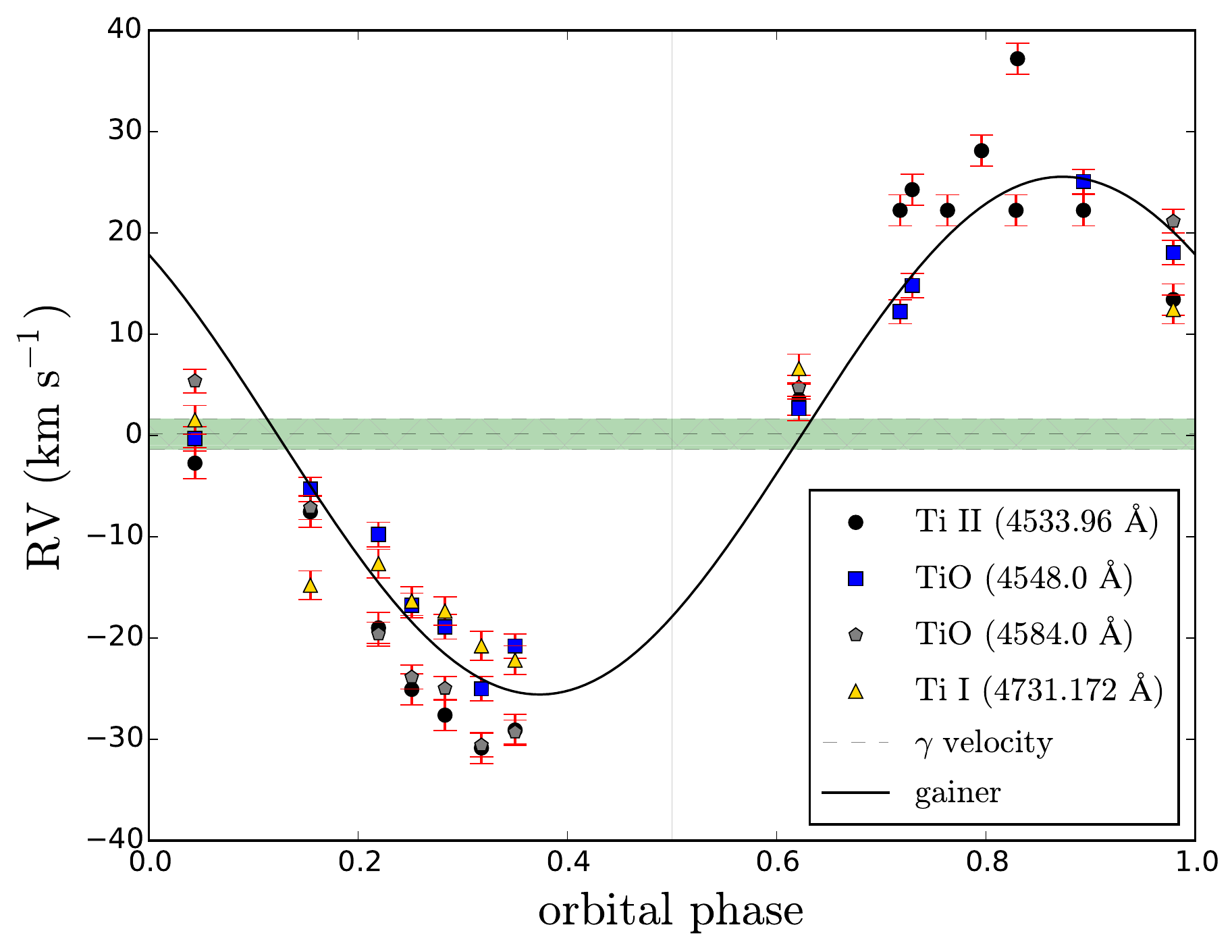}
	\end{center}
	\caption{Radial velocities for temperature indicators of gainer star. However, these are biased or they come from the same disk. Green dashed area corresponds to the system velocity $\gamma$.}
	\label{fig:Fig. 16}	
\end{figure}

The theoretical LTE model converged successfully for a minimum chi-square $\chi^2_{6}=2.1230$ for a effective temperature of $8000 \pm 125 ~\mathrm{K}$, a surface gravity of $1.0 \pm 0.25 ~\mathrm{dex}$, a macro-turbulence velocity of $1.0 \pm 0.25 ~\mathrm{km\,s^{-1}}$, the micro-turbulence of $2.0 ~\mathrm{km\,s^{-1}}$, a rotational velocity of $40 \pm 5 ~\mathrm{km\,s^{-1}}$, and a veiling factor $\eta_{7700-7800\,\AA{}}=0.9 \pm 0.05$, see Fig.\,\ref{fig:Fig. 15}. We notice the emission shoulders in the blue (orbital phases 0.6 and 0.7) and red (orbital phases 0.0 and 0.2) wings in Fig.\,\ref{fig:Fig. 14} and also in  Fig.\,\ref{fig:Fig. 15}. We interpret these emissions as true spectral features probably signatures of the circumstellar matter around the gainer. Similar emission wings with the same timing were reported in Balmer lines of the DPV V\,393 Scorpius. They were interpreted as due to photon scattering in a wind region \citep{2012MNRAS.427..607M}. In addition, the equivalent width for O\,I multiplet was measured directly on the average gainer spectrum and we obtained an $\mathrm{EW_{O\,I}=0.67\pm 0.03}$ and compared with HD\,47306 (AOII-type) provided by UVES-POP\footnote{\url{http://www.eso.org/sci/observing/tools/uvespop.html}}\citep{2003Msngr.114...10B}, with an effective temperature of $ 9738 \pm 145 ~\mathrm{K}$ provided by ExoFOP-TESS, and veiled for this purpose,  and we measured an equivalent width of $\mathrm{EW=0.65 \pm 0.01}$. In addition, it was noted that the pronounced in V and R intensities could be caused by the accretion disk, and the differences are possibly caused by emission on the hot spot on the accretion disk. Therefore, the temperature indicator could be affected by the flux contribution from the accretion disk, and for now, it is only possible to estimate the range of the effective temperature of the accretion disk since it could be creating its own atmosphere. The results for the temperature and surface gravity of the gainer star differ from the typical nature of the DPV systems, wherein the gainer star is always a B-type dwarf star. Despite that, this and other oxygen lines could explain the reason for the observed lag obtained in the Radial Velocities analysis, because although these lines follow the movement of the gainer star, they would belong to the accretion disk and create this observed lag in the RV of the gainer. Following this reasoning, we have measured a series of other absorption lines, such as Ti\,II (4536\,\AA{}), TiO\,(4548.0\,\AA{}), TiO\,(4584.0\,\AA{}), and Ti\,I (4731.172\,\AA{}), in order to improve our understanding about the gainer and its accretion disk. However, these lines are representative of single stars with a surface gravity and effective temperature less than a B-type dwarf, varying from spectral types A0 to M2 type. Therefore, the accretor star is being hidden by its accretion disk and at the same time, the movement that we observe from the RVs measurements corresponds to the accretion disk that follows the movement of its accretor (see Fig.\,\ref{fig:Fig. 16}).  

\ \\

%%%%%%%%%%%%%%%%%%%%%%%%%%%%%%%%%%%%%%%%%%%%%%%%%%%%%%%%%%%%%%%%%%%%%%%%%%%%%%%%%%%%%%%%%%%%%%%%%%%%%%%%%%%%%%%%%%%%%%%%%%%%%%%%%%%%%%%%%%%%%%%%%%%%%%%%%%%%%%
%%%%%%%%%%%%%%%%%%%%%%%%%%%%%%%%%%%%%%%%%%%%%%%%%%%%%%%%%%%%%%%%%%%%%%%%%%%%%%%%%%%%%%%%%%%%%%%%%%%%%%%%%%%%%%%%%%%%%%%%%%%%%%%%%%%%%%%%%%%%%%%%%%%%%%%%%%%%%%
%%%%%%%%%%%%%%%%%%%%%%%%%%%%%%%%%%%%%%%%%%%%%%%%%%%%%%%%%%%%%%%%%%%%%%%%%%%%%%%%%%%%%%%%%%%%%%%%%%%%%%%%%%%%%%%%%%%%%%%%%%%%%%%%%%%%%%%%%%%%%%%%%%%%%%%%%%%%%%
%%%%%%%%%%%%%%%%%%%%%%%%%%%%%%%%%%%%%%%%%%%%%%%%%%%%%%%%%%%%%%%%%%%%%%%%%%%%%%%%%%%%%%%%%%%%%%%%%%%%%%%%%%%%%%%%%%%%%%%%%%%%%%%%%%%%%%%%%%%%%%%%%%%%%%%%%%%%%%
\subsection{Mass ratio, Roche lobe radius and structure}
\label{Sec: Sec. 3.7}

\begin{figure}
	\begin{center}
		\includegraphics[trim=0.0cm 0.0cm 0.0cm 0.0cm,clip,width=0.5\textwidth,angle=0]{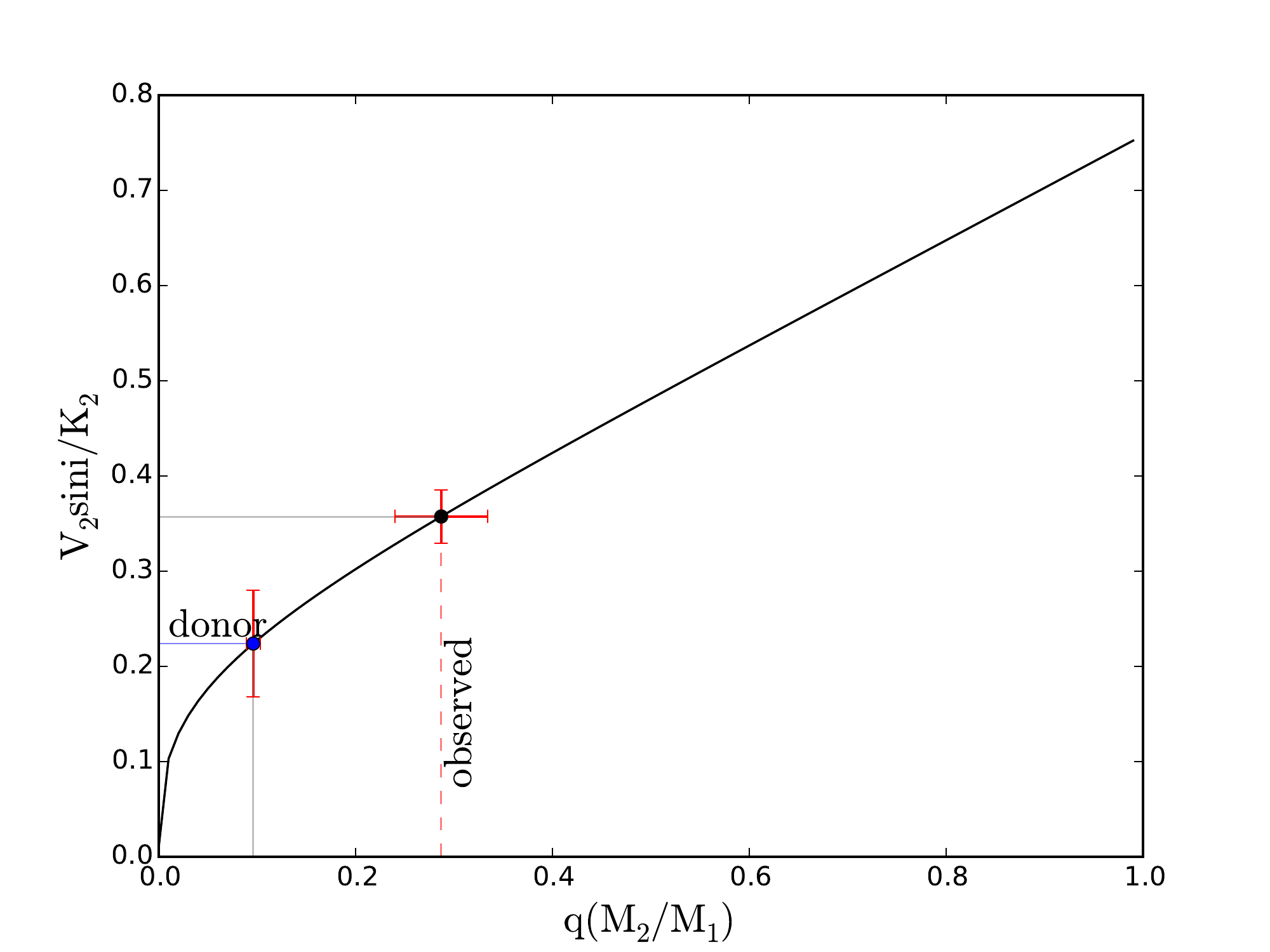}
	\end{center}
	\caption{A solid black line is given by the equation \ref{eq: eq. 6}. The dashed red line represents the obtained mass ratio ($q=0.287$) from RVs analysis (Synchronous) and the continuum blue line represents the suggested mass ratio ($q_\mathrm{theo.}=0.096$) given by the equation \ref{eq: eq. 6} (sub-synchronous).}
	\label{fig:Fig. 17}
\end{figure}

Since we have confirmed the interacting binary nature of V4142\,Sgr and the existence of an accretion disk, we can assume a semidetached nature of this system, which is a characteristic of the binaries Algol-type. Therefore, it is assumed that the less massive star, in this case, the donor star has filled the Roche lobe. Then, beyond the synchronism concept, we decided to analyze the ratio between the effective Roche lobe geometry of the donor star $R_\mathrm{L}$ concerning the orbital separation, using the following equation given by \citet{1983ApJ...268..368E}:

\ \\
\begin{equation}
\frac{R_\mathrm{d}}{a} = \frac{0.49q^{2/3}}{ 0.6q^{2/3}+ \ln(1+q^{1/3})}= 0.278 \pm 0.011,
\label{eq: eq. 5}
\end{equation}

\ \\
\noindent
for a mass ratio $q=0.287 \pm 0.047$, whose value is consistent with binary interactions of conservative process. Other relevant analysis for the DPVs is the study of the synchronism of the systems. Thus, to analyze the compatibility of a synchronously rotating donor star and filling the Roche Lobe  we used the composite equation between equations 3.5 and 3.9 provided by \citet{2006epbm.book.....E}, which is valid for system in synchronization with:

\ \\
\begin{equation}
\frac{v_\mathrm{d,rot}\sin i}{K}\approx (1+q)\frac{0.49q^{2/3}}{ 0.6q^{2/3}+ \ln(1+q^{1/3})},
\label{eq: eq. 6}
\end{equation}

\ \\
\noindent
Using the obtained values for the semi-amplitude of the donor $K_\mathrm{d}= 89.2 \pm 0.5 ~\mathrm{km\,s^{-1}}$, its rotational velocity $V_\mathrm{d,rot}\sin{i}=20 \pm 5 ~\mathrm{km\,s^{-1}}$ obtained in the Section \ref{Sec: Sec. 3.4} and using the above equation (eq.\,\ref{eq: eq. 6}), suggests a mass ratio $q_\mathrm{theo}=0.096 \pm 0.007$, i.e. the theoretical mass ratio should be one-third of the observed. This difference suggests that the system is non-synchronous (see Fig.\,\ref{fig:Fig. 17}). In addition, it is known that the mass ratio obtained from  section\,\ref{Sec: Sec. 3.2} comes from the kinematics of the binary system whose analysis does not consider the proper rotation of each star.

\ \\

\begin{figure}
	\begin{center}
		\includegraphics[trim=0.0cm 0.0cm 0.0cm 0.0cm,clip,width=0.5\textwidth,angle=0]{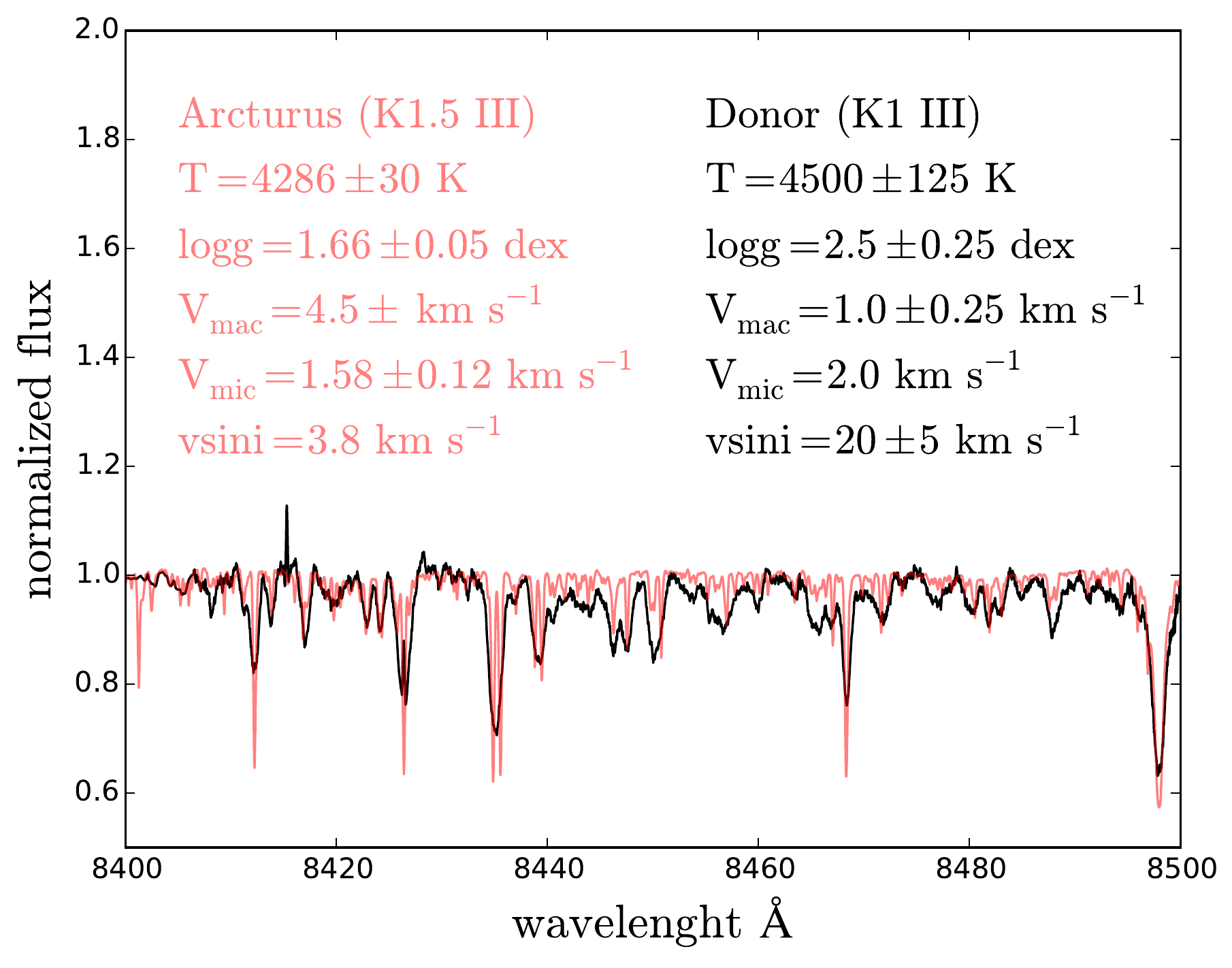}
	\end{center}
	\caption{A detailed comparison between the observed spectrum of Donor star of the Galactic DPV V4142\,Sgr (black line), and a single star Arcturus (red line) (previously veiled). Both show similar spectral types and parameters.}
	\label{fig:Fig. 18}
\end{figure}

\ \\
Despite that, the mean density $\bar{\rho}_\mathrm{d} ~\mathrm{(g\,cm^{-3})}$ of the donor star was quantified through the mass ratio from RVs analysis and the following relation $\bar{\rho}_\mathrm{d}= 3M_\mathrm{d}/4\pi R_{d}^{3}$, $M_\mathrm{d}=Mq/(1+q)$ and using the Kepler's third law to obtain the following equation: 

\ \\
\begin{equation}
\bar{\rho}_\mathrm{d}= \frac{3q}{(1+q)}\frac{1}{(R_{\mathrm{d}}/a)^{3}}\frac{\pi}{G P^{2}}=2.09\times10^{-4} \pm 1.58 \times 10^{-7} \mathrm{(g\,cm^{-3})},
\label{eq: eq. 7}
\end{equation}

\ \\
\noindent
the obtained value close to the expected one for a giant K1\,III type (see Fig.\,\ref{fig:Fig. 18}), which is complemented by the temperature and the rotational velocity obtained in the Section \ref{Sec: Sec. 3.4}. Another, interesting relation linked to the mean density of a star that fills its Roche Lobe and the effective radius $R_\mathrm{L}$ \citet{1983ApJ...268..368E,2006epbm.book.....E} is the critical period, which corresponds to the shortest possible period (days) for a binary system of given mass ratio into which a star of given mean density (in solar units, i.e. $M_\mathrm{d}=4\pi R^{3}_\mathrm{L}/3$), which can be fitted without overflowing its Roche Lobe:

\begin{equation}
P_\mathrm{cr}\sqrt{\bar{\rho}}= \left(\frac{3\pi}{G} \right)^{1/2} \left(\frac{q}{1+q}\right)^{1/2}x_\mathrm{L}^{-3/2}=0.444 \pm 0.001
\label{eq: eq. 8}
\end{equation}

\ \\
\noindent
Then, the critical period can be redefined using equations \ref{eq: eq. 7} and \ref{eq: eq. 8}. Following this line, we computed a critical period of $P_\mathrm{cr}= 36\fd5 \pm 0\fd1$, which is greater than to the currently observed in $5\fd867 \pm 0\fd003$. In order to constraint the system we decided to analyze the ratio between the stellar radius and the mass as determined by its internal structure, i.e. the response of the donor's Roche Lobe radius to the mass transfer depends on the mass ratio and the accretion efficiency $0\leq \epsilon \leq 1$ ($\epsilon=1$ conservative, $\epsilon=0$ non-conservative mass transfer), using the following dimensionless equation:

\ \\
\begin{equation}
R'_\mathrm{L}\equiv \frac{\partial \ln{R_\mathrm{L}}}{\partial \ln M_\mathrm{d}}= \frac{\partial \ln{a}}{\partial \ln{M_\mathrm{d}}} + \frac{\partial \ln {R_\mathrm{L}/a}}{\partial \ln {q}}{\frac{\partial \ln{q}}{\partial \ln{M_\mathrm{d}}}}.
\label{eq: eq. 9}
\end{equation}

\ \\
\noindent
In detail, the equation of the logarithmic derivate of the orbital separation is composed by a first-term related to the mass loss/transfer:

\ \\
\begin{equation}
\frac{\partial\ln{a}}{\partial\ln{M_\mathrm{d}}}= \frac{2M_\mathrm{d}^{2}-2M_\mathrm{g}^2-M_\mathrm{d}M_\mathrm{g}(1-\beta)}{M_\mathrm{g}(M_\mathrm{d}+M_\mathrm{g})}.
\label{eq: eq. 10}
\end{equation}

\ \\
\noindent
Since the second term of the eq.\,\ref{eq: eq. 9} is composed by two terms more, and in resume, it is the Roche Lobe's response to the change in mass ratio \citep{1983ApJ...268..368E}:

\ \\
\begin{equation}
\frac{\partial\ln{R_\mathrm{L}/a}}{\partial\ln{q}}=\frac{2}{3}-\frac{q^{1/3}}{3}\frac{1.2q^{1/3}+1/(1+q^{1/3}) }{0.6q^{2/3}+\ln{(1+q^{1/3})}},
\label{eq: eq. 11}
\end{equation}

\ \\
\noindent
The other part represents the response of the mass ratio to the change in donor mass, where $\epsilon$ corresponds to the accreted fraction:

\begin{equation}
{\frac{\partial \ln {q}}{\partial \ln{M_\mathrm{d}}}}=1+\epsilon\frac{M_\mathrm{d}}{M_\mathrm{g}}.
\label{eq: eq. 12}
\end{equation}
\\

\begin{figure}
	\begin{center}
		\includegraphics[trim=0.0cm 0.0cm 0.0cm 0.0cm,clip,width=0.4\textwidth,angle=0]{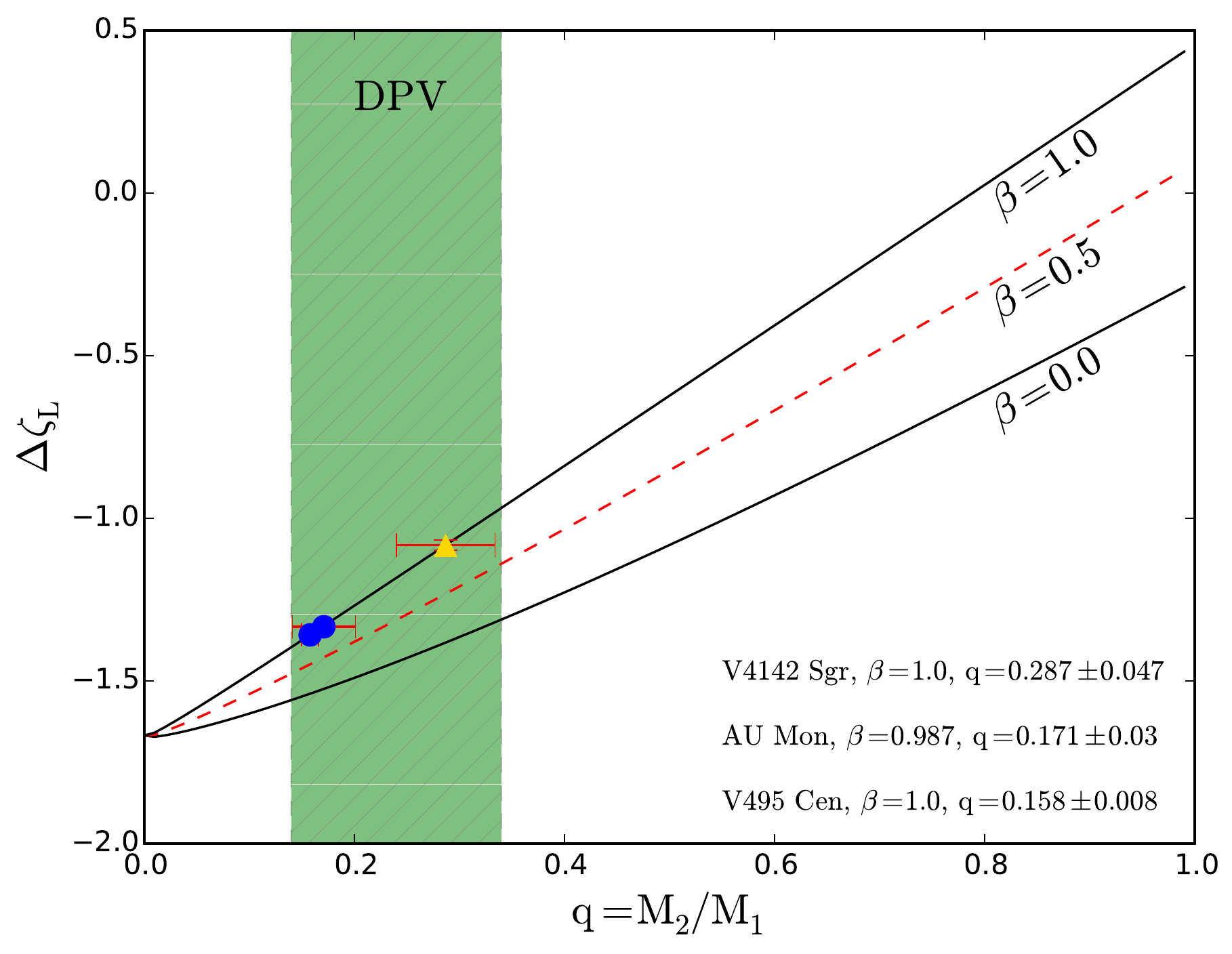}
	\end{center}
	\caption{Roche-lobe response for dynamical stability of mass transfer. The donor star is able to recover its hydrodynamic equilibrium while remaining within its Roche lobe with a mass ratio $q= 0.287 \pm 0.047$ to different accretion efficiency values from $\epsilon =0.0$ to 1.0 (Yellow triangle). Blue dots represent other well-studied DPVs. The dashed green rectangle spans from the minimum to maximum mass ratio of the DPVs known to date.}
	\label{fig:Fig. 19}
\end{figure}

\noindent
Thus, for an accretion efficiency value $\epsilon=1$, the donor star is able to recover its hydrodynamic equilibrium, while remaining within its Roche lobe with a mass ratio $q = 0.287 \pm 0.047$, i.e. V4142\,Sgr is inside to the zone of dynamical-stability of mass-transfer, see Fig.\,\ref{fig:Fig. 19}.

\subsection{Doppler Tomography}
\label{Sec: Sec. 3.x}

\ \\
\noindent
We performed a Doppler tomography (DT) to reconstruct the velocity space of the system and study the structure of the accretion disk and the possible circumstellar material. Doppler tomography has been a widely used method in the field of binary systems \citep{1988MNRAS.235..269M} and numerous studies using the DT method to date have revealed features of the accretion disk, stream impact region, etc.

For that references, the \texttt{DTTVM} code \citep{2015PASJ...67...22U}\footnote{https://home.hiroshima-u.ac.jp/uemuram/dttvm/} was used, which allows to build Doppler tomograms for the H$\alpha$ emission line. Briefly, the method reconstructs a Doppler map, i.e., the intensity map in the velocity space ($v_{x}$,$v_{y}$), based on the model of DT \citep{1988MNRAS.235..269M}, where the orbital phase is computed from observation ephemeris of the binary system and assumes that the noise follows a symmetric probability distribution, like a Gaussian distribution, and that all observations are yielded with the same variance. For further detail, the code is described by \citet{2015PASJ...67...22U}

\begin{figure*}[h!]
	\begin{center}
		\includegraphics[trim=0.0cm 0.0cm 0.0cm 0.0cm,clip,width=0.4\textwidth,angle=0]{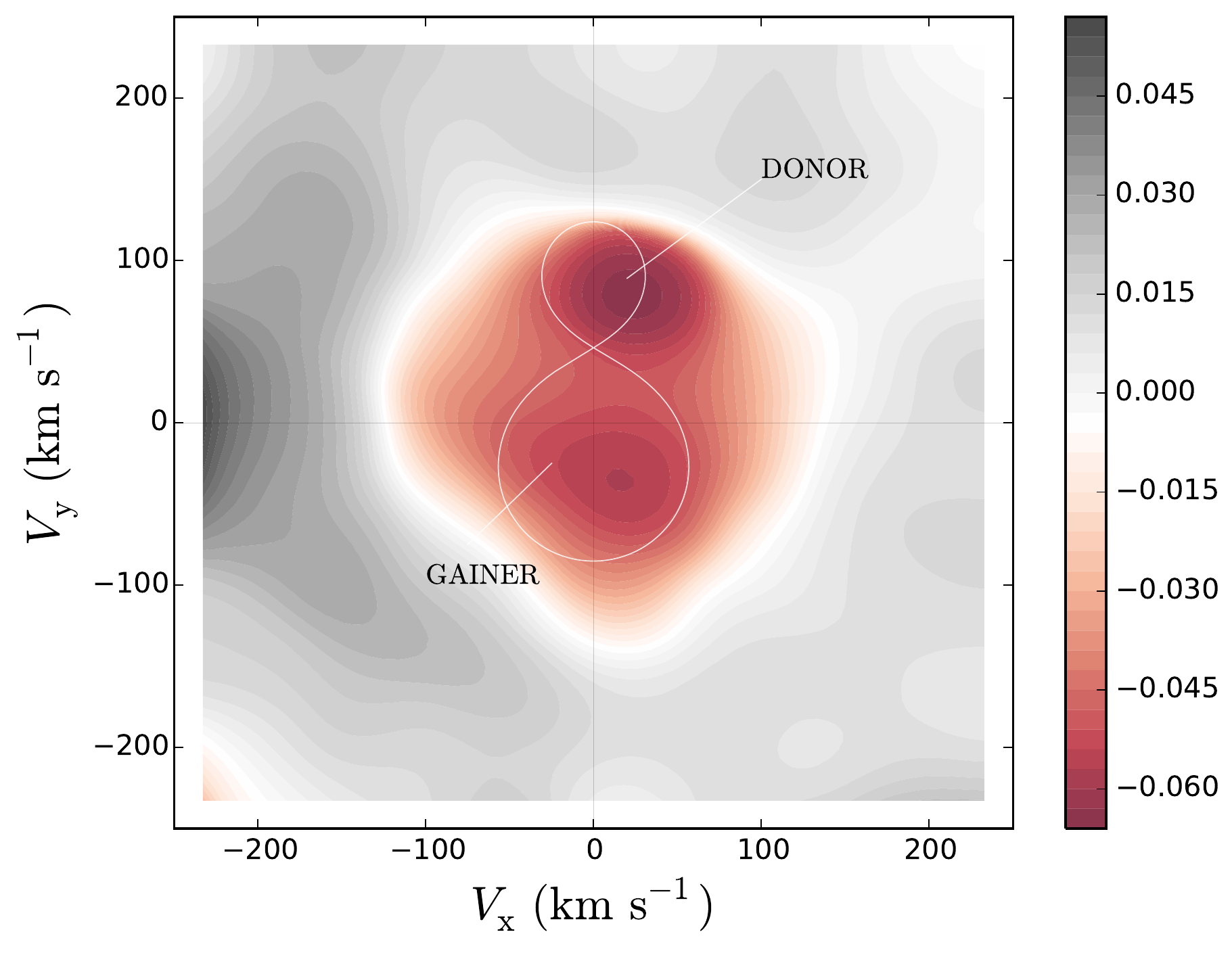}
		\includegraphics[trim=0.0cm 0.0cm 0.0cm 0.0cm,clip,width=0.4\textwidth,angle=0]{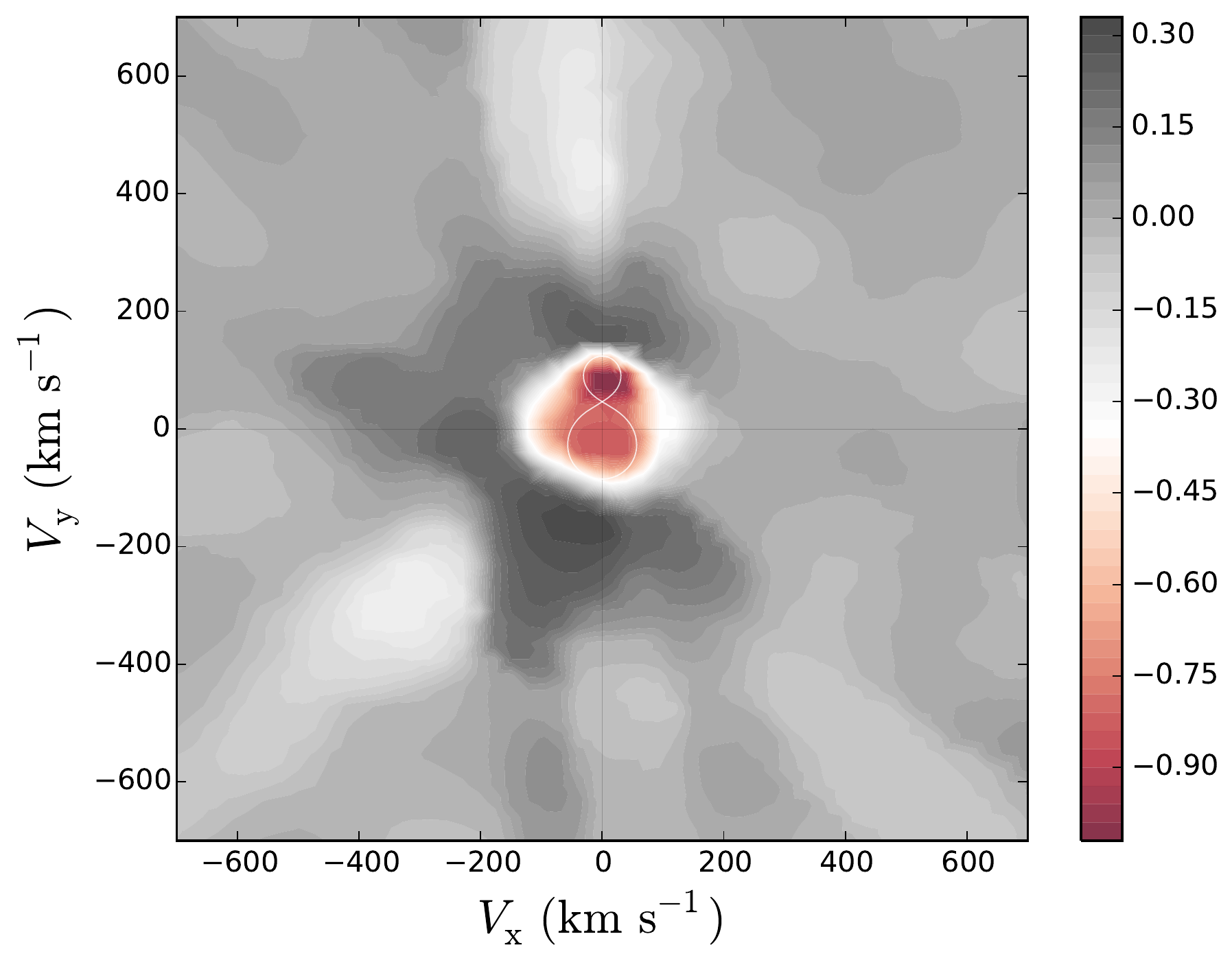}
		\includegraphics[trim=0.0cm 0.0cm 0.0cm 0.0cm,clip,width=0.4\textwidth,angle=0]{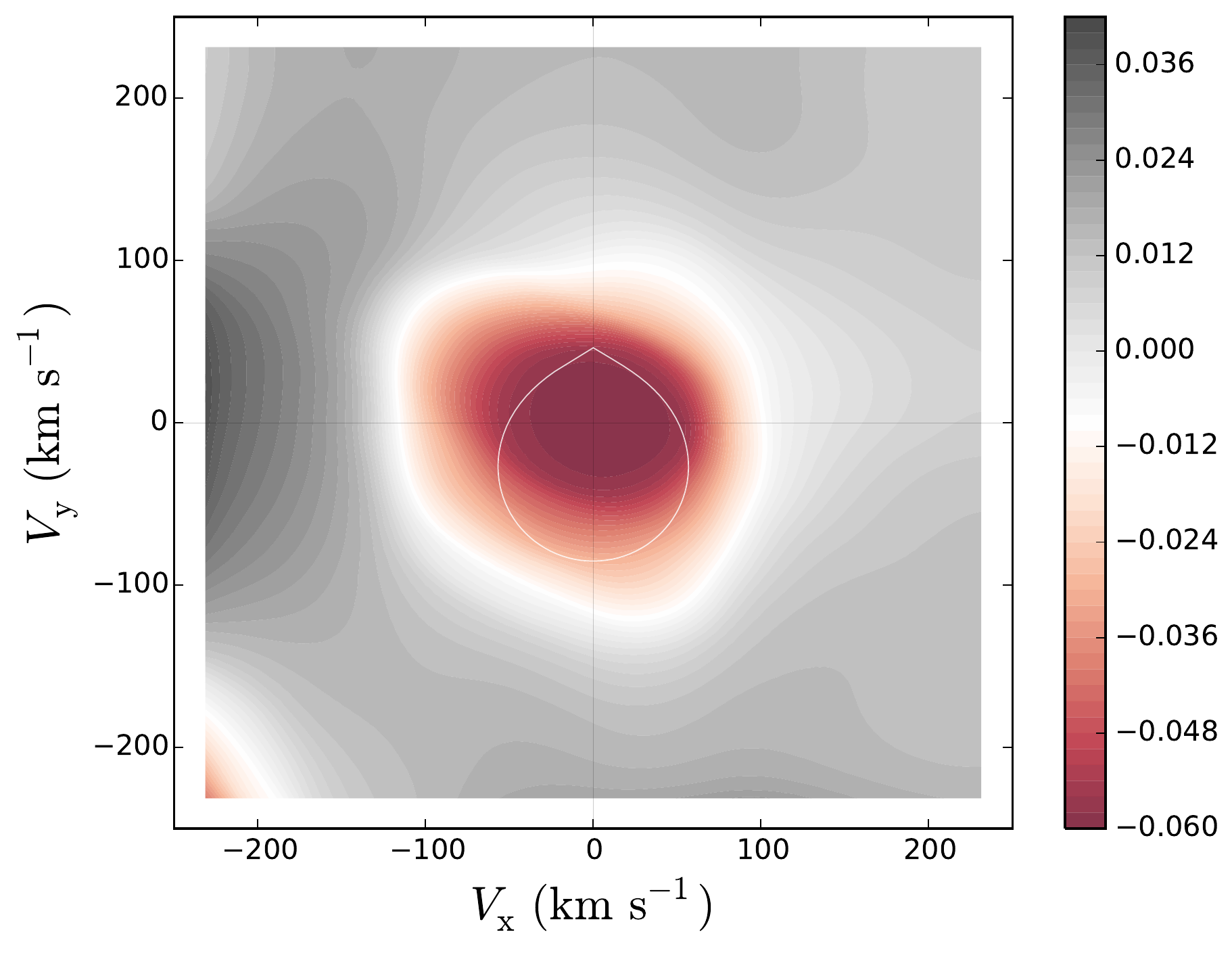}
		\includegraphics[trim=0.0cm 0.0cm 0.0cm 0.0cm,clip,width=0.4\textwidth,angle=0]{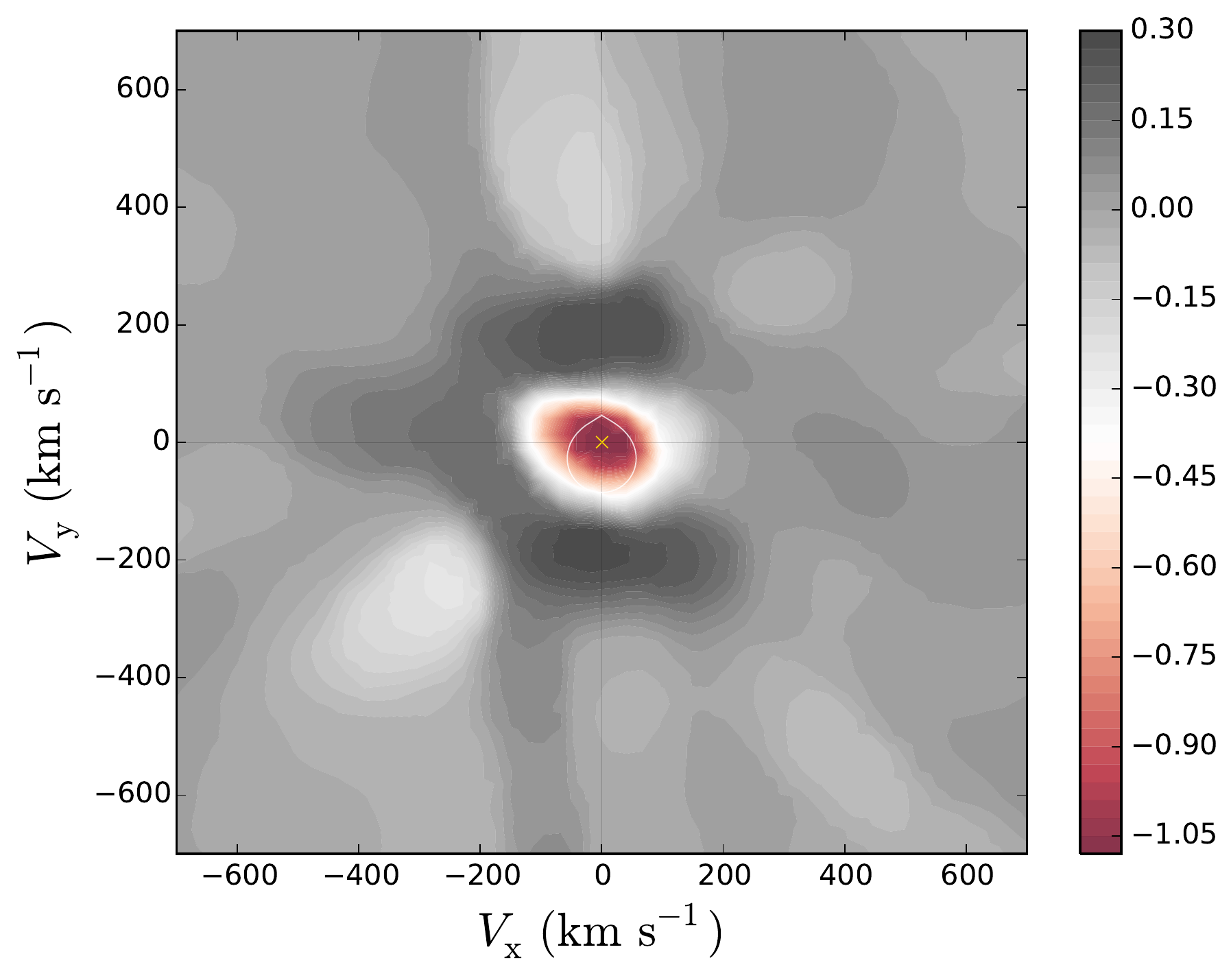}
	\end{center}
	\caption{(Top-left) Doppler map for V4142\,Sgr which shows the gainer almost at the center of mass together with its companion. (Top-right) The structure of the accretion disk shows slight/wide absorption zones. The contour lines represent levels of emission. (Bottom) Doppler map without donor contribution, showing diverse dense zones of absorption. The set of panels were calculated with different $\lambda$. Red zones indicate absorption and gray/black zones emission in the H$\alpha$ line.}
	\label{fig:Fig. 20}
\end{figure*}

The method considers different parameters such as the number of bins on side of the Doppler map $\mathrm{nbin}$, with a total number of $N= \mathrm{nbin} \times \mathrm{nbin}$, the resolution's velocity of the input spectra ($\mathrm{km\,s^{-1}}$) $\mathrm{resol}$, $\lambda$ is the hyper-parameter of the model, i.e. it is a weight for the total variation term, $\gamma$ is the system's radial velocity ($\mathrm{km\,s^{-1}}$) and finally the $\mathrm{thresh}$ is a step to terminate the iteration. Thus, the tomograms were produced using the previous mentioned parameters of the following way; $\mathrm{nbin}=90$, $\mathrm{resol}=2.96 ~\mathrm{km\,s^{-1}}$, $\gamma=0.2 ~\mathrm{km\,s^{-1}}$, and $\mathrm{thresh}=1E-5 ~\mathrm{km\,s^{-1}}$, and different $\lambda$ for 2 configurations close-up and far-off:

\ \\
\begin{itemize}
	\item $\lambda_{1}=2.33572$: close-up donor/gainer/accretion disk (up-left)
	\item $\lambda_{2}=2.33572$: close-up gainer/accretion disk (down-left)
	\item $\lambda_{3}=0.07848$: far-off donor+gainer+accretion disk (up-right)
	\item $\lambda_{4}=0.04833$: far-off gainer/accretion disk (down-right)
\end{itemize}

\ \\
\noindent
The implemented parameters for the Doppler converge successfully and show maps which suggests an optical accretion disk, consistent with the scenario of mass transfer. However, we noted that the Balmer line emission has horseshoe-shaped. In addition, the accretion disk shows inhomogeneous zones of low and high intensities, and two bright structures are detected at the second and third quadrant (Fig.\,\ref{fig:Fig. 20} up-right). On the other hand, for the Doppler map without donor contribution the same more defined structure was observed with this horseshoe-shaped and their bright structures are smaller and located between quadrants I-II and the following III-IV.

%%%%%%%%%%%%%%%%%%%%%%%%%%%%%%%%%%%%%%%%%%%%%%%%%%%%%%%%%%%%%%%%%%%%%%%%%%%%%%%%%%%%%%%%%%%%%%%%%%%%%%%%%%%%%%%%%%%%%%%%%%%%%%%%%%%%%%%%%%%%%%%%%%%%%%%%%%%%%%
%%%%%%%%%%%%%%%%%%%%%%%%%%%%%%%%%%%%%%%%%%%%%%%%%%%%%%%%%%%%%%%%%%%%%%%%%%%%%%%%%%%%%%%%%%%%%%%%%%%%%%%%%%%%%%%%%%%%%%%%%%%%%%%%%%%%%%%%%%%%%%%%%%%%%%%%%%%%%%
%%%%%%%%%%%%%%%%%%%%%%%%%%%%%%%%%%%%%%%%%%%%%%%%%%%%%%%%%%%%%%%%%%%%%%%%%%%%%%%%%%%%%%%%%%%%%%%%%%%%%%%%%%%%%%%%%%%%%%%%%%%%%%%%%%%%%%%%%%%%%%%%%%%%%%%%%%%%%%
%%%%%%%%%%%%%%%%%%%%%%%%%%%%%%%%%%%%%%%%%%%%%%%%%%%%%%%%%%%%%%%%%%%%%%%%%%%%%%%%%%%%%%%%%%%%%%%%%%%%%%%%%%%%%%%%%%%%%%%%%%%%%%%%%%%%%%%%%%%%%%%%%%%%%%%%%%%%%%
\section{Light curve model and system parameters}
\label{Sec: Sec. 4}

\subsection{The fitting procedure}
\label{Sec: Sec. 4.1}

From the spectroscopic analysis using the temperature indicator of the gainer star, it can be confirmed that the accretion disk is creating its own atmosphere surrounding its accretor, hiding it, and obstructing knowing their physical parameters such as temperature, radius, surface gravity, and others. To determine and characterize the DPV V4142\,Sgr, it is necessary to determine the fundamental parameters of each stellar component and the accretion disk. Therefore, a fit in V-band light curve was performed, using an algorithm developed by \citet{1992Ap&SS.197...17D,1996Ap&SS.243..413D}

The current algorithm is capable to solve the inverse problem for the disentangled orbital light curve, of close binary systems with a hot star surrounded by an accretion disk and incorporating a bright spot located on the edge of this, and a hot-spot placed where occurs the impact of the gas stream. This active region or hot-spot is described by the ratio of the hot-spot temperature and the unperturbed local temperature of the disk \citep{2003ARep...47..809B}, their angular/longitude dimension are measured in arc degree. The algorithm, provide the stellar and disk parameters, based on inverse problem solving, i.e., this method consists in the determination of the optimal values yielding the best fit between the observed light curve and the theoretical one, resolving an iterative cycle of corrections to the model elements based on the Nelder-Mead simplex algorithm \citep{1992nrfa.book.....P} with optimizations described by \citet{1991SIAM J.Optim..1..448}, which depends on some determined parameters previously in previous sections. In addition, the model uses the results of hydrodynamical simulation of gas dynamics for interacting close binary systems by \citet{1998MNRAS.300...39B,1999ARep...43..797B,2003ARep...47..809B} and has been tested in several studies \citep{2013MNRAS.428.1594G,2015MNRAS.448.1137M,2018MNRAS.476.3039R}. 

To obtain reliable estimates of the system we have decided restrict some free parameters fixing some of them, obtained previously in the above sections. We fixed the mass ratio $q=0.287$ and the temperature of the donor star to $T_\mathrm{d}= 4500 ~\mathrm{K}$ based on the spectroscopic analysis. Gravity-darkening of the donor $\beta_\mathrm{d}=0.08$ and gainer $\beta_\mathrm{d}=0.25$ was set, and the albedo of both components $A_\mathrm{d}=0.5$ and $A_\mathrm{g}=1.0$, according to von Zeipel\textquoteright{s} law \citep{1924MNRAS..84..702V} for radiative envelopes and  complete re-radiation \citep{1980MNRAS.193...79R}, while the limb-darkening for each component were computed by the way described by \citet{2010MNRAS.409..329D}.

In addition, the rotation of the donor as synchronous ($f_\mathrm{d}=1.00$) was also considered, i.e. the parameter $f_{\rm d} = {\rm P_{orb}/P_{d}} = 1$ means that the donor rotates synchronously with the orbital period. If $f_{\rm d}$ is different from 1, the rotation is non-synchronous. However, the present study is not capable to determine spectroscopically a synchronous or non-synchronous state for the gainer star, because it is hidden by the accretion disk, and their absorption lines are produced in the disk more than in the gainer. Despite that, the analysis of light curves allows to estimate the equivalent radius of the gainer, i.e. the radius that a spherical star of the same volume would have if a rotationally deformed gainer, so in this way the speed of rotation of the gainer can be indirectly estimated, at which the estimated radius is obtained. In other words, $f_{\rm h}$ can in some cases be treated as a model free parameter in solving the inverse problem of light curves analysis. This is exactly how $f_{\rm h}$ was estimated $f_{\rm h}= 23.8 \pm 1.1$ in Tab.\,\ref{Tab: Tab. 6} where $f_{\rm h}$ is the non-synchronous rotation coefficient of the more massive gainer.

%%%%%%%%%%%%%%%%%%%%%%%%%%%%%%%%%%%%%%%%%%%%%%%%%%%%%%%%%%%%%%%%%%%%%%%%%%%%%%%%%%%%%%%%%%%%%%%%%%%%%%%%%%%%%%%%%%%%%%%%%%%%%%%%%%%%%%%%%%%%%%%%%%%%%%%%%%%%%%
%%%%%%%%%%%%%%%%%%%%%%%%%%%%%%%%%%%%%%%%%%%%%%%%%%%%%%%%%%%%%%%%%%%%%%%%%%%%%%%%%%%%%%%%%%%%%%%%%%%%%%%%%%%%%%%%%%%%%%%%%%%%%%%%%%%%%%%%%%%%%%%%%%%%%%%%%%%%%%
%%%%%%%%%%%%%%%%%%%%%%%%%%%%%%%%%%%%%%%%%%%%%%%%%%%%%%%%%%%%%%%%%%%%%%%%%%%%%%%%%%%%%%%%%%%%%%%%%%%%%%%%%%%%%%%%%%%%%%%%%%%%%%%%%%%%%%%%%%%%%%%%%%%%%%%%%%%%%%
%%%%%%%%%%%%%%%%%%%%%%%%%%%%%%%%%%%%%%%%%%%%%%%%%%%%%%%%%%%%%%%%%%%%%%%%%%%%%%%%%%%%%%%%%%%%%%%%%%%%%%%%%%%%%%%%%%%%%%%%%%%%%%%%%%%%%%%%%%%%%%%%%%%%%%%%%%%%%%
\subsection{The best light-curve model}
\label{Sec: Sec. 4.2}

The described code in the above subsection (Sec.\,\ref{Sec: Sec. 4.1}) was used, computing the best fit, O-C residuals, individual flux contributions of the donor, gainer and disk of V4142\,Sgr and included the view of the optimal model at orbital phases 0.17, 0.45, 0.80, and 0.93. From residuals, we notice a scatter around $\Delta{V}\sim 0.2 ~\mathrm{(mag)}$ without dependencies related to orbital or long cycle phases. The best-fitting model contains an optically and geometrically thick accretion disk surrounding a non-synchronous gainer star with the presence of a bright and hot spot (see Table\,\ref{Tab: Tab. 6}). The accretion disk is assumed in physical contact with its accretor and it is characterized by its radius ${\cal R}_\mathrm{d}$, with a temperature distribution along the radius of the disk:

\begin{equation}
T(r)={\mathrm{T_{d}}}\left(\frac{\cal R_\mathrm{d}}{r} \right)^{a_\mathrm{T}}
\label{eq: eq. 13}
\end{equation}

\ \\
\noindent
where $\mathrm{T_{d}}$ is the disk temperature at its outer edge ($r={\cal R}_\mathrm{d}$) and $a_\mathrm{T}$ is the temperature exponent. The last term represents how close is the radial temperature profile to steady-state configuration. The best model shows a system under inclination $81.\!\!^{\circ}5 \pm 0.\!\!^{\circ}3$, with a cold and evolved donor star of mass $M_\mathrm{d}= 1.11 \pm 0.20 ~\mathrm{M_{\odot}}$, temperature $T_\mathrm{d}= 4500 \pm 125 ~\mathrm{K}$, and a radius $R_\mathrm{d}=19.4 \pm 0.2 ~\mathrm{M_{\odot}}$, and a surface gravity {$\log{g_\mathrm{d}}= 1.90 \pm 0.05$} dex
. In addition, the model shows a hot and rejuvenated companion (gainer) of $M_\mathrm{g}= 3.86 \pm 0.3 ~\mathrm{M_{\odot}}$, $T_\mathrm{g}=14380 \pm 700 ~\mathrm{K}$ corresponding to a BV6-type star. The radius is $R_\mathrm{g}= 6.35 \pm 0.2 ~\mathrm{R_{\odot}}$ and the surface gravity  $\log{g_\mathrm{g}}= 3.42 \pm 0.05$ dex. The model shows that the gainer is surrounded by an optically and geometrically thick accretion disk of radial extension ${\cal R}_\mathrm{d}= 22.8 \pm 0.3 ~\mathrm{R_{\odot}}$, contributing $\sim 1.4$ percent of the total luminosity of the system at the $V$-band at orbital phase 0.25. 

\begin{figure}
	\begin{center}	
		\includegraphics[trim=0.1cm 0.0cm 0.0cm 0.0cm,clip,width=0.38\textwidth,angle=0]{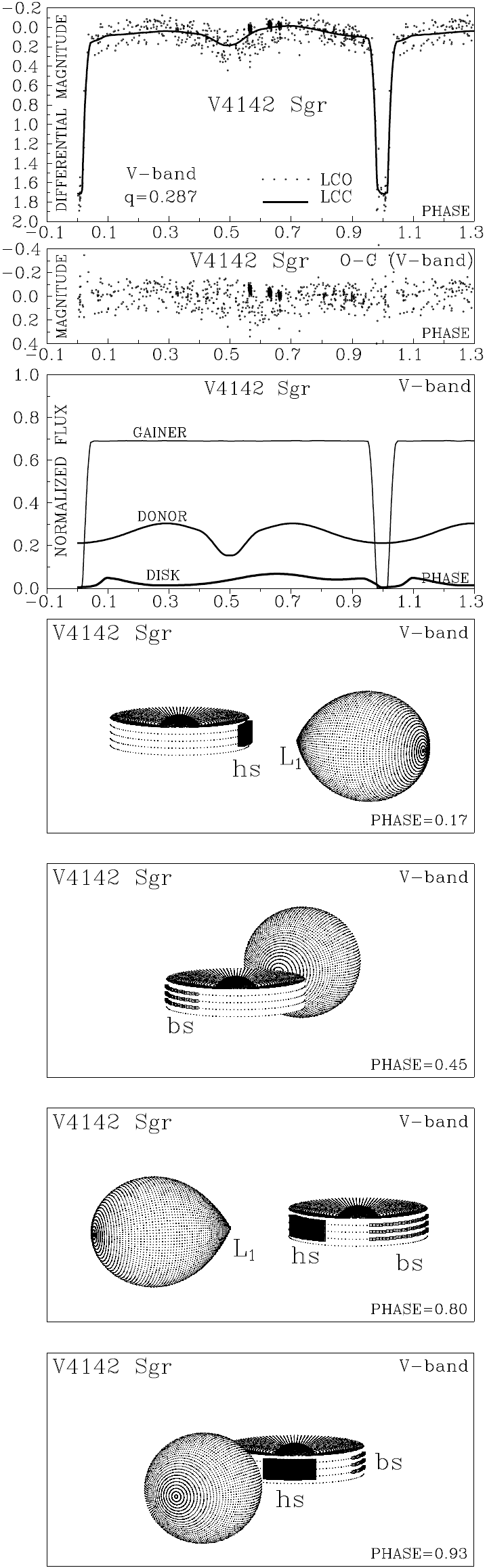}
	\end{center}
	\caption{Observed (LCO), synthetic (LCC) light-curves and the final O-C residuals between the observed and synthetic light curves of the V4142 Sgr; fluxes of gainer, donor, and the accretion disk, normalized to the total system flux at phase 0.25; the views of the model at orbital phases 0.17, 0.45, 0.80 and 0.93, obtained with parameters estimated by light curve analysis.}
	\label{fig:Fig. 21}	
\end{figure}

The disk is characterized by a hot-spot with an angular dimension of $19.\!\!^{\circ}7 \pm 2.\!\!^{\circ}6$, situated at longitude $\lambda_\mathrm{hs}= 339.\!\!^{\circ}0 \pm 6.\!\!^{\circ}2$ roughly placed where the stream hits the disk and an additional bright-spot separated $102.\!\!^{\circ}5 \pm 0.\!\!^{\circ}04$ degree apart along the disk edge rim in the direction of the orbital motion, with a temperature $T_\mathrm{hs}= 4440 \pm 235 ~\mathrm{K}$ which it is 43 percent higher than the disk edge temperature (see Fig.\,\ref{fig:Fig. 21}). The bright spot should be located beyond the disk and is due to the interaction between the envelope and the stream, according to \citet{2003ARep...47..809B}. i.e., it is a product of the interaction of the circumdisk halo and the stream, dubbed as \emph{hotline}. It is located at longitude $\lambda_\mathrm{bs}= 236.\!\!^{\circ}5 \pm 9.\!\!^{\circ}1$ degree, an angular dimension of $41.\!\!^{\circ}4 \pm 8.\!\!^{\circ}0$, and a temperature $T_\mathrm{bs}=4316 \pm 232 ~\mathrm{K}$.

\begin{table}
	\caption{Results of the analysis of {V4142 Sgr} ASAS V-band light-curve obtained by solving the inverse problem for the Roche model with an accretion disk around the more-massive (hotter) gainer in the critical non-synchronous rotation regime.}
	\label{Tab: Tab. 6}
	\[
	\begin{array}{llll}
	\hline
	\noalign{\smallskip}
	
	{\rm Quantity} & & {\rm Quantity} \\
	\noalign{\smallskip}
	\hline
	\noalign{\smallskip}
	n                                  & 806             & \cal M_{\rm_h} {[\cal M_{\odot}]} & 3.86  \pm 0.30 \\
	{\rm \Sigma(O-C)^2}                & 5.3250          & \cal M_{\rm_c} {[\cal M_{\odot}]} & 1.11  \pm 0.20 \\
	{\rm \sigma_{rms}}                 & 0.0813          & \cal R_{\rm_h} {\rm [R_{\odot}]}  & 6.35  \pm 0.20 \\
	i {\rm [^{\circ}]}                 & 81.5  \pm 0.3   & \cal R_{\rm_c} {\rm [R_{\odot}]}  & 19.4  \pm 0.2 \\
	{\rm F_d}                          & 0.656 \pm 0.03  & {\rm log} \ g_{\rm_h}             & 3.42  \pm 0.05\\
	{\rm T_d} [{\rm K}]                & 3105  \pm 100   & {\rm log} \ g_{\rm_c}             & 1.90  \pm 0.05\\
	{\rm d_e} [a_{\rm orb}]            & 0.132 \pm 0.005 & M^{\rm h}_{\rm bol}               & -3.19 \pm 0.20 \\
	{\rm d_c} [a_{\rm orb}]            & 0.031 \pm 0.005 & M^{\rm c}_{\rm bol}               & -0.57 \pm 0.10 \\
	{\rm a_T}                          & 0.46  \pm 0.05  & a_{\rm orb}  {\rm [R_{\odot}]}    & 70.2  \pm 0.3 \\
	{\rm f_h}                          & 23.8  \pm 1.1   & \cal{R}_{\rm d} {\rm [R_{\odot}]} & 22.8  \pm 0.3 \\
	{\rm F_h}                          & 1.000           & \rm{d_e}  {\rm [R_{\odot}]}       & 9.3   \pm 0.2 \\
	{\rm T_h} [{\rm K}]                & 14380 \pm 700   & \rm{d_c}  {\rm [R_{\odot}]}       & 2.2   \pm 0.2 \\
	{\rm A_{hs}=T_{hs}/T_d}            & 1.43  \pm 0.06  &                                                   \\
	{\rm \theta_{hs}}{\rm [^{\circ}]}  & 19.7  \pm 2.6   &                                                   \\
	{\rm \lambda_{hs}}{\rm [^{\circ}]} & 339.0 \pm 6.2   &                                                   \\
	{\rm \theta_{rad}}{\rm [^{\circ}]} & 28.4  \pm 5.5   &                                                   \\
	{\rm A_{bs}=T_{bs}/T_d}            & 1.39  \pm 0.06  &                                                   \\
	{\rm \theta_{bs}}{\rm [^{\circ}]}  & 41.4  \pm 8.0   &                                                   \\
	{\rm \lambda_{bs}}{\rm [^{\circ}]} & 236.5 \pm 9.1   &                                                   \\
	{\Omega_{\rm h}}                   & 13.789\pm 0.030  &                                                   \\
	{\Omega_{\rm c}}                   & 2.437 \pm 0.020  &                                                   \\
	\noalign{\smallskip}
	\hline
	\end{array}
	\]

	FIXED PARAMETERS: $q={\cal M}_{\rm c}/{\cal M}_{\rm h}=0.287$ - mass ratio of the components, ${\rm T_c=4500 K}$  - temperature of the less-massive (cooler)
	donor, ${\rm F_c}=1.0$ - filling factor for the critical Roche lobe of the donor, $f{\rm _{c}}=1.00$ - non-synchronous rotation coefficients of the donor, ${\rm F_h}=R_h/R_{zc}$ - filling factor for the critical non-synchronous lobe of the hotter, more-massive gainer (ratio of the stellar polar radius to the critical Roche lobe radius along z-axis for a star in critical non-synchronous rotation regime), ${\rm \beta_h=0.25}$, ${\rm \beta_c=0.08}$ - gravity-darkening coefficients of the components, ${\rm A_h=1.0}$, ${\rm A_c=0.5}$  - albedo coefficients of the components.
	
	\smallskip \noindent Note: $n$ - number of observations, ${\rm \Sigma (O-C)^2}$ - final sum of squares of residuals between observed (LCO) and synthetic (LCC) light-curves, ${\rm \sigma_{rms}}$ - root-mean-square of the residuals, $i$ - orbit inclination (in arc degrees), ${\rm F_d=R_d/R_{yc}}$ - disk dimension factor (the ratio of the disk radius to the critical Roche lobe radius along y-axis), ${\rm T_d}$ - disk-edge temperature, $\rm{d_e}$, $\rm{d_c}$, - disk thicknesses (at the edge and at the center of the disk, respectively) in the units of the distance between the components, $a_{\rm T}$ - alpha disk temperature distribution coefficient, $f{\rm _h}$ - non-synchronous rotation coefficient of the more massive gainer (in the critical non-synchronous rotation regime), ${\rm T_h}$ - temperature of the gainer, ${\rm A_{hs,bs}=T_{hs,bs}/T_d}$ - hot spot temperature coefficients, ${\rm \theta_{hs,bs}}$ and ${\rm \lambda_{hs,bs}}$ - spot angular dimension and longitude (in arc degrees), ${\rm \theta_{rad}}$ - angle between the line perpendicular to the local disk edge surface and the direction of the hot-spot maximum radiation, ${\Omega_{\rm h,c}}$ - dimensionless surface potentials of the hotter gainer and cooler donor, $\cal M_{\rm_{h,c}} {[\cal M_{\odot}]}$, $\cal R_{\rm_{h,c}} {\rm [R_{\odot}]}$ - stellar masses and mean radii of stars in solar units, ${\rm log} \ g_{\rm_{h,c}}$ - logarithm (base 10) of the system components effective gravity, $M^{\rm {h,c}}_{\rm bol}$ - absolute stellar bolometric magnitudes, $a_{\rm orb}$ ${\rm [R_{\odot}]}$, $\cal{R}_{\rm d} {\rm [R_{\odot}]}$, $\rm{d_e} {\rm [R_{\odot}]}$, $\rm{d_c} {\rm [R_{\odot}]}$ - orbital semi-major axis, disk radius and disk thicknesses at its edge and center, respectively, given in solar units.
\end{table}

\begin{figure*}
	\begin{center}	
		\includegraphics[trim=0.1cm 0.0cm 0.0cm 0.0cm,clip,width=0.9\textwidth,angle=0]{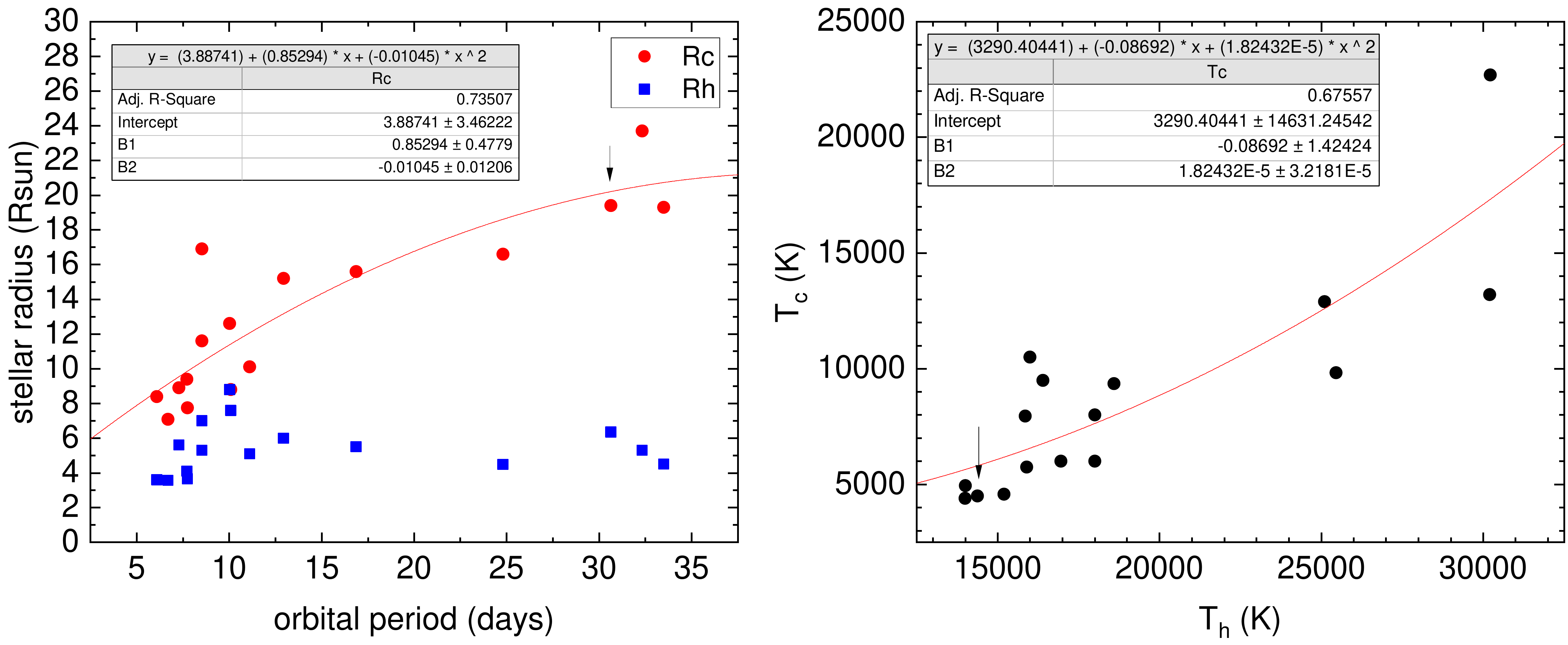}
	\end{center}
	\caption{Left: Stellar radius versus orbital period for DPVs. The arrow indicates the position of V\,4142\,Sgr and the solid line the best 2nd order polinomial fit for the donor star data. Right: Temperatures for the hot and cool stellar components of DPVs. The arrow indicate the position of V\,4142\,Sgr and the solid line is the best 2nd order polinomial fit to the data. }
	\label{fig:Fig. 22}	
\end{figure*}

%%%%%%%%%%%%%%%%%%%%%%%%%%%%%%%%%%%%%%%%%%%%%%%%%%%%%%%%%%%%%%%%%%%%%%%%%%%%%%%%%%%%%%%%%%%%%%%%%%%%%%%%%%%%%%%%%%%%%%%%%%%%%%%%%%%%%%%%%%%%%%%%%%%%%%%%%%%%%%
%%%%%%%%%%%%%%%%%%%%%%%%%%%%%%%%%%%%%%%%%%%%%%%%%%%%%%%%%%%%%%%%%%%%%%%%%%%%%%%%%%%%%%%%%%%%%%%%%%%%%%%%%%%%%%%%%%%%%%%%%%%%%%%%%%%%%%%%%%%%%%%%%%%%%%%%%%%%%%
%%%%%%%%%%%%%%%%%%%%%%%%%%%%%%%%%%%%%%%%%%%%%%%%%%%%%%%%%%%%%%%%%%%%%%%%%%%%%%%%%%%%%%%%%%%%%%%%%%%%%%%%%%%%%%%%%%%%%%%%%%%%%%%%%%%%%%%%%%%%%%%%%%%%%%%%%%%%%%
%%%%%%%%%%%%%%%%%%%%%%%%%%%%%%%%%%%%%%%%%%%%%%%%%%%%%%%%%%%%%%%%%%%%%%%%%%%%%%%%%%%%%%%%%%%%%%%%%%%%%%%%%%%%%%%%%%%%%%%%%%%%%%%%%%%%%%%%%%%%%%%%%%%%%%%%%%%%%%
\section{Discussion}
\label{Sec: Sec. 5}

We find that  V4142\,Sgr is a relatively typical DPV in some aspects, consisting of a B-type gainer surrounded by an accretion disk that receive matter from a K-type giant filling its Roche lobe. However, its relatively long orbital period of 30\fd6 make it interesting, since not many of these long period DPVs have been studied spectroscopically. In addition, some interesting features appear, like the presence of a wind and the thick disk partially hidding the gainer. In spite of this occultation, the gainer is so bright that the main eclipse is much deeper than the secondary one. A wind or general mass loss was detected in the DPVs OGLE 05155332-6925581 (P$_{\rm{o}}$ = 7\fd3), V393\,Sco (P$_{\rm{o}}$ = 7\fd7), AU\,Mon (P$_{\rm{o}}$ = 11\fd1),  $\beta$\,Lyr (P$_{\rm{o}}$ = 12\fd9), HD\,170582 (P$_{\rm{o}}$ = 16\fd9) and V495\,Cen (P$_{\rm{o}}$ = 33\fd5). Theoretical studies suggest that a radiatively supported mass outflow or wind could be formed in the hotspot region, especially in systems with high mass transfer rate \citep{2008A&A...487.1129V}. The changing aspect of the eclipse depth in ASAS and ASAS-SN data suggests a variable  disk structure. The complex structure of the mass flows are reflected in the Doppler tomograms, showing non-homogeneous emissivity distribution for H$\alpha$. The constancy of the orbital period on the other hand, suggests that the mass transfer rate is not so large, otherwise changes in the orbital period should be observed due to the re-distribution of the angular momentum among two bodies of changing masses orbiting their center of mass. The influence of the disk is observed in the continuum and the optical spectral lines. Double and asymetrical Balmer emission lines as well as sharp absorption lines of low ionized atoms, like Ti\,II, Ti\,O and Ti\,I are probably formed in low temperature, optically thin disk regions, not as the continuum light probably formed in the optically thick and denser part of the disk. The $\gamma$ and phase shifts observed in the OI\,8446.35 \AA\ line suggest that the photosphere of the gainer is hidden by the disk and that the absorption lines of the more massive star photosphere are filled with disk emission, difficulting the task of determining the spectral type of the gainer.    

We compiled stellar temperatures and radii for DPVs and show the position of V\,4142\,Sgr\ among their pairs in Fig. \ref{fig:Fig. 22}. While the radius of the DPV gainers remain between 4 and 8 R$_{\odot}$ consistent with B-type dwarfs, the donor radius increases with the orbital period, consisting with giant stars filling their Roche lobes.
The stellar temperatures also show a significant tendency, hotter gainer are accompaigned by hotter donors.  V\,4142\,Sgr\ is located in the cold extreme of the sample. The fits shown in this figure are intended to provide global tendencies in the data:
\begin{equation}
\rm R_{c}= 3.887 + 0.853 P_{o} - 0.01045 P_{o}^2,
\label{eq: eq. 14}
\end{equation}

\noindent
with standard deviation $2.40\,R_{\odot}$ and Adj. R-sqare 0.735 and,
\begin{equation}
\rm T_{c}= 3290.4 - 0.0869 T_{h} + 1.82432 P_{o}^2,
\label{eq: eq. 15}
\end{equation}

\noindent
with standard deviation 2490 K and Adj. R-sqare 0.676. The data used in these fits are from \citet{2016MNRAS.455.1728M} and references therein, and from more recent datasets published by \citet{2021A&A...653A..89M}, \citet{ 2021AJ....162...66R}, \citet{2020A&A...642A.211M}, \citet{2020A&A...641A..91M}, \citet{2019MNRAS.487.4169M} and \citet{2018MNRAS.476.3039R}.

%OGLE05155332-6925581
%AU Mon 

\begin{figure}
	\begin{center}	
		\includegraphics[trim=0.1cm 0.0cm 0.0cm 0.0cm,clip,width=0.45\textwidth,angle=0]{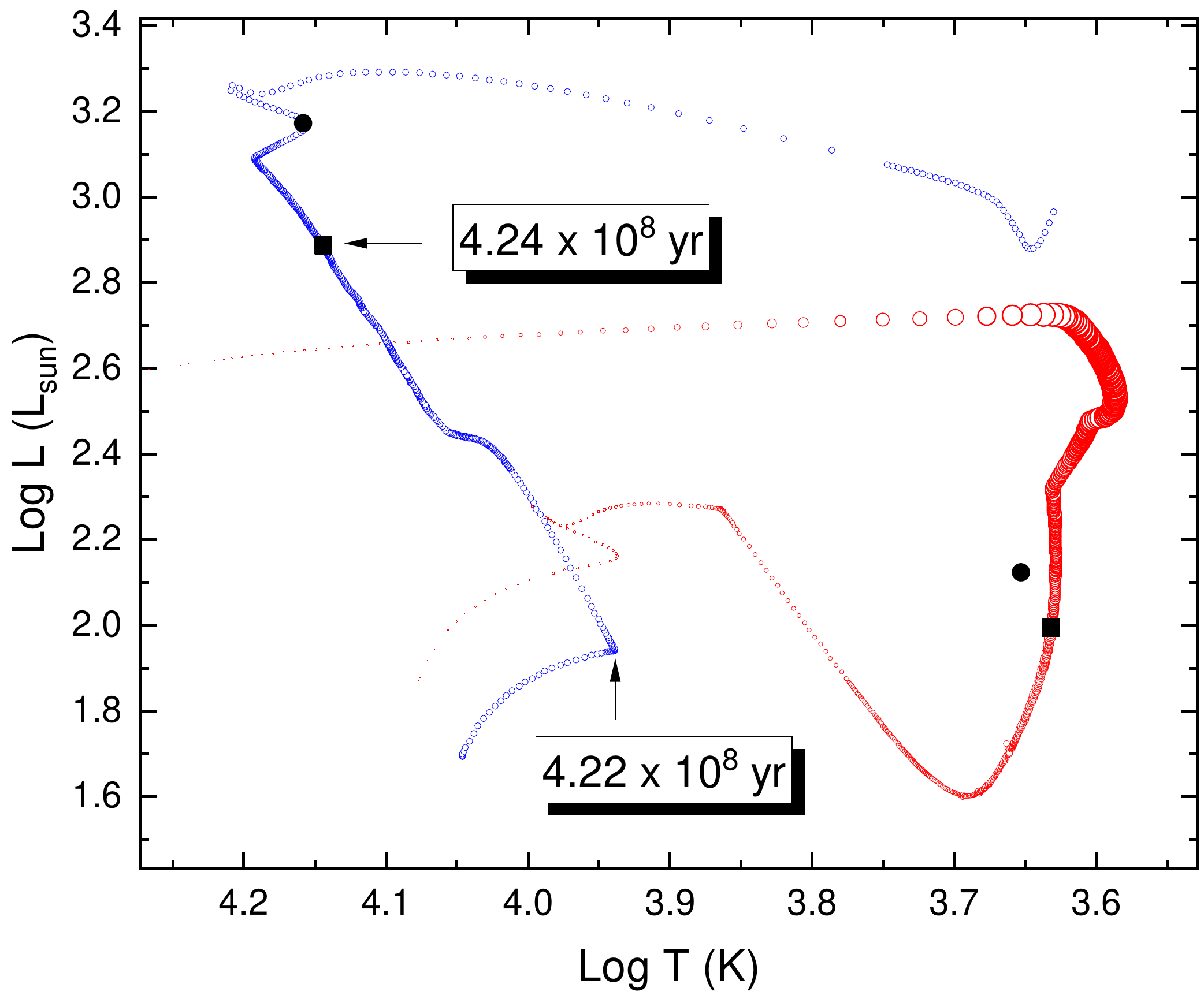}
	\end{center}
	\caption{Evolutionary paths for the donor (red circles) and gainer (blue circles). The system starts mass transfer around 4.22 $\times$ 10$^{8}$ yr and now it is found at an age of 4.24 $\times$ 10$^{8}$ yr. Squares indicate the best model parameters and the dots the parameters derived from observations.   }
	\label{fig:Fig. 23}	
\end{figure}

We searched for the best model and evolutive path  for V\,4142\,Sgr among a grid of binary star evolutionary models \citep{2008yCat..34871129V} and following a $\chi^2$ minimization method among observed and theoretical radii, luminosities, temperatures, masses and orbital period,  considering the errors of these parameters as weights in the process \citep{2012MNRAS.421..862M}. We find that the current position of the system can be reproduced by a binary system with initial period
5 days, initial masses 3 and 2.7 M$_{\odot}$, radii 2.0 and 1.9 R$_{\odot}$ that it is found at an age of 4.24 $\times$ 10$^{8}$ yr. At this age the model consists of a B-type star of temperature 13930 K and mass 4.78 M$_{\odot}$ and a K-type star of temperature 4285 K and mass 0.925 M$_{\odot}$ filling its Roche lobe and transferring mass onto the more massive star at a rate of 1.4 $\times$ 10$^{-7}$ M$_{\odot}$/yr in a conservative way, being the orbital period of the system 30\fd754. This results need to be interpreted with caution, since it is limited because of the resolution of the grid of models, that is rather coarse, and also because the efficience of mass transfer in the system is not well constrained, and is included only in a simplified way in the models. However, some general results can be inferred: (i) the initially less massive star turns to be rejuvenated because of the mass accreted from the donor star and now is the more luminous and massive star of the system (Fig. \ref{fig:Fig. 23}) and (ii) the donor has depleted  of hydrogen its core that consists now of 98\% of helium while the gainer has a core hydrogen fraction of 0.28 and helium fraction of 0.70. The above scenario, especially the hydrogen depleted donor, is consistent with investigations of other DPVs indicating case-B mass exchange systems \citep[][]{2016MNRAS.455.1728M}.

\section{Conclusions}
\label{Sec: Sec. 6}

We have investigated spectroscopically and photometrically the Galactic DPV V4142\,Sgr in detail, and from the photometric analysis, the main eclipsing times spanning 8.7 years were analyzed and an improved orbital period of 30\fd633(2) and a long cycle of $1201 \pm 14 ~\mathrm{days}$ were found. In addition, during the comparison between data from ASAS and ASAS-SN a morphological difference was observed in the shape and amplitude of the light curves during the primary eclipse.

We also from the analysis of the radial velocities found a circular orbit, a mass ratio $q=0.287 \pm 0.047$, a half amplitude  $K_\mathrm{d}= 89.2 \pm 0.5 ~\mathrm{km\,s^{-1}}$ and systemic velocity $\gamma=0.2 \pm 0.77 ~\mathrm{km\,s^{-1}}$. In addition, a semi-amplitude $K_\mathrm{g}=25.56 \pm 1.20 ~\mathrm{km\,s^{-1}}$ was calculated.

The next notable case we found is that the color-color analysis considering $\mathrm{J-H}=0.575 \pm 0.037 ~\mathrm{mag}$ and $H-K=0.192 \pm 0.032 ~\mathrm{mag}$ shows a color excess in JHK photometry, considering as reference a single star of temperature around 5000 K, and this excess could be interpreted as evidence for circumstellar material.

The inverse problem was solved to derive the parameters producing the best match between a theoretical light curve and the observed one. The best model of the V-band ASAS light curve includes a cold and evolved donor star of $M_\mathrm{d}=1.11 \pm 0.2 ~\mathrm{M_{\odot}}$, $T_\mathrm{d}=4500 \pm 125 ~\mathrm{K}$ and $R_\mathrm{d}=19.4 \pm 0.2 ~\mathrm{R_{\odot}}$, and a rejuvenated B type star of $M_\mathrm{g}=3.86 \pm 0.3 ~\mathrm{M_{\odot}}$, $T_\mathrm{g}=14380 \pm 700 ~\mathrm{K}$, and $R_\mathrm{g}= 6.35 \pm 0.2 ~\mathrm{R_{\odot}}$. We determined that the gainer star is surrounded by a concave and geometrically thick disk creating its atmosphere around the main component of a radial extension ${\cal R}_{\mathrm d} = 22.8 \pm 0.3 ~\mathrm{R_{\odot}}$, contributing $\sim 1.4$ percent of the total luminosity of the system at the V-band at orbital phase 0.25. The disk is characterized by a hot spot roughly placed where the stream hits the disk and an additional bright spot separated $102.\!\!^{\circ}5 \pm 0.\!\!^{\circ}04$ degree apart along the disk edge rim in the direction of the orbital motion. In addition, the accretion disk atmosphere is detected in absorption lines Ti\,II 4533.96 \AA, Ti\,O 4548.0 \AA, Ti\,O 4584.0 \AA, and Ti\,I 4731.172 \AA. The double peak H$\alpha$ emission with persistent $V$ $\leq$ $R$ asymmetry might indicate optically thin regions of a wind emerging from the system.

Considering that Doppler maps converged successfully, these suggest an optically thick accretion disk, consistent with the scenario of mass transfer of DPVs. In addition, we noted that the Balmer line emissivity distribution has a horseshoe shape and shows inhomogeneous zones of low and high intensities, and two bright structures are detected in the second and third quadrant, wherein the first one would correspond to the hot spot and the second one to the bright spot detected by the light curve analysis.

Improving our understanding of the evolution, we determined that the system is the result of the evolution of an initially shorter orbital period binary, that inverted its mass ratio as a result of the mass transfer among the initially more massive star onto the initially less massive star. At present, the donor star is an early K-type giant that depleted its core of hydrogen while the companion has gained about 2 M$_{\odot}$ and increased its luminosity by one order of magnitude or more in a process lasting about 2 Myr. These results are consistent with the evolutionary histories of other investigated DPVs \citep{2016MNRAS.455.1728M}.

%\begin{eqnarray}
%(m_\mathrm{d,g}-M_\mathrm{d,g})_{0} & = &  5\log(R_\mathrm{d,g}/R_{\odot})+(m_\mathrm{d,g}-A_\mathrm{V}) \nonumber\\
%& &- M_\mathrm{bol\odot}+ 10\log(T_\mathrm{d,g}/T_{\odot})\nonumber\\
%& & + BC_\mathrm{d,g},
%\label{eq: eq. 17}
%\end{eqnarray}
%\\

%\begin{equation}
%m_\mathrm{d,g}-m_\mathrm{t}=-2.5\log\left(\frac{f_\mathrm{d,g}}{f_\mathrm{t}}\right),
%\label{eq: eq. 18}
%\end{equation}
%\\

%\begin{equation}
%d_\mathrm{d,g} (pc)= 10^{((m_\mathrm{d,g}-M_\mathrm{d,g})_{0}+5)/5},
%\label{eq: eq. 19}
%\end{equation}
%\\

%%%%%%%%%%%%%%%%%%%%%%%%%%%%%%%%%%%%%%%%%%%%%%%%%%%%%%%%%%%%%%%%%%%%%%%%%%%%%%%%%%%%%%%%%%%%%%%%%%%%%%%%%%%%%%%%%%%%%%%%%%%%%%%%%%%%%%%%%%%%%%%%%%%%%%%%%%%%%%
%%%%%%%%%%%%%%%%%%%%%%%%%%%%%%%%%%%%%%%%%%%%%%%%%%%%%%%%%%%%%%%%%%%%%%%%%%%%%%%%%%%%%%%%%%%%%%%%%%%%%%%%%%%%%%%%%%%%%%%%%%%%%%%%%%%%%%%%%%%%%%%%%%%%%%%%%%%%%%
%%%%%%%%%%%%%%%%%%%%%%%%%%%%%%%%%%%%%%%%%%%%%%%%%%%%%%%%%%%%%%%%%%%%%%%%%%%%%%%%%%%%%%%%%%%%%%%%%%%%%%%%%%%%%%%%%%%%%%%%%%%%%%%%%%%%%%%%%%%%%%%%%%%%%%%%%%%%%%
%%%%%%%%%%%%%%%%%%%%%%%%%%%%%%%%%%%%%%%%%%%%%%%%%%%%%%%%%%%%%%%%%%%%%%%%%%%%%%%%%%%%%%%%%%%%%%%%%%%%%%%%%%%%%%%%%%%%%%%%%%%%%%%%%%%%%%%%%%%%%%%%%%%%%%%%%%%%%%
\section{Acknowledgments}

We thank the referee Dr. Mikhail Kovalev for useful comments on the first version of this manuscript. G.D. and J.P.  gratefully acknowledge the financial support of the Ministry of Education, Science and Technological Development of the Republic of Serbia through contract No. 451-03-68/2020-14/200002. R.E.M.  and D.S. gratefully acknowledge support by PFB--06/2007, ANID BASAL projects ACE210002 and FB210003, FONDECYT Regular 1190621 and FONDECYT Regular 1201280.
J.R.G. acknowledges support by BASAL Centro de Astrof{\'{i}}sica y Tecnolog{\'{i}}as Afines (CATA) FB210003. I.A. and M.C. acknowledge support from FONDECYT project Nº 1190485. I. A. is also grateful for the support from FONDECYT project Nº 11190147. This work has made use of data from the European Space Agency (ESA) mission {\it Gaia} (\url{https://www.cosmos.esa.int/gaia}), processed by the {\it Gaia} Data Processing and Analysis Consortium (DPAC, \url{https://www.cosmos.esa.int/web/gaia/dpac/consortium}). Funding for the DPAC has been provided by national institutions, in particular the institutions participating in the {\it Gaia} Multilateral Agreement. This paper includes data collected by the TESS mission, which are publicly available from the Mikulski Archive for Space Telescopes (MAST).

%%%%%%%%%%%%%%%%%%%%%%%%%%%%%%%%%%%%%%%%%%%%%%%%%%%%%%%%%%%%%%%%%%%%%%%%%%%%%%%%%%%%%%%%%%%%%%%%%%%%%%%%%%%%%%%%%%%%%%%%%%%%%%%%%%%%%%%%%%%%%%%%%%%%%%%%%%%%%%
%%%%%%%%%%%%%%%%%%%%%%%%%%%%%%%%%%%%%%%%%%%%%%%%%%%%%%%%%%%%%%%%%%%%%%%%%%%%%%%%%%%%%%%%%%%%%%%%%%%%%%%%%%%%%%%%%%%%%%%%%%%%%%%%%%%%%%%%%%%%%%%%%%%%%%%%%%%%%%
%%%%%%%%%%%%%%%%%%%%%%%%%%%%%%%%%%%%%%%%%%%%%%%%%%%%%%%%%%%%%%%%%%%%%%%%%%%%%%%%%%%%%%%%%%%%%%%%%%%%%%%%%%%%%%%%%%%%%%%%%%%%%%%%%%%%%%%%%%%%%%%%%%%%%%%%%%%%%%
%%%%%%%%%%%%%%%%%%%%%%%%%%%%%%%%%%%%%%%%%%%%%%%%%%%%%%%%%%%%%%%%%%%%%%%%%%%%%%%%%%%%%%%%%%%%%%%%%%%%%%%%%%%%%%%%%%%%%%%%%%%%%%%%%%%%%%%%%%%%%%%%%%%%%%%%%%%%%%

\section*{Data Availability}
The data underlying this article will be shared on reasonable request to the corresponding author.

%%%%%%%%%%%%%%%%%%%%%%%%%%%%%%%%%%%%%%%%%%%%%%%%%%%%%%%%%%%%%%%%%%%%%%%%%%%%%%%%%%%%%%%%%%%%%%%%%%%%%%%%%%%%%%%%%%%%%%%%%%%%%%%%%%%%%%%%%%%%%%%%%%%%%%%%%%%%%%
%%%%%%%%%%%%%%%%%%%%%%%%%%%%%%%%%%%%%%%%%%%%%%%%%%%%%%%%%%%%%%%%%%%%%%%%%%%%%%%%%%%%%%%%%%%%%%%%%%%%%%%%%%%%%%%%%%%%%%%%%%%%%%%%%%%%%%%%%%%%%%%%%%%%%%%%%%%%%%
%%%%%%%%%%%%%%%%%%%%%%%%%%%%%%%%%%%%%%%%%%%%%%%%%%%%%%%%%%%%%%%%%%%%%%%%%%%%%%%%%%%%%%%%%%%%%%%%%%%%%%%%%%%%%%%%%%%%%%%%%%%%%%%%%%%%%%%%%%%%%%%%%%%%%%%%%%%%%%
%%%%%%%%%%%%%%%%%%%%%%%%%%%%%%%%%%%%%%%%%%%%%%%%%%%%%%%%%%%%%%%%%%%%%%%%%%%%%%%%%%%%%%%%%%%%%%%%%%%%%%%%%%%%%%%%%%%%%%%%%%%%%%%%%%%%%%%%%%%%%%%%%%%%%%%%%%%%%%

\end{document}